\documentclass[a4paper,fleqn,usenatbib]{mnras}

\usepackage{newtxtext,newtxmath}

\usepackage{graphicx}	
\usepackage{amsmath}	
\usepackage{xcolor}           
\usepackage{xfrac}
\usepackage{gensymb}
\usepackage{pdflscape}
\usepackage{longtable}
\usepackage{subfig}
\usepackage[utf8]{inputenc} 

\newcommand{\Msun}{M$_{\odot}$}
\newcommand{\Mbh}{$M_{\rm BH}$}
\newcommand{\Mnsc}{$M_{\rm NSC}$}

\newcommand{\Mbulge}{$M_{\rm bulge}$}

\newcommand{\Lsun}{L$_\odot$}

\newcommand{\ml}{$M/L$}

\newcommand{\hst}{\emph{HST}}
\newcommand{\kms}{km~s$^{-1}$}

\mathchardef\mhyphen="2D

\title[MBHBM$^{\star}$ Project -- II.\ Weighing the SMBH in NGC~3593]{The MBHBM$^{\star}$ Project -- II.\ Molecular Gas Kinematics in the Lenticular Galaxy NGC~3593 Reveal a Supermassive Black Hole} 

\author[Dieu D.\ Nguyen et al.]{
Dieu D.\ Nguyen,$^{1,2}$ \thanks{E-mail: nddieuphys@gmail.com}
Martin Bureau,$^{3,4}$ 
Sabine Thater,$^{5}$ 
Kristina Nyland,$^{6}$
Mark den Brok,$^{7}$ 
\newauthor{
Michelle Cappellari,$^{3}$ 
Timothy A.\ Davis,$^{8}$ 
Jenny E.\ Greene,$^{9}$ 
Nadine Neumayer,$^{10}$ }
\newauthor{
Masatoshi Imanishi,$^{2,11}$ 
Takuma Izumi,$^{2,11}$ 
Taiki Kawamuro,$^{11}$ 
Shunsuke Baba,$^{2}$ }
\newauthor{
Phuong M.\ Nguyen,$^{13}$ 
Satoru Iguchi,$^{2,11}$ 
Takafumi Tsukui,$^{2,11}$ 
Lam N.\ T.,$^{14}$ 
Than Ho$^{15}$ } 
\\
$^{1}$Department of Physics, International University - Vietnam National University, Quarter 6, Linh Trung Ward, Thu Duc City, Ho Chi Minh City, Vietnam\\
$^{2}$National Astronomical Observatory of Japan (NAOJ), National Institute of Natural Sciences (NINS), 2-21-1 Osawa, Mitaka, Tokyo 181-8588, Japan\\
$^{3}$Sub-department of Astrophysics, Department of Physics, University of Oxford, Denys Wilkinson Building, Keble Road, Oxford OX1 3RH, UK\\
$^{4}$Yonsei Frontier Lab and Department of Astronomy, Yonsei University, 50 Yonsei-ro, Seodaemun-gu, Seoul 03722, Republic of Korea\\
$^{5}$Department of Astrophysics, University of Vienna, T\"urkenschanzstrasse 17, 1180 Wien, Austria\\
$^{6}$National Research Council, Resident at the Naval Research Laboratory, Washington, DC 20375, USA\\
$^{7}$Leibniz-Institut f\"ur Astrophysik Potsdam (AIP), An der Sternwarte 16, 14482 Potsdam, Germany\\
$^{8}$School of Physics and Astronomy, Cardiff University, Queens Buildings, The Parade, Cardiff, CF24 3AA, UK\\
$^{9}$Department of Astrophysics, Princeton University, Princeton, NJ 08540, USA\\  
$^{10}$Max Planck Institut f\"ur Astronomie (MPIA), K\"onigstuhl 17, D-69121 Heidelberg, Germany\\
$^{11}$Department of Astronomical Science, The Graduate University for Advanced Studies (SOKENDAI), 2-21-1 Osawa, Mitaka, Tokyo 181-8588, Japan\\
$^{12}$Nu\'{c}leo de Astronom\'{i}a de la Facultad de Ingenier\'{i}a, Universidad Diego Portales, Av. Ej\'{e}ercito Libertador 441, Santiago, Chile\\
$^{13}$Department of Physics, Quy Nhon University, 170 An Duong Vuong, Quy Nhon, Binh Dinh, Vietnam\\
$^{14}$LESIA, Observatoire de Paris, Universit\'e PSL, CNRS, Sorbonne Universit\'e, Univ. Paris Diderot, Sorbonne Paris Cit\'e, 5 place Jules Janssen, F-92195\\ 
\hspace{5mm} Meudon, France\\
$^{15}$Training Management Office, Mientrung University of Civil Engineering, Ha Huy Tap, Tuy Hoa, Phu Yen, Vietnam
}

\date{Accepted 2021 October 15. Received 2021 October 6; in original form 2021 March 11}

\pubyear{2021}

\begin{document}
\label{firstpage}
\pagerange{\pageref{firstpage}--\pageref{lastpage}}
\maketitle

\begin{abstract}
\normalsize  As part of the Measuring Black Holes in Below Milky Way-mass (M$^\star$) galaxies (MBHBM$^\star$) Project, we present a dynamical measurement of the supermassive black hole (SMBH) mass in the nearby lenticular galaxy NGC~3593, using cold molecular gas $^{12}$CO(2-1) emission observed at an angular resolution of $\approx0\farcs3$ ($\approx10$~pc) with the Atacama Large Millimeter/submillimeter Array (ALMA). Our ALMA observations reveal a circumnuclear molecular gas disc (CND) elongated along the galaxy major axis and rotating around the SMBH. This CND has a relatively low velocity dispersion ($\lesssim10$~\kms) and is morphologically complex, with clumps having higher integrated intensities and velocity dispersions ($\lesssim25$~\kms). These clumps are distributed along the ridges of a two-arm/bi-symmetric spiral pattern surrounded by a larger ring-like structure (radius $r\approx10\arcsec$ or $\approx350$~pc). This pattern likely plays an important role to bridge the molecular gas reservoirs in the CND and beyond ($10\arcsec\lesssim r\lesssim35\arcsec$ or $350$~pc~$\lesssim r\lesssim1.2$~kpc). Using dynamical modelling, the molecular gas kinematics allow us to infer a SMBH mass $M_{\rm BH}=2.40_{-1.05}^{+1.87}\times10^6$~\Msun\ (only statistical uncertainties at the $3\sigma$ level). We also detect a massive core of cold molecular gas (CMC) of mass $M_{\rm CMC}=(5.4\pm1.2)\times10^6$~\Msun\ and effective (half-mass) radius $r_{\rm CMC,e}=11.2\pm2.8$~pc, co-spatial with a nuclear star cluster (NSC) of mass $M_{\rm NSC}=(1.67\pm0.48)\times10^7$~\Msun\ and effective radius $r_{\rm NSC,e}=5.0\pm1.0$~pc (or $0\farcs15\pm0\farcs03$). The mass profiles of the CMC and NSC are well described by S\'{e}rsic functions with indices $1-1.4$. Our  \Mbh\ and \Mnsc\ estimates for NGC~3593 agree well with the recently compiled \Mbh--\Mnsc\ scaling relation. Although the $M_{\rm NSC}$ uncertainty is twice the inferred $M_{\rm BH}$, the rapid central rise of the rotation velocities of the CND (as the radius decreases) clearly suggests a SMBH. Indeed, our dynamical models show that even if \Mnsc\ is at the upper end of its allowed range, the evidence for a black hole does not vanish, but remains with a lower limit of $M_{\rm BH}>3\times10^5$~\Msun.

\end{abstract}

\begin{keywords}
galaxies: nuclei -- galaxies: ISM -- galaxies: kinematics and dynamics -- galaxies: disc -- galaxies: supermassive black holes -- galaxies: elliptical and lenticular 
\end{keywords}



\section{Introduction}\label{sec:intro} 

The co-evolution of supermassive black holes (SMBHs) and their host galaxies is one of the most important puzzles in galaxy formation and evolution \citep[e.g.][]{Schawinski07}. Galaxies with spheroids (or more generally bulges) ubiquitously harbour SMBHs at their centres, and the SMBH masses (\Mbh) correlate surprisingly well with macroscopic properties of the bulges, e.g.\ central stellar velocity dispersion ($\sigma_\star$; \citealt{Gebhardt00, Ferrarese00}) and stellar mass (\Mbulge; \citealt{Kormendy95, Magorrian98, Haring04, Marconi03}), despite bulges extending far beyond the sphere of influence (SOI) of even the largest black hole (BH\footnote{In this work, we use the abbreviations SMBH and BH interchangeably.}).

The tightest scaling relation among these relationships is the \Mbh--$\sigma_\star$ correlation \citep[e.g.][]{Gebhardt00, Ferrarese00}, but there is growing evidence of divergence between galaxies of different morphological types or bulge masses, especially towards the low-mass regimes of both BHs and their hosts (see, e.g. Fig.~1 of \citealt{Krajnovic18} and Fig.~17 of \citealt{Nguyen19}). To fully understand the extent of the co-evolution among all these galaxy properties, it is essential to gather a larger, more diverse sample of low-mass galaxies and perform more reliable measurements of their SMBH masses \citep[e.g.][]{McConnell13, vandenBosch16, GrahamMT18, Nguyen18, Nguyen19}.

Recently, the number of known $\lesssim10^6$~\Msun\ BHs has increased dramatically, with masses inferred from a variety of methods including (1) the velocity widths of broad optical emission lines \citep{Barth04, Greene07, Thornton08, Dong12, Reines13, Baldassare15, Reines15, Chilingarian18, Woo18, Woo19, Baldassare20b}, (2) the accretion signatures of narrow-line emission \citep[e.g.][]{Moran14} and coronal emission in the mid-infrared (MIR; \citealt{Satyapal09}), (3) tidal-disruption events (TDEs; e.g.\ \citealt{Maksym13, Stone17}), (4) hard X-ray emission \citep[e.g.][]{Gallo08, Desroches09, Gallo10, Miller15, She17a, She17b}, (5) detection of other emission lines, such as Br$\gamma$, that might be associated with X-ray radiation from the accreting region \citep{Osterbrock89, Panessa06, Cresci10, Reines11, Nguyen14}, (6) the dynamics of accretion discs containing megamasers \citep{Miyoshi95, Lo05, Kuo11, vandenBosch16} and (7) the dynamics of stars and warm/ionised gas (\citealt{Verolme02, Valluri05, Neumayer07, vandenBosch10, Seth10a, denBrok15, Nguyen17b, Nguyen17, Nguyen18, Nguyen19, Thater17, Thater19a, Krajnovic18}) and dynamically cold molecular gas (\citealt{Combes19, Davis20}; this work) in low-mass galaxies ($5\times10^{8}<M_\star\lesssim10^{10}$~\Msun) and ultracompact dwarfs (UCDs; $1\times10^7<M_\star\leq5\times10^8$~\Msun; \citealt{Seth14, Ahn17, Afanasiev18, Ahn18, Voggel18}). The importance of the $\lesssim10^6$~\Msun\ BH population is discussed in detail in \citet{Nguyen17, Nguyen18, Nguyen19}.

We started the project ``Measuring Black Holes in Below Milky Way-mass (M$^\star$) galaxies'' (MBHBM$^\star$ Project; \citealt{Nguyen19c, Nguyen20}) to gather a large sample of gas-rich galaxies with reliably-measured SMBH masses in the regime $M_\star\lesssim5\times10^{10}$~\Msun\ \citep[e.g.][]{Baldry12, Cautun20} and $\sigma_\star<120$~\kms, where stellar kinematics of sufficiently high spatial and spectral resolutions are hard to obtain. Instead, we use cold molecular gas tracers observed with Atacama Large Millimeter/submillimeter Array (ALMA) to measure their central kinematics and thus dark central masses, that are likely BHs. The outcomes of this project will reveal the demographics of the $\lesssim10^{5-7}$~\Msun\ BHs (i.e.\ the regime with currently sparse or limited observations) and provide enough measurements to accurately constrain the scatters and slopes of BH--galaxy scaling relations at the low-mass end. The molecular gas method is also the most reliable method to precisely measure the occupation fraction ($f_{\rm occ}$) of central BHs among low-mass galaxies, an important parameter to constrain the possible BH seed formation mechanisms in the early Universe \citep[e.g.][]{Greene12, Reines15, Greene20, Neumayer20}, either the direct collapse of gas clouds ($f_{\rm occ}<60\%$; e.g.\ \citealt{Lodato06, Gallo08, Bonoli14, Miller15}) or the death of the first stars ($f_{\rm occ}>60\%$; e.g.\ \citealt{Volonteri08, vanWassenhove10, Volonteri10, Volonteri12b, Volonteri12a, Reines16, Inayoshi20, Haemmerle20}).

Many recent works have used molecular gas tracers to weigh central BHs dynamically, proving this method can be applied to a variety of galaxy types and masses. The method was first pioneered with Combined Array for Research in Millimetre-wave Astronomy (CARMA) observations of the galaxy NGC~4526 \citep{Davis13} and has now been applied to both active and non-active as well as early-type (ETGs; \citealt{Barth16a, Barth16b, Davis13, Davis17, Davis18, Onishi17, Boizelle19, Boizelle21, Combes19, Nagai19, North19, Smith19, Smith20, Thater19b, Cohn21}; Thater et al.\ in preparation; Nguyen et al.\ in preparation]) and late-type (LTGs; \citealt{Onishi15, Combes19, Nguyen20, Nguyen21a, Boizelle21}) galaxies with ALMA and CARMA. The SMBH masses have in fact now been shown to correlate with the molecular gas line widths \citep{Smith21} and there are thousands of potential targets \citep{Davis14}. This method has also allowed some of the most accurate \Mbh\ measurements to date, in the radio galaxy NGC~0383 ($M_{\rm BH}=(4.2\pm0.2)\times10^9$~\Msun; \citealt{North19}) and the non-active elliptical galaxy NGC~3258 ($M_{\rm BH}=(2.249\pm0.004)\times10^9$~\Msun; \citealt{Boizelle19}). These measurements rival the best megamaser measurements, up to now the ``gold standard'' of extragalactic $M_{\rm BH}$ measurements. The molecular gas method also extends accurate measurements towards the regime of $\lesssim10^6$~\Msun\ BHs, for example in NGC~404 with $M_{\rm BH}=5^{+1}_{-2}\times10^5$~\Msun\ \citep{Davis20} and now NGC~3593 in this work. All of these works prove that the cold-gas dynamical method combined with ALMA observations at high angular resolutions can now be effective over a range of BH masses covering six orders of magnitude ($10^{5-10}$~\Msun).

This article is the second of a series from the MBHBM$^\star$ Project \citep{Nguyen19c}, following the first measurement in the nearby double-bar LTG NGC~3504 ($M_{\rm BH}=1.6^{+0.6}_{-0.4}\times10^7$~\Msun; \citealt{Nguyen20}). The paper is organised into eight sections. The properties of the target galaxy NGC~3593 are presented in Section~\ref{sec:ngc3593}. In Section~\ref{sec:data}, we present {\it Hubble Space Telescope} (\hst) images of the galaxy and ALMA observations of the nuclear $^{12}$CO(2-1) emission, discussing in detail our data reduction and analysis. We describe the Kinematic Molecular Simulation (KinMS; \citealt{Davis13}) model that we use to constrain the mass profile of the galaxy in Section~\ref{sec:massmodel} and the inferred central \Mbh\ in Section~\ref{sec:bh}. We also determine the masses and sizes of the nuclear star cluster (NSC) and massive core of cold molecular gas (CMC) in Section~\ref{sec:cores}. We further discuss our results in Section~\ref{sec:discussion} and conclude in Section~\ref{sec:conclusions}.

Throughout this work, we (1) quote all quantities using a foreground extinction correction $A_V=0.053$~mag \citep{Schlafly11} and the \citet{Cardelli89} interstellar extinction law and (2) adopt a Tully-Fisher (TF) distance\footnote{The choice of distance $D$ does not influence our conclusions but merely sets the scale of our models in physical units. In particular, lengths and dynamically-derived masses such as \Mbh\ scale as $D$, luminosity-derived masses scale as $D^2$, and mass-to-light ratios scale as $D^{-1}$.} to NGC~3593 of $7\pm2$~Mpc \citep{Wiklind92}, with the uncertainty based on the spread of TF distances in the National Aeronautics and Space Administration (NASA) Infrared Processing and Analysis Center (IPAC) Extragalactic Database (NED\footnote{\url{https://ned.ipac.caltech.edu/}.}), yielding a physical scale of $\approx35$~pc~arcsec$^{-1}$ assuming a current Hubble constant $H_0=70.3\pm1.6$~\kms~Mpc$^{-1}$, matter density (with respect to the critical mass density) $\Omega_{\rm m,0}=0.277\pm0.019$ and dark energy density (with respect to the critical mass density) $\Omega_{\rm \Lambda,0}=0.723\pm0.019$ from the {\it Wilkinson Microwave Anisotropy Probe} (WMAP; \citealt{Moura-Santos16, Verschuur16, Calabrese17}) and PLANCK Collaboration \citep{PlanckCollaboration14}. All the maps presented in this article are plotted with north up and east to the left. Tables and Figures labelled with numbers only appear in order in the main text, while those labelled with both letters and numbers appear in the corresponding appendices.

\vspace{-6mm}
\section{NGC~3593}\label{sec:ngc3593}

We summarise the known properties of NGC~3593 in Table~\ref{tab:ngc3593property} and further discuss these properties in detail in this section.

NGC~3593 (UGC~6272) is classified as a lenticular galaxy \citep{Buta07}, with a morphological classification of SA(s)0/a or numerical Hubble type $T_{\rm Hubble}=-0.4\pm0.9$ (NED) and it is a member of the Leo Group \citep{Stierwalt09}. It has a dust disc obscuring the galaxy central regions to the north of the major axis \citep{Sandage94}.

NGC~3593 is known to contain two distinct stellar populations that are rotating in opposite directions \citep{Bertola96, Corsini98, Garcia-Burillo00, Coccato13}. The main stellar component is slightly older and more metal-rich (luminosity-weighted age of $3.6\pm0.6$~Gyr and metallicity ${\rm [Z/H]}=-0.04\pm0.03$) and is rotating slightly slower (rotation velocity around the nucleus of $\approx100$~\kms), while the secondary (counter-rotating) stellar component is slightly younger and more metal-poor (age $2.0\pm0.5$~Gyr and metallicity ${\rm [Z/H]}=-0.15\pm0.07$) and is rotating slightly faster ($\approx120$~\kms; \citealt{Coccato13}). Such a configuration can arise from the merger of a dwarf galaxy \citep{Balcells98, Jesseit07, Eliche-Moral11, Bois11}, although an alternative explanation is that the galaxy nucleus accreted gas on retrograde orbits from an external source, that then underwent star formation. The stellar velocity dispersions of these two stellar components are in the range $30$--$80$~\kms, while \citet{Bertola96} measured $\sigma_\star\approx60$~\kms\ in the galaxy centre, suggesting a $\approx1.7_{-1.1}^{+3.2}\times10^6$~\Msun\ central BH based on the \citet{Kormendy13} \Mbh--$\sigma_\star$ relation for massive ETGs.

The nucleus of NGC~3593 is red ($F$555W--$F$814W$\approx2.5$~mag), due to the presence of dust and a luminous and massive NSC ($I$-band luminosity $L_{I,\,\rm NSC}=2.89\times10^7$~\Lsun\ and stellar mass $M_{\rm NSC}=1.58\times10^8$~\Msun), that has a S\'{e}rsic index $n_{\rm NSC}=1.4\pm0.14$ and an effective (half-light) radius $r_{\rm NSC,e}=5.50\pm0.23$~pc (\citealt{Pechetti20}; although see our improved measurements in Sections~\ref{ssec:nsccmc}).

\citet{Bertola96} also performed a photometric decomposition using an $r$-band stellar surface-brightness map obtained from the spectral decomposition of European Southern Observatory (ESO) 1.5-m spectroscopic telescope data, and found two stellar discs with the same radially-constant ellipticity ($\epsilon=0.55$) and position angle (${\rm P.A.}=90\degr$). The two stellar discs can be parameterised by infinitely thin exponential discs with different scale lengths ($h$), central surface brightnesses ($\mu$) and thus total stellar masses ($M_\star$): $h_1=40\arcsec$ ($1.4$~kpc), $\mu_1=19.9$~mag~arcsec$^{-2}$ and $M_{\star,1}=1.2\times10^{10}$~\Msun, and $h_2=10\arcsec$ ($350$~pc), $\mu_2=18.5$~mag~arcsec$^{-2}$ and $M_{\star,2}=2.7\times10^9$~\Msun. The masses were obtained by fitting the ionised gas rotation curve, allowing the mass-to-light ratios of the two discs to vary independently. NGC~3593 has a total stellar mass of $M_{\star}\approx1.5\times10^{10}$~\Msun\ \citep{Bertola96} and thus can be classified as a sub-$M^\star$ galaxy \citep{Baldry12}.

\begin{table}
  \caption{Properties of NGC~3593.}
  \begin{tabular}{lcr}
    \hline\hline   
    Parameter (Units) & Value & References\\
    \hline
    Morphology                                       & SA(s)0/a & (1, 2)\\
    R.A.\ (J2000)                               & $11^{\rm h}14^{\rm m}37\fs1$ & (3)\\
    Decl.\ (J2000)                              & $+12\degr49\arcmin05\farcs6$ & (3)\\
    Position angle ($^{\circ}$)                      & $90$                 & (4)\\
    Inclination angle ($^{\circ}$)                   &  $67$                & (5)\\
    Systemic velocity (\kms)                         & $629$               & (3)\\
    Distance (Mpc)                                   & $7$                  & (6)\\
    Linear scale (pc~arcsec$^{-1}$)                  & $35$                 & (7, 8)\\
    $\log(L_{\rm 2-10\;keV}/{\rm erg\,s^{-1})}$          & $<38.23$           & (9)\\
    $\log(L_{\rm 0.3-8\;keV}/{\rm erg\,s^{-1}})$         & $38.33$              & (10)\\
    Total stellar mass (\Msun)                       & $1.5\times10^{10}$  & (11)\\
    Total \ion{H}{i} mass (\Msun)  & $1.3\times10^8$     & (12)\\
    Total dust mass (\Msun)                          & $1.0\times10^6$     & (13)\\
    Dust temperature (K)                             & $40$               & (13)\\
    Stellar velocity dispersion (\kms)             & $60$               & (11)\\
    $\langle SFR \rangle_{\rm CND}$ (\Msun~yr$^{-1}$)& $3$                & (14)\\
    {\it Stellar properties:}                        & (age, [Z/H])        &    \\
    $\;\;$Main component                             & ($3.6$~Gyr, $-0.04$)  & (15)\\
    $\;\;$Secondary component                        & ($2.0$~Gyr, $-0.15$)  & (15)\\
    {\it NSC properties:}                            &                    &    \\
    $\;\;$Effective radius (pc)          & $5.50\pm0.23$               & (16)\\
    $\;\;$S\'ersic index                    & $1.40\pm0.14$               & (16)\\
    $\;\;$Mass (\Msun)                   & $1.58\times10^8$           & (16)\\
    $\;\;L_I$ (\Lsun)                       & $2.89\times10^7$           & (16)\\
    \hline
  \end{tabular}
  \parbox[t]{0.472\textwidth}{\textit{Notes:} (1) \citet{Sandage81}; (2) \citet{deVaucouleurs91}; (3) \citet{Garcia-Burillo00}; (4) \citet{Sandage81}; (5) \citet{Rubin85}; (6) \citet{Wiklind92}; (7) \citet{Moura-Santos16}; (8) \citet{PlanckCollaboration14}; (9) \citet{She17a}; (10) \citet{Zhang09}; (11) \citet{Bertola96}; (12) \citet{Pogge93}; (13) \citet{Fich93a}; (14) \citet{Ho97}; (15) \citet{Coccato13}; (16) \citet{Pechetti20}.}
      \label{tab:ngc3593property}
\end{table}

Narrow-band H$\alpha$+[\ion{N}{ii}] ($\lambda\lambda$~$654.80$, $658.34$~nm) and [\ion{S}{ii}] ($\lambda\lambda$~$671.65$, $673.08$~nm) observations reveal a ring of ionised gas in the circumnuclear region of NGC~3593, that extends to a radius $r\approx17\arcsec$ \citep[or $\approx595$ pc;][]{Corsini98}. The ionised gas kinematics show that this material rotates in the same sense as the secondary/counter-rotating stellar component and has a velocity dispersion $\lesssim30$~\kms.

Using Institut de Radioastronomie Millim\'{e}trique (IRAM) Plateau de Bure Interferometer (PdBI) $^{12}$CO(1-0) observations at an angular resolution of $4\arcsec\times3\arcsec$, \citet{Garcia-Burillo00} found that NGC~3593 has an inner disc of molecular gas extending to $r\approx35\arcsec$ (or $\approx1.2$ kpc), counter-rotating at all radii with respect to the most massive/primary stellar disc. Half of the $^{12}$CO(1-0) emission (and hence half of the associated mass) arises from an elongated circumnuclear disc (CND) within a region of radius $r\approx10\arcsec$ (or $\approx350$ pc), with an outer ring-like structure and a two-arm/bi-symmetric spiral pattern within it. This CND connects to an outer gas reservoirs containing the remaining half of the $^{12}$CO(1-0) emission (and thus of its associated mass) and extending out to $r\approx35\arcsec$, that allows the gas to flow to the northern half of the disc \citep{Garcia-Burillo00}. \citet{Pogge93} report a total neutral hydrogen (\ion{H}{i}) gas mass of $\approx1.3\times10^8$~M$_\odot$, which is a compilation from the RC3 catalogue \citep{Corwin94}. 

The nucleus of NGC~3593 is classified as an \ion{H}{ii} star-forming nucleus \citep{Hunter89}, with an upper limit on the total star formation rate (SFR) of the CND alone (${\rm SFR}_{\rm CND}\lesssim3$~\Msun~yr$^{-1}$; \citealt{Ho97}). This upper limit is due to H$\beta$ emission being almost absent from the centre of the galaxy. A similar behaviour is seen in other high-resolution observations of star formation tracers such as H$\alpha$ and Pa$\alpha$ \citep{Garcia-Burillo00}, fuelling the second/counter-rotating stellar disc. Optical and near-infrared (NIR) recombination lines suggest a $V$-band extinction $A_V\approx1$~mag in the CND, while the CO and $100$~$\mu$m fluxes suggest $A_V>5$~mag \citep{Garcia-Burillo00}.

The total mass and temperature of the dust of NGC~3593 were estimated to be $T_{\rm dust}\approx40$~K and $M_{\rm dust}\approx10^6$~\Msun\ using observations at $1.1$~mm, $800$~$\mu$m and $450$~$\mu$m \citep{Fich93a}, yielding a gas-to-dust mass ratio of $\approx300$. This ratio is two times higher than the canonical value of $\approx150$ derived from CO lines associated with high extinction regions and widely used for the Galaxy \citep{Spitzer78, Hildebrand83, Draine84}.

The X-ray detection in the nucleus of NGC~3593 with {\it Chandra} is debated. \citet{Martinez-Garcia17} and \citet{She17b} found no X-ray emission, and \citet{She17a} reported an upper limit on the X-ray luminosity of $\log(L_{2-10\,\rm{keV}}/{\rm erg\,s^{-1}})<38.23$. On the other hand, \citet{Zhang09} do report an X-ray detection and a nuclear luminosity of $\log(L_{0.3-8\,\rm{keV}}/{\rm erg\,s^{-1}})=38.33$. The reason behind these different conclusions is as yet unknown, but it may be the different detection criteria adopted. In any case, the SMBH of NGC~3593, as inferred from \Mbh--$\sigma_\star$ correlations, must be accreting at an extremely low rate, $\dot{M}_{\rm BH}/M_{\rm BH}\lesssim10^{-7}$ of the Eddington limit. Thus, there is no evidence for an active galactic nucleus (AGN) in NGC~3593.

\section{Data and Data Reduction}\label{sec:data}

\subsection{{\it HST} images}\label{ssec:hst}

We use \hst\ Wide-Field Planetary Camera~2 (WFPC2) images in the F450W and F814W bands taken on 2007 November 20 (GO-11128, PI: Fisher) to create a central stellar mass model of NGC~3593 (see Section~\ref{ssec:massinner}), that will be used as an input to our dynamical models in Section~\ref{sec:bh}. More details of these images are listed in Table~\ref{tab:hst}.

\begin{figure}
  \centering\includegraphics[scale=0.35]{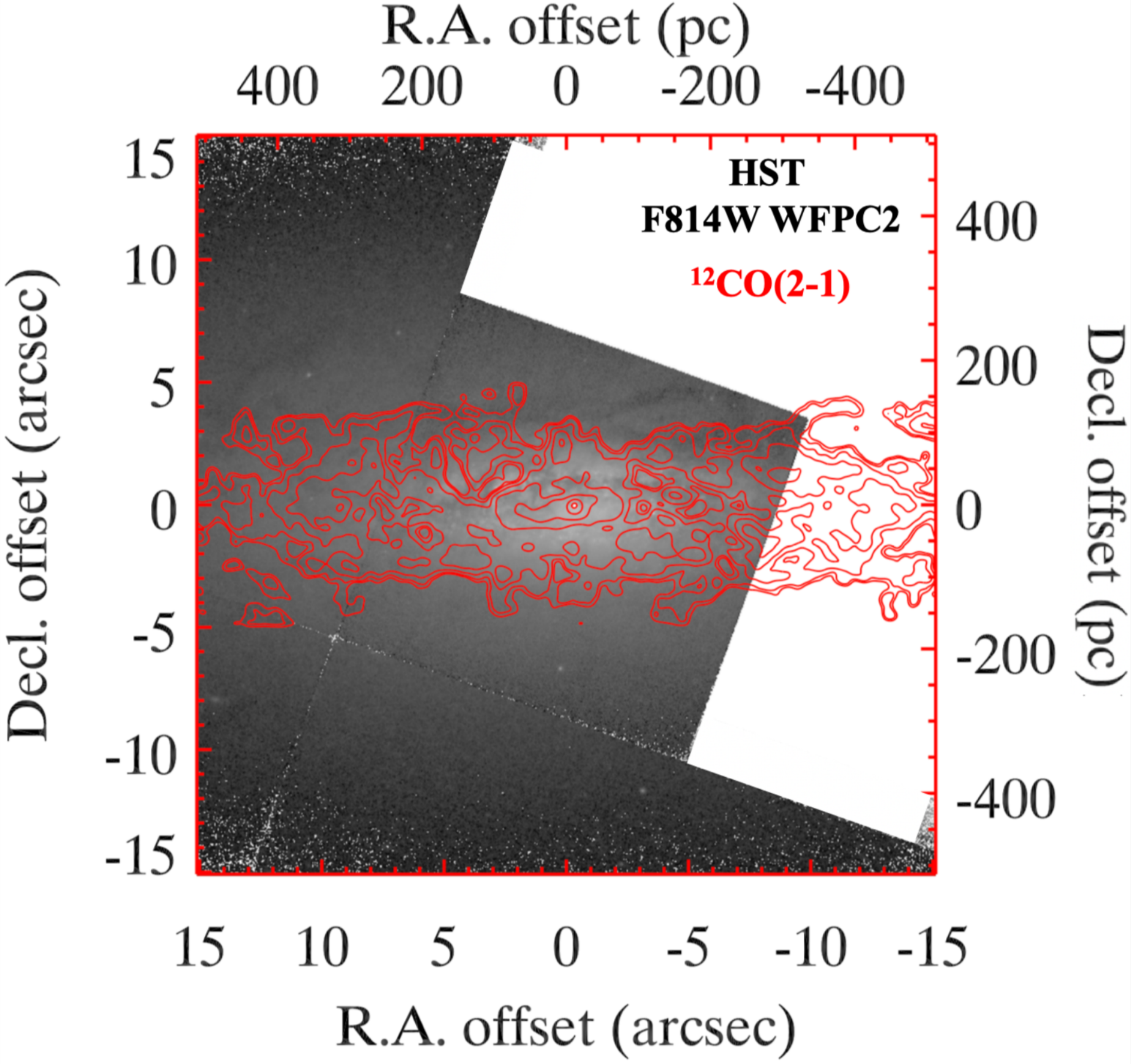}
  \caption{\hst/WFPC2 $F$814W image of NGC~3593 with a field-of-view of $30\arcsec\times30\arcsec$ ($1.05\times1.05$~kpc$^2$), overlaid with the $^{12}$CO(2-1) integrated intensity contours from our ALMA observations. The \hst~image is in an arbitrary logarithmic scale. Grey dust lanes are clearly visible on the northern side of the nucleus.}
  \label{fig:hstimage}   
\end{figure}

The photometric centre of the galaxy in the \hst/WFPC2 images is offset by ($-0\fs13$, $-0\farcs11$) with respect to the photometric and kinematic centre of NGC~3593, as determined from our own high-resolution $^{12}$CO(2-1) data and discussed in more details in Section~\ref{ssec:centre}. This is within the positional uncertainty of the \hst\ images, so we align the \hst\ images to this $^{12}$CO(2-1) centre to correct for the astrometric mismatch, and show the (offset) $F$814W image overlaid with the $^{12}$CO(2-1) iso-intensity contours in Fig.~\ref{fig:hstimage}.

We used \texttt{Tiny Tim} point spread functions (PSFs; \citealt{Krist95, Krist11}) of the WFPC2 $F$450W and $F$814W images to create (1) a $F$555W--$F$814W colour map and (2) a multi-Gaussian expansion (MGE; \citealt{Emsellem94a, Cappellari02}) stellar light model in Section~\ref{ssec:massouter}.

\begin{table}
  \caption{\hst/WFPC2 data.}
  \begin{tabular}{ccccc}
    \hline\hline   
    Filter&Pixel scale$^a$&Exposure\ time&Zero point$^b$&$A_\lambda^c$\\
            &(arcsec~pixel$^{-1}$)&  (s)    &   (mag)  &    (mag)   \\
    (1)   &    (2)   &   (3)   &     (4)  &     (5)  \\
    \hline
    $F$450W &0.0455&$2\times200$&24.11&0.066\\
    $F$814W &0.0455&$2\times130$&23.76&0.030\\
    \hline
  \end{tabular}
  \parbox[t]{0.472\textwidth}{\textit{Notes:} $^a$ \citet{Holtzman95}. $^b$Vega System. $^c$ Foreground extinction correction assuming a Milky-Way interstellar extinction law from ultraviolet to NIR \citep{Cardelli89, Schlafly11}.}
  \label{tab:hst}
\end{table}

\subsection{$^{12}$CO(2-1) ALMA observations}\label{ssec:alma}

Our ALMA observations of NGC~3593 were carried out on 2018 September 23$^{\rm rd}$ (PID: 2017.1.00964.S, PI: Nguyen, Dieu). The $^{12}$CO(2-1) emission line was observed for a total of $90$~min ($54$~min on source) using $49$ ALMA $12$-m antennae in the C43-5 configuration (baseline range $15$--$1,400$~m), resulting in a maximum recoverable scale (MRS) of $\approx30\arcsec$ in diameter and a synthesised beam full-width-at-half-maximum (FWHM) of $0\farcs33\times0\farcs29$ ($11.6\times10.2$~pc$^2$) oriented at ${\rm P.A.}=35\degr$. The correlator was set up using four spectral windows, including one window covering the $^{12}$CO(2-1) line in frequency division mode (FDM; $1875$~MHz bandwidth with $1.13$~MHz or $\approx1.5$~\kms\ channels) and three windows to probe continuum emission in time division mode (TDM; $2$~GHz bandwidth with $31.25$~MHz or $\approx40.7$~\kms\ channels). The raw ALMA data were calibrated by ALMA Regional Center staff using the standard ALMA pipeline. Flux and bandpass calibrations were carried out using the quasars J1037-2934, J0854+2006 and J1118+1234, while the atmospheric phase offsets were determined using the quasar J1103+1158.

No continuum emission is detected across the primary beam of $25\farcs3$ diameter, so we report an upper limit on the continuum emission of $\approx30$~$\mu$Jy~beam$^{-1}$, resulting from a summation of all line-free channels across all four spectral windows. Also as a result, we created a three-dimensional (3D; R.A., decl., velocity) datacube directly from the calibrated measurement set without continuum subtraction using the \texttt{clean} task of the \texttt{Common Astronomy Software Applications} (\texttt{CASA}; \citealt{McMullin07}) package version 5.1.1, following the same successful strategy employed in our previous studies \citep[e.g.][]{Davis17, Nguyen20}. Specifically, we created the ALMA $^{12}$CO(2-1) datacube of NGC~3593 with a pixel size of $0\farcs1$, a binned channel width of $10$~\kms\ (this is several times the raw channel width of $\approx1.5$~\kms, such that the channels are effectively independent) and Briggs weighting with a robust parameter of $0.5$. For the \texttt{clean} task, we used the interactive masking mode to further reduce the sidelobes of the data, estimating the root-mean-square (RMS) noise in a few channels of the residual cube and setting $3\times{\rm RMS}$ as the cleaning threshold in regions of source emission in dirty channels. We performed primary beam correction after cleaning. Our final fully calibrated and cleaned $^{12}$CO(2-1) datacube has a RMS noise of $\approx1$~mJy~beam$^{-1}$ per 10~\kms\ binned channel and emission is detected from $\approx500$ to $\approx750$~\kms\ with a mean (systemic) velocity of $629$~\kms. 

\subsection{$^{12}$CO(2-1) moment maps}\label{ssec:moments}

Fig.~\ref{fig:co21moms} shows the $^{12}$CO(2-1) integrated intensity (moment~0), intensity-weighted mean line-of-sight (LOS) velocity (moment~1) and intensity-weighted LOS velocity dispersion (moment~2) maps of the CND region, roughly matching the $12$-m antennae primary beam. We created these maps using the moment-masking technique \citep{Dame01, Dame11}. The mask was thus created by first spatially smoothing each channel by a factor of $\alpha$ by convolving each channel by a Gaussian of FWHM $\alpha$ times the FWHM of the synthesised beam, where we varied $\alpha$ and gauged the spatial and velocity coherence of the signal. Spatial smoothing increases the sensitivity while decreasing the angular resolution, helping to expunge noise peaks.  Second, we performed ``$\beta\times\sigma$-clipping'', where $\beta$ is a positive factor and $\sigma$ the RMS noise. Here, at any given position, all channels with intensities below $\beta\sigma$ were set to zero. We note that this mask created from the smoothed cube is only used to identify and mask out emission-free regions of the original cube; the latter is used to create the moment maps at full spatial and velocity resolutions. We experimented with appropriate choices of the smoothing and masking parameters to obtain the best moment maps, finally adopting $\alpha=3$ and $\beta=0.75$.

The $^{12}$CO(2-1) emission is significant within a $\approx30\arcsec\times10\arcsec$ rectangular central region roughly corresponding to the CND (see panel~A of Fig.~\ref{fig:co21moms}) and peaking at (R.A., decl.)$\,=(11^{\rm h}14^{\rm m}37\fs1$, $+12\degr49\arcmin05\farcs6$), that is identified as the galaxy centre (see Section~\ref{ssec:centre}). Interestingly, there is a distinct massive core of cold molecular gas (CMC; radius $r\lesssim0\farcs5$) embedded within the dense CND ($r\lesssim10\arcsec$), itself surrounded by a more diffuse and extended gas disc ($10\arcsec\lesssim r\lesssim35\arcsec$). The CND has an outer ring-like structure (radius $\approx10\arcsec$) and a two-arm/bi-symmetric spiral pattern within it, apparently extending down to the CMC. This pattern seems to match the ridge of high-velocity dispersion regions in the moment~2 map.

The intensity-weighted mean LOS velocity map in panel~C of Fig.~\ref{fig:co21moms} confirms that the central molecular gas is consistent with a rotating disc, with a total velocity width $\Delta V\approx250$~\kms. This rotation is consistent with both that of the  counter-rotating stellar component measured using Very Large Telescope  (VLT) Visible Multi-object Spectrograph (VIMOS) integral-field observations  \citep{Coccato13} and that of the counter-rotating ionised-gas component measured using ESO's 1.5~m spectroscopic telescope \citep{Bertola96}. The rotational velocities of the molecular gas are however higher than those  of the primary stellar component ($\Delta V\approx200$~\kms).

The intensity-weighted LOS velocity dispersion map in panel~D of Fig.~\ref{fig:co21moms} is quite flat, with a roughly constant dispersion $\sigma\approx10$~\kms, except for a few regions of higher velocity dispersion ($25$--$33$~\kms) coincident with bright emission clumps. The velocity dispersion of the molecular gas is consistent with that of the ionised gas ($\lesssim30$~\kms; \citealt{Bertola96}). The high velocity dispersion at the very centre is consistent with the existence of the CMC.

\begin{figure*}
  \includegraphics[scale=0.59]{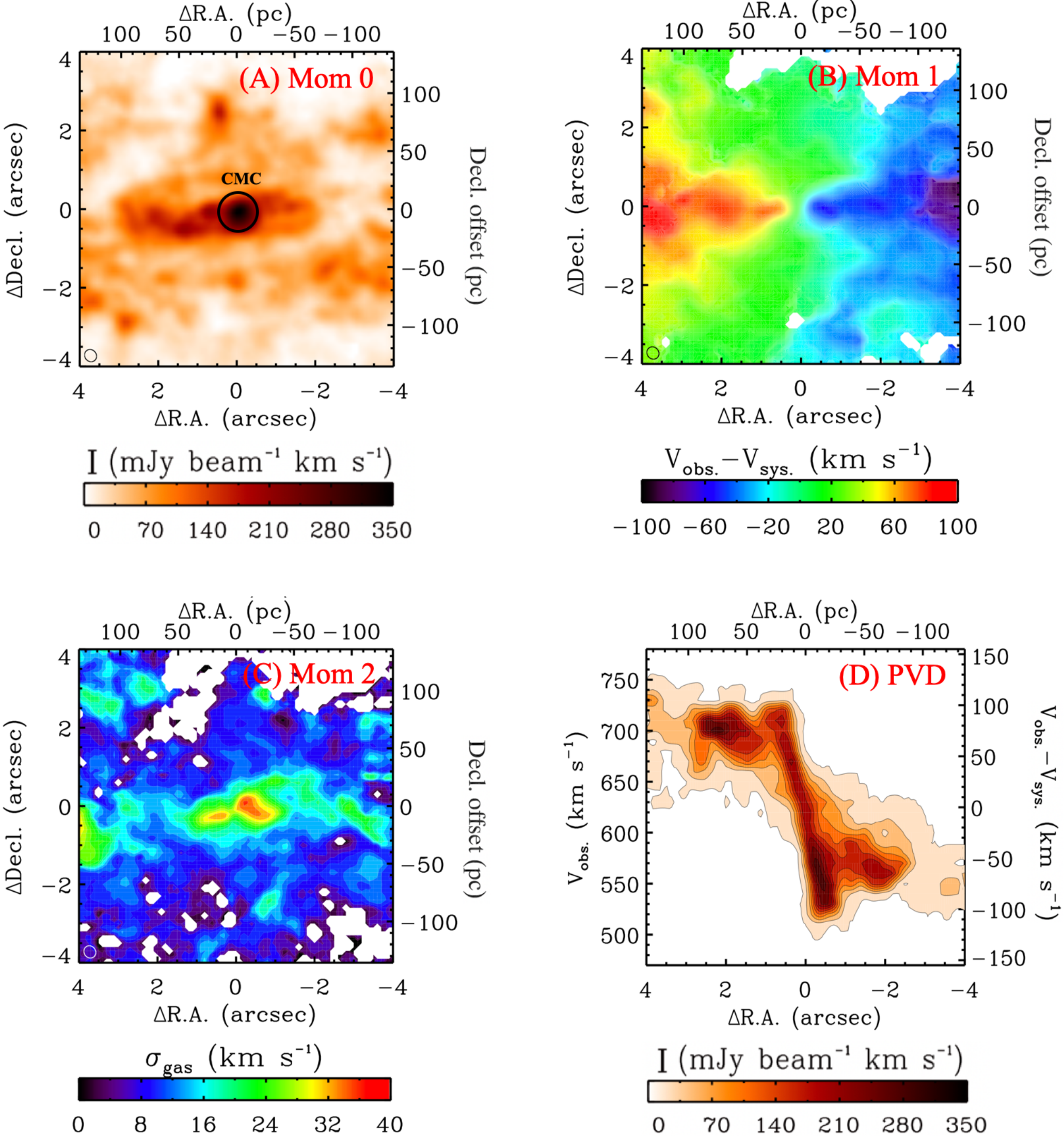}
  \caption{Zoomed-in moment maps and PVD of the $^{12}$CO(2-1) emission of NGC~3593 with a field-of-view of $\approx4\arcsec\times4\arcsec$ ($\approx140\times140$~pc$^2$), illustrating the morphology and kinematics of the CND in the vicinity of the central SMBH. As for Figs.~\ref{fig:co21moms} and \ref{fig:co21pvd0}, the panels include the integrated intensity map (panel~A; the CMC is clearly visible at the galaxy centre and is indicated by a black circle), intensity-weighted mean LOS velocity map (panel~B) and intensity-weighted LOS velocity dispersion map (panel~C), as well as the kinematic major-axis PVD (panel~D). The synthesised beam ($0\farcs33\times0\farcs29$ or $11.6\times10.2$~pc$^2$) is shown as a tiny black or white ellipse in the bottom-left corner of panels~A--C.}
  \label{fig:co21moms2} 
\end{figure*} 

Fig.~\ref{fig:co21spec} shows the $^{12}{\rm CO(2-1)}$ integrated spectrum of NGC~3593, with the classic double-horn shape of a rotating disc. We also plot the position-velocity diagram (PVD) extracted from a cut along the kinematic major axis of the disc (${\rm P.A.}=90\degr$) in Fig.~\ref{fig:co21pvd0}. There is a sharp increase of the rotation towards the galaxy centre ($r\lesssim1\arcsec$ or 35 pc). We later interpret this rotation as being caused by the massive cores (NSC and CMC) and/or a SMBH at the centre of the galaxy.

In Fig.~\ref{fig:co21moms2}, we show a zoom (inner $\approx4\arcsec\times4\arcsec$ or $\approx140\times140$~pc$^2$) of the integrated intensity map (panel~A), intensity-weighted mean LOS velocity map (panel~B), intensity-weighted LOS velocity dispersion map (panel~C) and PVD extracted along the kinematic major axis of the CND (panel~D). Specifically, panel~A illustrates the detailed $^{12}$CO(2-1) morphology of the CND, with the bright CMC at its centre. Panel~B shows the central molecular gas kinematics under the influence of the compact central massive objects (i.e.\ NSC, CMC and SMBH) at radii $r\lesssim1\arcsec$, and as seen in the PVD in panel~D. Some non-circular motions may be present in the blue-shifted half of the velocity map away from the kinematic major axis. 

\subsection{Galaxy centre}\label{ssec:centre}

As can be seen from the different panels of Fig.~\ref{fig:co21moms2}, the photometric (integrated intensity peak or CMC) and kinematic centre of our high-resolution $^{12}$CO(2-1) data are consistent with each other. Within the stated uncertainties, these centres also agree with the kinematic centre derived by \citet{Garcia-Burillo00} from lower-resolution $^{12}$CO(1-0) data, and with the optical photometric centre derived from Sloan Digital Sky Survey (SDSS) data release 14 (DR14) data \citep{Abolfathi18}. We therefore adopt this centre (R.A.$\,=11^{\rm h}14^{\rm m}37\fs1$, decl.$\,=+12\degr49\arcmin05\farcs6$, $V_{\rm sys}=629$~\kms) as the centre of NGC~3593.

\section{Mass Model}\label{sec:massmodel}

In this section, we first use our $^{12}$CO(2-1) kinematics (see Section~\ref{ssec:alma}) and a dynamical model to constrain the outer part ($4\arcsec\leq r\lesssim30\arcsec$ or $140\leq r\lesssim1,050$~pc) of our galaxy mass model (Section~\ref{ssec:massouter}). Second, we use the \hst\ imaging data (see Section~\ref{ssec:hst}) to constrain the inner part ($r<4\arcsec$ or $r<140$~pc) of our galaxy mass model (Section~\ref{ssec:massinner}). The combination of the inner and outer mass model yields a complete mass model of NGC~3593, from the centre to a radius of $30\arcsec$ (Section~\ref{ssec:mass}), a key ingredient to estimate the central \Mbh\ through dynamical modelling.

\subsection{KinMS model}\label{ssec:kinms}

The KinMS tool we use for dynamical modelling comprises two main elements. First, for a given set of model parameters, it creates a simulated data cube for comparison to observations. Second, it explores parameter-space in an efficient manner to identify the best-fitting model. 

To simulate a data cube, KinMS adopts a parametric function (specified with some free parameters) describing the distribution of the (massless) kinematic tracer (here $^{12}$CO(2-1) emission). Here, we also assume the tracer moves on circular orbits governed by a circular velocity curve, calculated from the \texttt{mge\_circular\_velocity} procedure within the \texttt{Interactive Data Language} (\texttt{IDL}) Jeans Anisotropic Modelling (JAM\footnote{\url{https://purl.org/cappellari/software}}; \citealt{Cappellari08}) package, that itself uses as an input an (axisymmetric) mass model specified via MGE parametrisation (see Sections~\ref{ssec:massouter} and \ref{ssec:massinner}), that can include any number of mass components (here stars, gas, dust and the putative SMBH; see \citealt{Davis13}).

The KinMS tool simulates the whole cube, then compares it to the data via a likelihood function \citep{Davis17, Davis18, Onishi17, North19, Smith19, Smith21, Thater20, Nguyen20, Nguyen21a}. During the fit, the model, walks through parameter space using a Markov Chain Monte Carlo (MCMC) method controlled by the \texttt{emcee} algorithm \citep{Foreman-Mackey13} and an affine-invariant ensemble sampler \citep{Goodman10} in a Bayesian framework. At each step, the relative likelihood is calculated and used to determine the next move through parameter space. The best-fitting model is then determined from the full posterior distribution. In practice, this is all achieved by using the \texttt{python} code \texttt{KINMSpy\_MCMC}\footnote{\url{https://github.com/TimothyADavis/KinMS\_MCMC}}.

\subsection{Outer mass model}\label{ssec:massouter}

We model the outer part of the galaxy mass distribution ($4\arcsec\leq r\lesssim30\arcsec$ or $140\leq r\lesssim1,050$~pc) with two mass components described below: a stellar one with free mass normalisation and an interstellar medium (ISM) one that is fixed.

For the stellar component, we first average the mass surface-density profiles of the primary and secondary/counter-rotating stellar discs from \citet{Coccato13}, convert the average into MGE form, and scale it by a (free) mass surface density at a radius of $4\arcsec$ ($\Sigma_{\star,4\arcsec}$). Here, we exclude the inner region ($r<4\arcsec$) of the averaged mass surface-density profile to avoid doubly counting its mass later, when scaling the inner part (Section~\ref{ssec:massinner}) to the outer part ($4\arcsec\leq r\lesssim30\arcsec$) at this radius of $r=4\arcsec$ (Section~\ref{ssec:mass}).   

Second, because interstellar material (i.e.\ gas and dust) within the fitting region contributes significantly to the total mass and thus has a large impact on the fitting results, we must also include it in our mass model. Since the nucleus of NGC~3593 was classified as an \ion{H}{ii} star-forming region ionised by young massive stars \citep[i.e.\ undergoing a burst of star formation;][]{Hunter89}, we convert the $^{12}$CO(2-1) integrated intensity map to a molecular gas surface-density map by assuming a line ratio (in temperature units) $^{12}$CO(2-1)/$^{12}$CO(1-0)$\,=0.8$ \citep{Bigiel08} and a CO-to-H$_2$ conversion factor for starburst galaxies $X_{\rm CO}=(1.0\pm0.3)\times10^{20}$~cm$^{-2}$~(K~\kms)$^{-1}$ \citep{Kuno00, Kuno07, Bolatto13}. This yields a total molecular gas mass $M_{\rm H_2}=(2.8\pm1.2)\times10^8$~\Msun, $\approx40$ times smaller than the total stellar mass of the galaxy (see Sections~\ref{sec:ngc3593} and \ref{ssec:bhnsc}). \citet{Hunter89} reported a ratio of total atomic-to-molecular gas mass $M_{\rm \ion{H}{i}}/M_{\rm H_2}=0.5$ (thus also consistent with the \citealt{Pogge93} \ion{H}{i} measurement), that we adopt here (adding the atomic to the molecular hydrogen). For the dust, we adopt the total mass mentioned in Section~\ref{sec:ngc3593} ($M_{\rm dust}=10^6$~\Msun) and again add it to the molecular hydrogen. Lastly, we assume that the H$_2$, \ion{H}{i} and dust are all distributed according to the $^{12}$CO(2-1) integrated intensity. Next, we again utilise the MGE formalism to decompose this total ISM (molecular hydrogen, atomic hydrogen and dust) map into individual Gaussian components, that are listed in Table~\ref{tab:gas_mges} and are fixed (no free parameter).

\begin{table}
  \caption{ISM MGE model.}
  \begin{tabular}{cccc}
    \hline\hline  
    $j$ & $\log(\Sigma_{{\rm ISM},j}/{\rm M}_\odot\,{\rm pc}^2)$ & $\sigma_j\;({\rm arcsec})$ & $q_j$\\
    (1) & (2) & (3) & (4)\\
    \hline
    1 & 3.85 & \phantom{1}0.35 & 0.72\\
    2 & 3.13 & \phantom{1}2.42 & 0.64\\
    3 & 2.04 & 13.00 & 0.60\\
    \hline
  \end{tabular}
  \parbox[t]{0.472\textwidth}{\textit{Notes:} Central ISM mass surface density ($\Sigma_{{\rm ISM},j}$), width ($\sigma_j$) and axis ratio ($q_j$) of each deconvolved Gaussian component $j$.}
  \label{tab:gas_mges}
\end{table}

Because the NGC~3593 molecular gas surface brightness cannot be described by a simple analytic function (with few free parameters), and although this leads to a slight inconsistency between the ISM mass model above and the $^{12}$CO(2-1) surface-brightness model, we use the {\tt SkySampler\footnote{\url{https://github.com/Mark-D-Smith/KinMS-skySampler}}} tool \citep{Smith19} for the fit. As the {\tt SkySampler} clouds are created from the clean components of the cube, this is essentially equivalent to fitting only the kinematics of the molecular gas but not its distribution. The model thus matches the observed gas distribution with a single free parameter, the total flux $f$, used to rescale the entire cube (as the clouds are only assigned relative intensities by {\tt SkySampler}). As the clean components do not include cleaning residuals, the total flux of the clean components is slightly lower than that of the cube, and $f$ also allows the model to recover this missing flux (although assuming that its distribution matches that of the clean components). The parameter $f$ should then simply be equal to the integrated flux of the (fitted region of the) cube, which serves as a useful sanity check on the model. 

\begin{figure}
    \centering\includegraphics[scale=0.25]{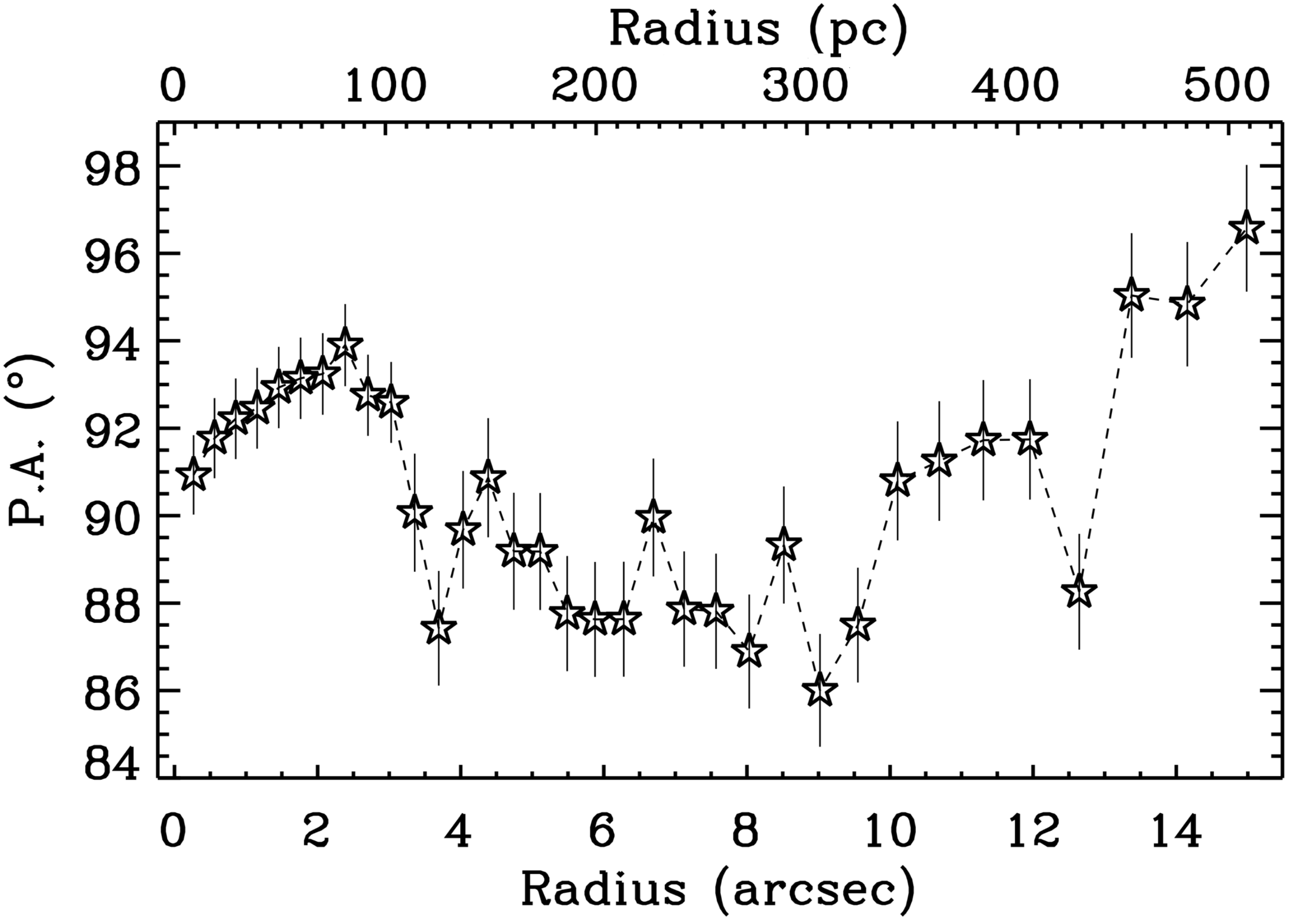}
     \caption{Radial position angle profiles of the $^{12}$CO(1-0) emission, derived using the {\tt Kinemetry} code of \citet{Krajnovic06}.}
    \label{radialPA} 
\end{figure}

Lastly, we try to account for the small kinematic twist present in the $^{12}$CO(1-0) intensity-weighted mean LOS velocity map (panel~B of Fig.~\ref{fig:co21moms2} and panel~C of Fig.~\ref{fig:co21moms}), by first extracting the radial profile of the kinematic P.A.\ using the {\tt Kinemetry}\footnote{\url{http://davor.krajnovic.org/idl/\#kinemetry}} code of \citet{Krajnovic06}, then using it as an additional model input along with the circular velocity curve. This kinematic P.A.\ profile varies only slightly ($86^\circ$--$96^\circ$) across $15\arcsec$, as shown in Fig.~\ref{radialPA}, but accounting for it in our dynamical model does help to reproduce the kinematic twist observed \citep[see also][]{Nguyen20, Nguyen21a}.

Overall, the model thus optimises the fit to the observations with seven free parameters: stellar mass surface density at a radius of $4\arcsec$ (stellar mass normalisation) $\Sigma_{\star,4\arcsec}$, total flux of the fitted region (ISM mass normalisation) $f$, inclination $i$, and galaxy centre in space ($x_{\rm c}$, $y_{\rm c}$) and velocity ($v_{\rm off}$; all defined with respect to the previously determined centre; see Section~\ref{ssec:centre}), to which we add a spatially-constant tracer (turbulent) velocity dispersion ($\sigma_0$). The kinematic tracer is assumed to lie in a thin disc (by fixing the disc thickness $d_{\rm t}=0$; \citealt{Davis20}). At this point, while modelling the outer part of the $^{12}$CO(2-1) gas kinematics only, we ignore the gravitational potential contributed by the central SMBH (but see Section~\ref{ssec:massinner}). 

We run the KinMS fit in an area of $300$~spaxels $\times$ $100$~spaxels ($30\arcsec\times10\arcsec$) but exclude the inner $4\arcsec$ ($40$~spaxels $\times$ $40$~spaxels) so that the central masses (NSC, CMC and SMBH) do not affect the outer mass surface-density profile. We will constrain the inner mass surface density independently in Section \ref{ssec:massinner}. We select a velocity range of $30$ channels ($-150$ to $150$~\kms) to cover the whole CND and adopt flat priors over reasonable parameter ranges, except for $\Sigma_{\star,4\arcsec}$ for which we adopt a flat prior in logarithmic space (to ensure efficient sampling of the posterior). The chain performs $10^5$ calculations, the first $40\%$ of which are considered a ``burn-in phase'' and are excluded from the full MCMC. The remaining $60\%$ of the iterations are used to produce the final posterior probability distributions of the free parameters.

The best-fitting parameters and their uncertainties are identified directly from this Bayesian analysis, relying on the likelihood probability distribution functions (PDFs) generated via MCMC. We adopt the median of each posterior PDF as the best fit for that parameter, nearly identical to the minimum $\chi^2$ ($<3\%$ difference in all cases).

As discussed by \citet{vandenBosch09}, the statistical uncertainties can be severely underestimated when working with very large datasets, as systematic uncertainties starts to dominate over statistical ones. They thus proposed an approximate correction, based on the assumption that systematic uncertainties are similar to statistical ones. Accordingly, they suggested increasing the $\chi^2$ difference ($\Delta\chi^2$) required to define a given confidence level by the standard deviation of the $\chi^2$ itself, namely $\sqrt{2(N-P)}\approx\sqrt{2N}$ (where $N$ is the number of constraints and $P$ is the number of free model parameters, here $P=7$; see e.g.\ Section~15.1 of \citealt{Press07}). When working with Bayesian methods rather than $\chi^2$ statistics, an equivalent effect can be achieved by dividing the model $\log$ likelihood by $\sqrt{2N}$ or equivalently multiplying the measurement uncertainties (RMS) by $(2N)^{1/4}$, as done by \citet{Mitzkus17}. The dominance of systematic uncertainties over statistical ones is a generic issue with ALMA data cubes, due to the very large number of high signal-to-noise ratio constraints. This rescaling approach was therefore adopted in a number of recent papers using KinMS and ALMA data \citep[e.g.][]{Nagai19, North19, Smith19, Smith21, Davis20, Nguyen20, Nguyen21a}, to yield more realistic uncertainties.

The best-fitting model has a reduced $\chi^2$ ($\chi^2_{\rm red}$) of $\approx1.15$ for $N=(300\times100-40\times40)\times30=852,000$ constraints (multiplying the uncertainties by $(2N)^{1/4}\approx36$ when calculating $\chi^2_{\rm red}$). This relatively high $\chi^2_{\rm red}$ (given the large number of constraints) is primarily due to the assumption of axisymmetry of the ISM mass distribution and mismatches between model and data in the high-velocity wings of the observed LOS velocity distributions at $|\Delta {\rm Position}|=4\arcsec-8\arcsec$ (see e.g.\ the observed PVD overlaid with the best-fitting model in the inset at the top-right of Fig.~\ref{fig:posterio_pvd}). However, the fit provides an adequate description of the $^{12}$CO(2-1) emission distribution and kinematics in the outer parts of the CND ($r\ge4\arcsec$), and thus of the total mass distribution of the galaxy in that region. The seven best-fitting outer mass model parameters and their statistical uncertainties are listed in Table~\ref{tab:fit}, while their PDFs and two-dimensional (2D) marginalisations are shown in the corner plot of Fig.~\ref{fig:posterio_pvd}. All parameters are well constrained and there is no strong covariance. 

\subsection{Inner mass model}\label{ssec:massinner}

Significant colour variation is seen in the nucleus of NGC~3593 ($r<4\arcsec$ or $r<140$~pc) due to the two stellar populations aligned along the major axis \citep{Coccato13} and dust extinction (primarily to the north of the nucleus). We therefore construct a \hst/WFPC2 $F$450W--$F$814W colour map to create a mass-to-light ratio (\ml) map based on the approximation of $F$450W and $F$814W to $g$- and $i$-band, respectively. The images are first astrometrically-aligned and spatially-convolved to match the PSFs, to mitigate potential spurious colour gradients near the galaxy centre \citep[see][]{Seth10a, Nguyen17, Nguyen18, Nguyen19}. The background level of each image is then estimated in small regions as far away from the galaxy centre as possible (radial range $13\arcsec$--$15\arcsec$ or 455--525 pc) and subtracted off. The inner $10\arcsec\times10\arcsec$ ($350\times350$~pc$^2$) of the resulting colour map is shown in panel~A of Fig.~\ref{fig:colourmaps}. There is a clear colour dichotomy between the northern and southern halves (with respect to the galaxy centre). 

As the stellar populations (e.g. ages and metallicity) in the nucleus of NGC~3593 are similar to those in the nucleus of NGC~5206 \citep[see Figs 7 and 12 of][]{Kacharov18}, we follow the procedure of \citet{Nguyen17, Nguyen18} and use the \citet{Roediger15} colour--\ml\ relation derived from stellar population synthesis models to estimate galaxy stellar masses \citep{Nguyen19}. Such a correlation between colour and \ml\ allows to calculate the \ml\ (and then the stellar mass) based on colour information without knowing the detailed stellar populations and internal ISM extinction. Here, to convert the $F$450W--$F$814W (taken as $\approx g-i$) colour map to a \ml$_{F{\rm814W}}$ map (panel~B of Fig.~\ref{fig:colourmaps}), we adopt the relationship that assumes a Chabrier initial mass function (IMF) and a dust attenuation in molecular clouds and the ambient ISM described by \citet{Charlot00}. Correcting for Galactic foreground extinction and utilising the photometric zero points listed in Table~\ref{tab:hst} and the \hst/WFPC2 $F$814W Vega magnitude system\footnote{\url{http://mips.as.arizona.edu/~cnaw/sun.html}}, the multiplication of this \ml$_{F{\rm 814W}}$ map by the $F$814W-band luminosity surface-density map (panel~C of Fig.~\ref{fig:colourmaps}) yields our desired stellar-mass surface-density map (panel~D of Fig.~\ref{fig:colourmaps}). As clearly seen in the colour map, the central pixels at the putative NSC's location are redder than the surrounding galaxy, resulting in a NSC \ml\ (\ml$_{\rm NSC}$) a factor $5$--$10$ higher than that of the surrounding galaxy.

As the \citet{Charlot00} dust and ISM attenuation prescription is fixed, such a high \ml$_{\rm NSC}$ is most easily understood as high obscuration at short wavelengths in the nucleus. However, it is possible that some of this attenuation is not accounted for accurately, leading to a biased \ml$_{\rm NSC}$. Indeed, we caution that the resulting \ml\ map is dependent on the aforementioned assumptions and is thus subject to uncertainties. According to \citet{Roediger15}, the error budget of the \ml\ map (and thus the resulting stellar-mass surface density) is dominated by the stellar population modelling assumptions, with a colour-dependent bias of up to $0.3$~dex in the optical. Precise constraints on the \ml\ and ISM attenuation ($\tau_V$) require optical long-slit \citep{Nguyen17, Nguyen19} or integral-field \citep{Mitzkus17, Thater19a} spectroscopic data, but no such datum is publicly available for NGC~3593.

In any case, we then again describe (i.e.\ parametrise) the resulting stellar-mass surface-density map using a MGE model. Here  we use the procedure {\tt mge\_fit\_sectors\_regularized} (\citealt{Cappellari02}; see footnote~4) and constrain the allowable axis ratio ($q$) range to $0.39$--$0.95$, to avoid over-constraining the inclination of the $^{12}$CO(2-1) CND during modelling. Due to the significant dust extinction on the northern side of NGC~3593, the mass surface-density distribution remains highly asymmetric. We thus exclude all the pixels on the northern side during the MGE axisymmetric fit. A comparison of the mass surface density (black contours) and its MGE parametrisation (red contours) is shown in Fig.~\ref{fig:massmodelcontour} for the southern half of the galaxy. 

\subsection{Combined stellar-mass and total mass models}\label{ssec:mass}

Now that we have both inner ($r<4\arcsec$) and outer ($4\arcsec\le r\lesssim30\arcsec$) stellar-mass surface-density models, we scale the inner model (i.e.\ panel~D of Fig.~\ref{fig:colourmaps}; see Section~\ref{ssec:massinner}) to match the outer model (i.e.\ the two stellar discs averaged and scaled by the normalisation factor $\Sigma_{\star,4\arcsec}$; see Section~\ref{ssec:massouter}) at the boundary ($r=4\arcsec$ or $140$~pc), as the outer model was constrained more accurately by modelling the outer $^{12}$CO(2-1) kinematics. This simultaneously allows to (1) recalibrate the sky backgrounds previously subtracted from the \hst\ images, that were necessarily contaminated by galaxy light due to the small fields of view, and (2) avoid counting twice the inner stellar mass within $4\arcsec$ mentioned in Section~\ref{ssec:massouter}. We note that the outer stellar-mass surface-density model therefore affects the inner mass surface-density model, that in turn has a strong influence on the inferred SMBH mass (see Section~\ref{sssec:stellarmasserror}). Indeed, the scaling of the inner stellar-mass surface-density model to the outer stellar-mass surface-density model at the radius of $4\arcsec$ yields a combined stellar-mass surface-density model extending to at least $\approx30\arcsec$, that we use for all subsequent dynamical modelling. 

The top panel of Fig.~\ref{fig:massmodel1d} shows a major-axis cut of this combined stellar-mass surface-density model in the form of two truncated MGE models (black and blue open squares), overlaid with its best-fitting MGE parametrisation (red solid line). We note that this combined stellar-mass surface-density cut is not an observable, but was rather constructed from the outer (blue open squares; Section~\ref{ssec:massouter}) and inner  (black open squares; Section~\ref{ssec:massinner}) stellar-mass surface-density maps. The fractional difference between the cut and its best-fitting MGE model is also shown in the bottom panel of Fig.~\ref{fig:massmodel1d} and is $\lesssim10\%$ at all radii. The individual components of this combined stellar-mass surface-density MGE model are listed in Table~\ref{tab:star_mges}.

Given an inclination, any MGE Gaussian component can be deprojected analytically. Applying this to both our combined stellar (Table~\ref{tab:star_mges}) and ISM (Table~\ref{tab:gas_mges}) MGE mass models yields a 3D total (stars + ISM) mass volume-density model of NGC~3593. 

\begin{table}
  \caption{Combined stellar-mass $F$814W MGE model of NGC~3593.}
  \begin{tabular}{cccc}
    \hline\hline  
    $j$ & $\log(\Sigma_{\star,j}/{\rm M}_\odot\,{\rm pc}^2)$ & $\sigma_j\;({\rm arcsec})$ & $q_j$\\
      (1) & (2) & (3) & (4)\\
    \hline
        & Inner component &     &   \\
    \hline
    1 & 5.15 & \phantom{1}0.06 & 0.95\\
    2 & 4.89 & \phantom{1}0.13 & 0.95\\
    3 & 4.57 & \phantom{1}0.18 & 0.95\\
    \hline
        & Outer component & &   \\
    \hline    
    4 & 4.22 & \phantom{1}0.38\phantom{1} & 0.39\\
    5 & 3.61 & \phantom{1}1.25\phantom{1} & 0.87\\
    6 & 3.54 & \phantom{1}2.52\phantom{1} & 0.39\\
    7 & 3.79 & \phantom{1}3.96\phantom{1} & 0.95\\
    8 & 3.72 & 12.56\phantom{1}           & 0.45\\
    \hline
  \end{tabular}
 \parbox[t]{0.472\textwidth}{\textit{Notes:} All quantities as in Table~\ref{tab:gas_mges}.}
  \label{tab:star_mges}
\end{table}

\section{Black Hole Mass Measurement}\label{sec:bh}

\subsection{Results}\label{ssec:results}

We henceforth use our total MGE mass model (ISM and stars; Tables~\ref{tab:gas_mges} and \ref{tab:star_mges}) to constrain the mass of the SMBH in the nucleus of NGC~3593, utilising KinMS kinematic modelling analogous to that described in Section~\ref{ssec:massouter}. However, the stellar mass surface density at $4\arcsec$ ($\Sigma_{\star,4\arcsec}$) is no longer a free parameter (it is now fixed by the fitting of the outer kinematics; see Section~\ref{ssec:massouter}), and we must introduce two new free parameters: the central SMBH mass (\Mbh, with a flat prior in logarithmic space) and a stellar-mass scaling factor $\Gamma$, effectively the ratio of the dynamical and stellar population mass-to-light ratios ($\Gamma\equiv(M/L)_{\rm dyn}/(M/L)_{\rm pop}$), thus scaling the stellar potential of the galaxy. We also leave the disc inclination $i$ to vary to explore any possible variation associated with the inclusion of the inner $4\arcsec$ CND. Thus, this final KinMS model used to constrain the central SMBH mass (and optimised to fit the molecular gas observations at all radii) has eight free parameters: \Mbh, $\Gamma$, $f$, $i$, $x_{\rm c}$, $y_{\rm c}$, $v_{\rm off}$, $\sigma_0$. We run the model in the same manner as in Section~\ref{ssec:massouter}, with a total number of iterations of $10^5$ and the first $40\%$ of the iterations considered as the burn-in phase, yielding our final posterior PDFs.

\begin{figure*}
  \centering\includegraphics[scale=0.5]{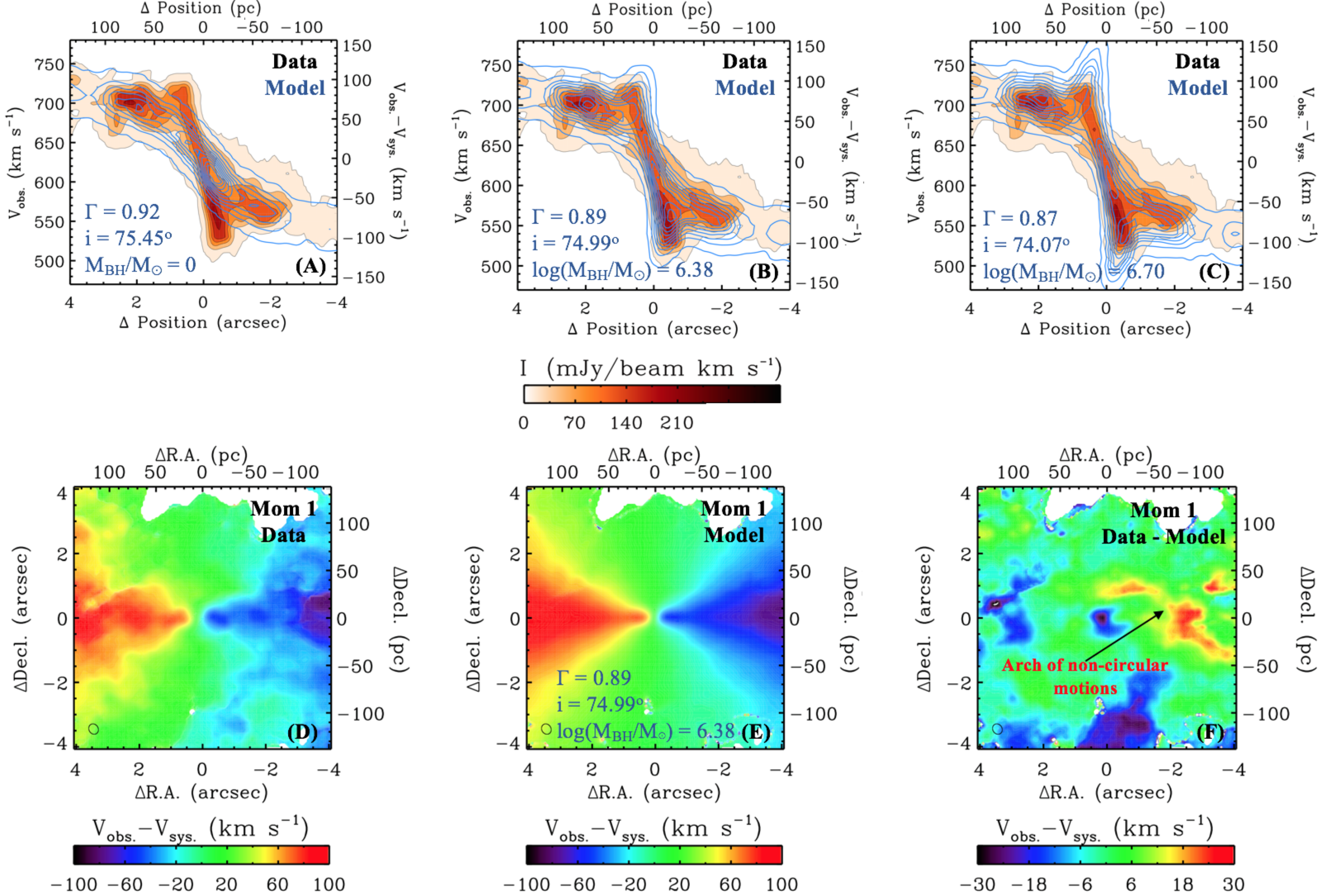} 
  \caption{As the first two rows of Fig.~\ref{fig:bestfitmaps2}, but for a much smaller field-of-view of $4\arcsec\times4\arcsec$ ($\approx140\times140$~pc$^2$).}
  \label{fig:bestfitmaps}   
\end{figure*}

We infer a SMBH mass that causes increasing rotation toward the centre as the radius decreases, although the NSC and CMC also contribute significant mass there. In fact, the significant contributions of these two compact components can be seen in the PVDs of the inset in Fig.~\ref{fig:posterio_pvd} (model with the CMC but no NSC nor SMBH) and panel~A of Fig.~\ref{fig:bestfitmaps2} (model with the CMC and NSC but no SMBH), that show models without a SMBH overlaid on the data. The best-fitting model without a SMBH (\Mbh$\,=0$~\Msun, $\Gamma=0.92$ and $i=75\fdg45$; with $\chi^2_{\rm red}\approx1.04$) does not fit the data well in the centre, and it is clear that the observed molecular gas kinematics call for a central SMBH. 

As listed in Table~\ref{tab:fit2}, the best-fitting KinMS model with a SMBH has (\Mbh, $\Gamma$, $i$)~= ($2.40_{-1.05}^{+1.87}\times10^6$~\Msun, $0.89_{-0.03}^{+0.06}$, $74\fdg99_{-0.55}^{+0.50}$) with $\chi_{\rm red}^2\approx1.01$ (all uncertainties are quoted at the $3\sigma$ statistical level), and the central molecular gas kinematics are now well reproduced. The resulting PVD is shown in panel~B of Fig.~\ref{fig:bestfitmaps2}, overlaid on the data. For comparison, we also show in panel~C of Fig.~\ref{fig:bestfitmaps2} a KinMS model with an overly massive SMBH ($M_{\rm BH}=5.02\times10^6$~\Msun, $\Gamma=0.87$ and $i=74\fdg07$) with $\chi_{\rm red}^2\approx1.06$, that again does not compare well to the data in the centre. We note that for the two models shown in panels~A and C of Fig.~\ref{fig:bestfitmaps2}, we varied \Mbh, $\Gamma$ and $i$ only while keeping the gas CND and other nuisance parameters fixed to those of the best-fitting model (see Table~\ref{tab:fit2}). The observed, best-fitting model and residual ({\tt data-model}) $^{12}$CO(2-1) mean LOS velocity maps are also shown in panels~D, E and F of Fig.~\ref{fig:bestfitmaps2}, respectively, to illustrate how well the model reproduces the data.

Additionally, the panels~G, H, and I of Fig.~\ref{fig:bestfitmaps2} show the observed, best-fiting model and residual maps of the $^{12}$CO(2-1) LOS velocity dispersion, respectively, while the the panels~J, K, and L of Fig.~\ref{fig:bestfitmaps2} show the analogous integrated intensity maps. At least some of the turbulent/non-circular motions visible in the residual velocity map (panel~F of Fig.~\ref{fig:bestfitmaps2}) and unaccounted for by our (axisymmetric) model are associated with regions of high velocity dispersion (panel~G of Fig.~\ref{fig:bestfitmaps2}) and high surface brightness (panel~J of Fig.~\ref{fig:bestfitmaps2}), possibly indicating the presence of streaming gas (inflow and/or outflow), shocks, turbulence and/or filaments.

As only the very central region of $4\arcsec\times4\arcsec$ (or $\approx140\times140$~pc$^2$) of the $^{12}$CO(2-1) kinematics matters to constrain the SMBH mass, we show a zoomed-in version of Fig.~\ref{fig:bestfitmaps2} (PVDs and mean LOS velocity maps only) in Fig.~\ref{fig:bestfitmaps}. The central rapidly rising velocities (as the radius decreases) of the molecular gas disc due to the central SMBH dominate within a radius $r\approx0\farcs5$ ($\approx17.5$~pc). However, the velocity residuals are significant (up to $\approx20$~\kms) in an arc on the west side of the nucleus, that will be discussed further in Section~\ref{ssec:morpho}.

Fig.~\ref{fig:posterial_full} shows the PDF and 2D marginalisations of each of the eight free parameters of our SMBH fit. The best-fitting model parameter is indicated by a vertical solid line in each PDF.  The uncertainties resulting from the PDFs are indicated by vertical dashed lines at the $1\sigma$ ($16$--$84\%$) confidence levels, while the contours in the 2D marginalisations show $0.5\sigma$ ($31$--$69\%$), $1\sigma$ ($16$--$84\%$), $2\sigma$ ($2.3$--$97.7\%$) and $3\sigma$ ($0.14$--$99.86\%$) confidence levels. Most of the parameters are well-constrained by the data, although as expected there is a significant covariance between \Mbh\ and $\Gamma$, arising from the degeneracy between the potential of the SMBH and that of the stars and ISM when the observations do not adequately spatially resolve the SMBH's SOI.

\begin{figure}
  \centering\includegraphics[scale=0.45]{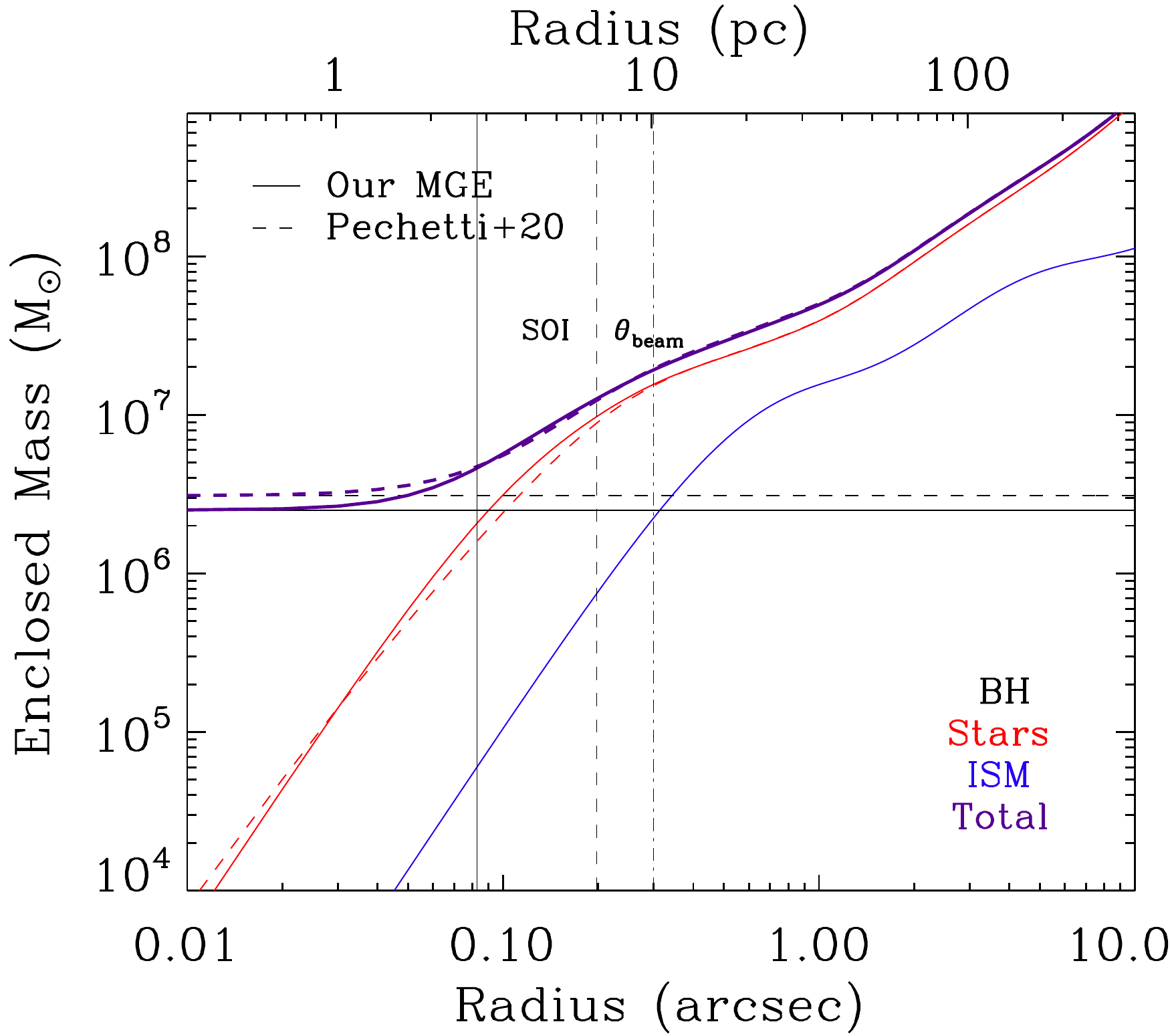}
  \caption{Cumulative mass distribution of the stars, ISM and BH of NGC~3593 for our best-fitting model (solid curves) and the NSC stellar-mass model modified according to \citet{Pechetti20} (dashed curves). The ISM (primarily the molecular gas) does not dominate the mass at any radius, but it does contribute significantly, especially at larger radii. The $R_{\rm SOI}$ (as defined in the text) of our best-fitting model and the NSC-modified model are indicated by the vertical solid and dashed lines, respectively. The synthesised beam of $\approx0\farcs3$ is indicated by the the vertical dot-dashed line.}
  \label{fig:cummass}   
\end{figure}

We show in Fig.~\ref{fig:cummass} the enclosed mass distributions (stars, ISM and BH) of our best-fitting model, and return to it later to gauge the robustness of this model.

Given the best-fitting $M_{\rm BH}\approx2.4\times10^6$~\Msun\ and the central stellar velocity dispersion $\sigma_\star\approx60$~\kms\ \citep[$r\lesssim5\arcsec$ or $r\lesssim175$ pc;][]{Bertola96}, the BH in NGC~3593 has a nominal SOI radius $R_{\rm SOI}\equiv GM_{\rm BH}/\sigma_\star^2\approx3.0$~pc ($\approx0\farcs09$). The $R_{\rm SOI}$ is thus $\approx3.5$ times smaller than what our ALMA observations (i.e.\ our synthesised beam of $\approx0\farcs30$) can spatially resolve.

\citet{Davis14}, \citet{Barth16a, Barth16b}, \citet{Boizelle19, Boizelle21}, and \citet{Nguyen20} demonstrated that the angular resolution $\theta_{\rm FWHM}$ required to perform reliable \Mbh\ measurements should satisfy $\theta_{\rm FWHM}\lesssim2\times\theta_{\rm R_{\rm SOI}}$, where $\theta_{\rm R_{\rm SOI}}$ is the angle subtended by $R_{\rm SOI}$. Measurements using data with poorer angular resolutions (i.e.\ larger synthesised beams) are more susceptible to systematic biases from stellar mass uncertainties. Our ALMA observations of NGC~3593 thus belong to the majority of \Mbh\ measurements with ALMA and CARMA, that have $\theta_{\rm FWHM}\gtrsim2\times\theta_{\rm R_{\rm SOI}}$ \citep{Davis13, Davis17, Onishi15, Onishi17, Smith19, Smith21, Nagai19, Nguyen20, Nguyen21a, Thater19b}. This suggests our \Mbh\ estimate would benefit from observations at higher angular resolutions, to further reduce the uncertainties arising from our stellar-mass model (but see Section~\ref{sssec:stellarmasserror} below).     

In addition to our limited synthesised beam (compared to the SMBH SOI), the (stellar) mass of the NSC is likely to be the greatest source of uncertainty on the SMBH mass, as it could be degenerate with \Mbh\ (see items (iii) and (iv) in Section~\ref{sssec:stellarmasserror}). However, it is worth noting here that the NSC is spatially resolved by the {\it HST} observations, which partially suppresses this stellar-mass uncertainty/degeneracy and leads to a statistically-significant rejection of the \Mbh$=0$ hypothesis. Indeed, the FWHM of the  all three MGE components of the NSC mass model are greater than the PSF of the HST data ($\theta_{\rm HST}\approx0\farcs08$; ${\rm FWHM}_{j=1,2,3}=2.35\,\sigma_{j=1,2,3}>\theta_{\rm HST}$; see Table~\ref{tab:star_mges}) and the NSC is almost spatially-resolved ($r_{\rm NSC}\approx2\theta_{\rm HST}$; see Table~\ref{tab:sersics}). 

\begin{table}
  \caption{Model parameters best fitting the inner part of the $^{12}$CO(2-1) disc.} 
    \begin{tabular}{lc cccc} 
    \hline\hline       
    Parameter & Search range & Best fit & $1\sigma$ uncertainty & $3\sigma$ uncertainty\\
              & & & ($1$--$84\%$) & ($0.14$--$99.86\%$) \\  
       (1) & (2) & (3) & (4) & (5)\\  
    \hline 
    {\it Black hole:} & & & &\\
    $\log(M_{\rm BH}/{\rm M}_\odot)$& $(1\to9)$  & $6.38$ & $-0.08$, $+0.07$ & $-0.25$, $+0.25$\\
    $\Gamma$ &$(0.1\to2.0)$  & $0.89$ &$-0.01$, $+0.02$ &$-0.03$, $+0.06$\\[1mm]
    {\it Gas CND:} & & & &\\     
    $f$ (Jy~\kms) & ($10^2-5\times10^3$)  & $1216.00$ & $-0.17$, $+0.18$ & $-0.57$, $+0.59$\\
    $\sigma_0$ (\kms)  & $(1\to50)$ & $15.01$ & $-0.14$, $+0.15$ & $-0.45$, $+0.45$\\
    $i$ ($\degr$) & $(70\to90)$ & $74.99$ &  $-0.14$, $+0.12$ & $-0.55$, $+0.50$\\[1mm]
    {\it Nuisance:} & & && \\
    $x_{\rm c}$ ($\arcsec$) & $(-1.0\to1.0)$   & $0.00$ & $-0.04$, $+0.04$ & $-0.12$, $+0.12$\\
    $y_{\rm c}$ ($\arcsec$) & $(-1.0\to1.0)$  & $0.00$ & $-0.03$, $+0.03$ & $-0.09$, $+0.09$\\
    $v_{\rm off}$ (\kms)       & $(-50\to50)$  & $32.98$& $-0.16$, $+0.14$ & $-0.50$, $+0.45$\\
     \hline
  \end{tabular}
 \parbox[t]{0.472\textwidth}{\textit{Notes:} Same as Table~\ref{tab:fit} but for the inner part of the $^{12}$CO(2-1) CND, and with the central SMBH mass (\Mbh) and mass-scaling factor ($\Gamma$) instead of the stellar-mass surface density at $4\arcsec$ ($\Sigma_{\star,4\arcsec}$). Here, we also list the uncertainties at the $1\sigma$ confidence level.}
  \label{tab:fit2}
\end{table}

\begin{figure*}
  \centering\includegraphics[scale=0.5]{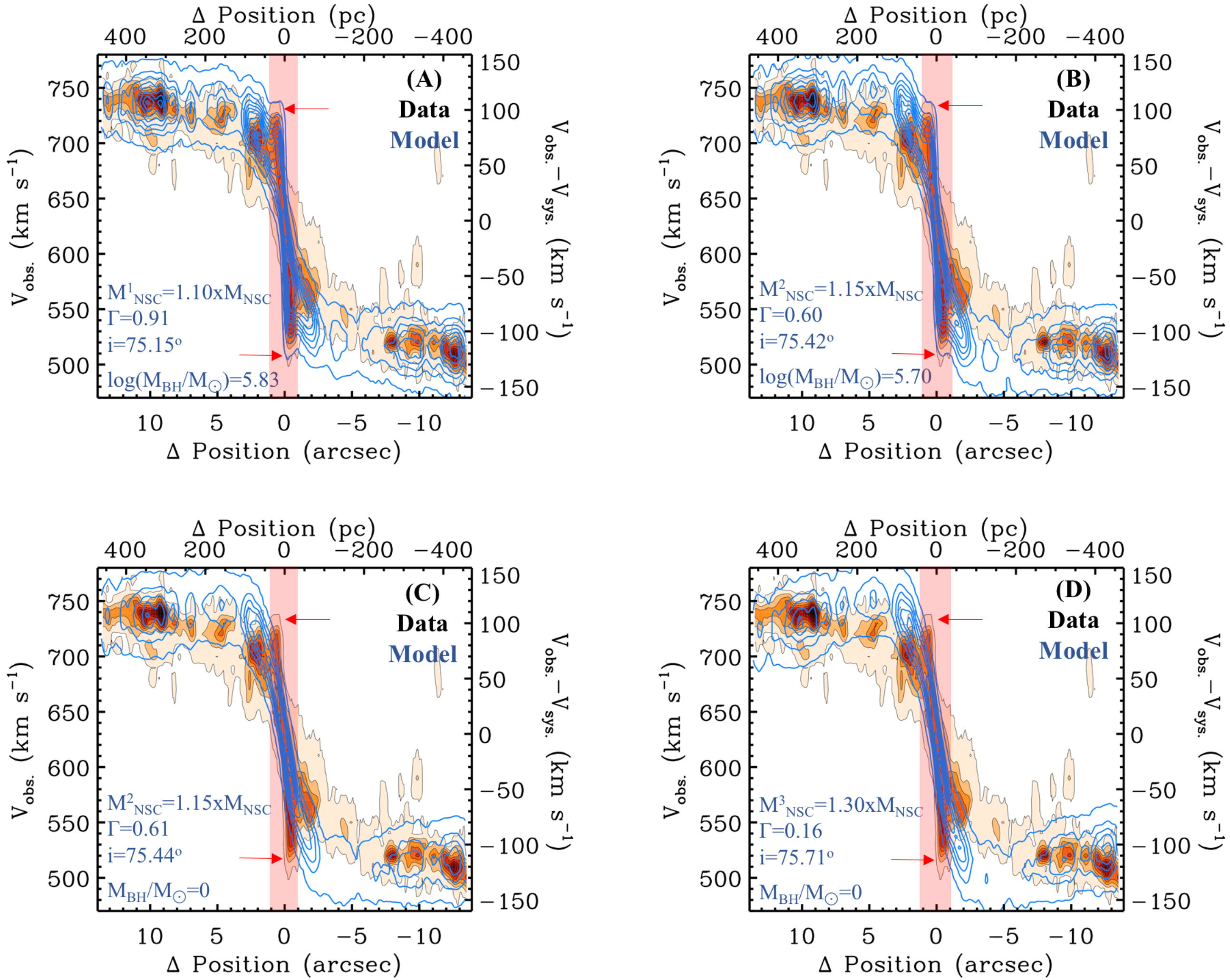} 
  \caption{PVD of the $^{12}$CO(2-1) emission of NGC~3593 extracted along the kinematic major-axis (orange scale and grey contours), overlaid with the modelled PVDs (blue contours) of the best-fitting models with different assumptions of NSC mass. The parameters of the best-fitting models are shown in the legends, while the red arrows indicate the rapidly rising rotation velocities (as the radius decreases) in the centre,  putatively caused by a central SMBH. Although \Mnsc\ and $\Gamma$ dictate the shapes of the PVDs at large radii, the central rising velocities require a SMBH.} 
      \label{fig:lowerlimit_BH}   
\end{figure*}

\subsection{Uncertainties}\label{ssec:masserror}

We test here the robustness of our dynamical model (and the inferred best-fitting parameters) under the influence of sources of errors other than the uncertainties in the ALMA kinematics, $i$ and $\Gamma$ (discussed in Section~\ref{ssec:results}). In these fits, except for \Mbh, $i$ and $\Gamma$, we fix all (CND and other nuisance) parameters to their best-fitting values listed in Table~\ref{tab:fit2}.

\subsubsection{Stellar mass models}\label{sssec:stellarmasserror}

Our inner mass model constructed from \hst\ images under the assumption of the colour--\ml\ relation of \citet{Roediger15} (that assumes the Chabrier IMF and the \citealt{Charlot00} dust attenuation correction) has a large impact on our model results. We therefore further examine the associated uncertainties, by considering other independent models constructed with different assumptions.

\begin{enumerate}

\item {\it Mass model without mask:} To create our inner stellar-mass model, we masked out most of the northern side of the nucleus, that is strongly extincted by dust. We therefore create here another stellar-mass MGE model without masking.

\item {\it Mass model using F450W filter:} We also test a mass model created from the \hst\ $F$450W (rather than $F$814W) image. As in Section~\ref{ssec:massinner}, we use the \citet{Roediger15} colour--\ml\ relation to derive a \ml$_{F{\rm450W}}$ map and in turn a $F$450W-based stellar-mass map, and then parametrise the resulting stellar-mass map using MGE. We then use this $F$450W-based MGE stellar-mass model (rather than that in Table~\ref{tab:star_mges}) in conjunction with KinMS for the kinematic modelling.

\item {\it Degeneracy between our NSC mass model and \Mbh:} In Section~\ref{ssec:nsccmc}, we will identify the summation of the three innermost MGE components of Table~\ref{tab:star_mges} as our NSC. That NSC is then $10\%$ smaller in size ($\approx R_{\rm SOI}$) but $10$ times smaller in mass than the NSC of \citet{Pechetti20}. As seen in panels~A of Figs~\ref{fig:bestfitmaps2} and \ref{fig:bestfitmaps}, our KinMS model with a NSC ($M_{\rm NSC}=1.67\times10^7$~\Msun) but no SMBH does not fit the data well in the centre (which requires a $\sim10^6$~\Msun\ SMBH; see panels~B of Figs~\ref{fig:bestfitmaps2} and \ref{fig:bestfitmaps}). This suggests that the masses of the NSC and SMBH of NGC~3593 may not be as degenerate as originally feared, at least on the spatial scale of $R_{\rm SOI}$ ($R_{\rm SOI}\approx3.0$~pc, see Section~\ref{ssec:results}; $r_{\rm NSC,e}\approx5.0$~pc, see Section~\ref{ssec:nsccmc} and Table~\ref{tab:sersics}). 
  
We nevertheless test this with models with a SMBH and slightly increased NSC masses (or equivalently stellar-mass surface densities, i.e.\ increasing $\Sigma_{\star,j=1,2,3}$ but keeping $\sigma_{j=1,2,3}$ and $q_{j=1,2,3}$ fixed in Table~\ref{tab:star_mges}), without increasing the more extended stellar-mass distribution (i.e.\ keeping the $j=4$--$8$ components fixed in Table~\ref{tab:star_mges}). Specifically, we increase the NSC mass by factors of $1.10$ ($10\%$, $M^1_{\rm NSC}\approx1.9\times10^7$~\Msun), $1.15$ ($15\%$, $M^2_{\rm NSC}\approx2.0\times10^7$~\Msun) and $1.30$ ($30\%$, $M^3_{\rm NSC}\approx2.2\times10^7$~\Msun) and check for variations of the best-fitting $M_{\rm BH}$.

While the first new NSC mass ($M^1_{\rm NSC}$) is chosen to yield a slightly smaller $M_{\rm BH}$, the second ($M^2_{\rm NSC}$) is chosen to explore the possibility of a SMBH mass close to zero. The best-fitting results are listed in Table~\ref{tab:bhbest} and shown in panels~A and B of Fig.~\ref{fig:lowerlimit_BH}, indicating as expected that in both cases a good fit to the $^{12}$CO(2-1) kinematics in the central $2\arcsec\times2\arcsec$ region (see the red arrows in Fig.~\ref{fig:lowerlimit_BH}) requires a SMBH mass smaller than that of our default best-fitting model (see Section~\ref{ssec:results} and Table~\ref{tab:fit2}). However, in both cases there are also significant mismatches between the data and the best-fitting model outside the centre ($1\arcsec\lesssim r\lesssim7\arcsec$), due to the increasing effect of the NSC's gravitational potential, and indeed the \Mbh\ PDF robustly excludes \Mbh$=0$ in both cases.  

To further test the presence of a SMBH in our models, the third new NSC mass ($M^3_{\rm NSC}$) is chosen to entirely remove the need for a SMBH. To achieve this, we choose a mass ($M^3_{\rm NSC}\approx2.2\times10^7$~M$_\odot$) equal to the highest NSC mass allowed by our default best-fitting model (see Table~\ref{tab:sersics}), nearly $10$ times the default best-fitting SMBH mass. We show the best-fitting $M^3_{\rm NSC}$ model in panel~D of Fig.~\ref{fig:lowerlimit_BH}, imposing \Mbh$=0$, and revealing that the rapidly rising rotation velocities of the $^{12}$CO(2-1) gas in the centre (as the radius decreases) cannot be fit without a SMBH. Imposing \Mbh$=0$ for $M^2_{\rm NSC}$ as well, panel~C of Fig.~\ref{fig:lowerlimit_BH} shows that, as expected, the problem becomes more acute for smaller $M_{\rm NSC}$. In fact, focussing exclusively on the central $2\arcsec\times2\arcsec$ region (and ignoring the increasing mismatch at large radii), dynamical models with $M_{\rm NSC}=M^3_{\rm NSC}$ still require a SMBH with \Mbh$>3\times10^5$~\Msun\ to match the the rapidly rising rotational velocities in the centre (see Table~\ref{tab:bhbest}).

Given the accuracy of our nuclear stellar-mass model based on the \citet{Roediger15} colour--$M/L$ relation (see Section~\ref{ssec:massinner} and Table~\ref{tab:star_mges}) and the S\'ersic fit to the NSC discussed in Section~\ref{ssec:nsccmc}, these tests put a firm lower limit on the SMBH mass in the nucleus of NGC~3593 ($M_{\rm BH}>3\times10^5$~M$_\odot$, including all possible uncertainties), and thus make our claim of a detection of a SMBH at the heart of NGC~3593 strong, this despite a synthesised beam size that is $3.5$ times larger than $R_{\rm SOI}$ (see Section~\ref{ssec:results}).
  
\item {\it Mass model of the NSC:} In item (iii) above, we tested the impact of the NSC on our \Mbh\ estimate by increasing its surface density and thus mass only (i.e.\ we varied $\Sigma_{\star,j}$ but kept $\sigma_j$ and $q_j$ fixed). Here, we further test this impact by considering variations of the NSC shape (i.e.\ $\sigma_j$ and $q_j$). For this, we replace the first three Gaussians of our (inner) stellar-mass MGE model (our NSC defined in item (iii) above; see Table~\ref{tab:star_mges}) by the MGE components obtained from the product of the NSC S\'{e}rsic light profile and mass-to-light ratio of \citet{Pechetti20} (listed in Table~\ref{tab:ngc3593property}; see Section~\ref{ssec:nsccmc} for the reasons behind an order of magnitude difference between the mass of this assumed \Mnsc\ and our own) and run our KinMS model again. The enclosed mass distributions (stars, ISM and BH) of this modified model is shown in Fig.~\ref{fig:cummass}, compared to that of our best-fitting model. The two profiles are nearly indistinguishable (including statistically consistent \Mbh), clearly demonstrating that our $M_{\rm BH}$ and $M/L_{F{\rm814W}}$ (i.e.\ $\Gamma$) estimates are robust against systematic but realistic changes to our NSC mass model (see also Section~\ref{ssec:nsccmc} and Table~\ref{fig:colourmaps}).

\end{enumerate}
 
The best-fitting parameters of the six modified models above are listed in Table~\ref{tab:bhbest}, along with those of our best-fitting model (Section~\ref{ssec:results} and Table~\ref{tab:star_mges}), demonstrating that our results are robust against reasonable changes of the stellar-mass model.

\subsubsection{Constant $M/L$ and variable dust extinction}\label{sssec:constml}

We note that our approach so far has essentially assigned all colour variations to stellar population (and thus \ml) variations (although the \citealt{Roediger15} colour--\ml\ relation does have a prescription for dust attenuation in the ISM). However, there is significant dust extinction near the major axis in the nucleus ($|\Delta{\rm Decl.}|\lesssim1\arcsec$; see panel~A of Fig.~\ref{fig:colourmaps}), leading to \ml\ variations of a factor of a few in that region and up to a factor of $\approx10$ in the NSC. It is unclear if these variations are indeed real, or if they could instead be due to dust that has been misaccounted for in the \citet{Roediger15} model. This could lead to a significant overestimate of the stellar-mass surface densities in the nucleus, and in turn an underestimate of the inferred SMBH mass.

Given that the galaxy likely contains two stellar populations in co-spatial counter-rotating discs (see Section~\ref{sec:ngc3593}), it may be that the stellar populations (and intrinsic colours and \ml) are rather uniform across the nucleus. Here, we therefore assume that the stellar populations are indeed uniform and adopt a single intrinsic colour (and thus \ml\ according to the \citealt{Roediger15} colour--\ml\ relation using the Chabrier IMF and \citealt{Charlot00} attenuation prescription) across the whole field-of-view (inner part), in effect assigning all colour variations to dust extinction.

The $F$450W--$F$814W and \ml$_{F{\rm814W}}$ maps (panels~A and B of Fig.~\ref{fig:colourmaps}) suggest that the southern half of the nucleus is largely dust free. First, we therefore adopt the typical colour and \ml\ of that region ($F$450W--$F$814W~$\approx1.7$~mag and \ml$_{F{\rm814W}}\approx1.7$~${\rm M}_\odot/{\rm L}_{\odot,\;F\rm814W}$) as our unique colour and \ml\ for the entire FOV (inner part).

Next, we assume $F$450W~$\approx B$ and $F$814W~$\approx I$, adopt an intrinsic colour $F$450W--$F$814W~$\approx(B-I)_0\approx1.7$~mag, and use another form of the Milky Way extinction law\footnote{\url{http://www.astro.sunysb.edu/metchev/PHY517\_AST443/extinction\_lab.pdf}}, i.e.\ $A_I=0.572\times E(B-I)=0.572\times[(B-I)-(B-I)_0]$, to correct our $F$814W map for this dust extinction {\em pixel-by-pixel} over the entire $F$450W--$F$814W colour map (panel~A of Fig.~\ref{fig:colourmaps}). For example, in the central region co-spatial with the NSC and CMC, $(B-I)\approx3.6$~mag, yielding an $I$-band dust extinction $A_I\approx1.1$~mag. This process yields an $I$-band image corrected pixel-by-pixel for dust extinction, that we multiply by our unique \ml$_I$ pixel-by-pixel to get the stellar-mass surface-density map (corrected pixel-by-pixel for dust extinction), that we finally parametrise with MGE as usual.

Using this arguably extinction-free stellar-mass model, we re-run our KinMS kinematic model and obtain $M_{\rm BH}=1.54_{-0.57}^{+0.49}\times10^6$~\Msun, $\Gamma=0.80_{-0.15}^{+0.16}$ and $i=73\fdg86_{-0.57}^{+0.78}$ (also listed in Table~\ref{tab:bhbest}). This suggests that colour variations purely due to dust extinction (i.e.\ a fixed stellar population and thus \ml) lead to a central SMBH mass $\approx36\%$ smaller than that of our default best-fitting model assigning colour variations primarily to stellar population (and thus \ml) variations. The mass scaling factor $\Gamma$ (or equivalently the stellar \ml) is also correspondingly smaller by $\approx10\%$. Given our adopted colour ($F$450W--$F$814W~$\approx1.7$~mag) and \ml\ (\ml$_{F{\rm814W}}\approx1.7$~${\rm M}_\odot/{\rm L}_{\odot,\;F\rm814W}$) for the entire inner part of the FOV ($r<4\arcsec$ or $r<140$ pc), this lighter SMBH is expected. Indeed, the dust-extinction correction in the nucleus is then larger than that of our default model, leading to higher corrected (i.e.\ intrinsic) surface brightnesses and thus masses (including for the NSC), and thus to a smaller \Mbh.

\subsubsection{ISM disc}\label{sssec:ismerror}

Given our inferred SMBH mass and associated $R_{\rm SOI}$ ($\approx3.5$ times smaller than the synthesised beam), the ISM mass contained within the central beam ($\approx0\farcs3$ or $\approx10.5$ pc) is dynamically significant, $\approx2\times10^6$~\Msun\ (see Fig.~\ref{fig:cummass}) or $\approx83\%$ of \Mbh\ and slightly greater than the $3\sigma$ \Mbh\ uncertainty (see Table~\ref{tab:bhbest}). We thus also test the impact of the ISM on the inferred \Mbh, removing the ISM whose mass model is listed in Table~\ref{tab:gas_mges} (see Section~\ref{ssec:massouter}) and turning on the \texttt{gasGrav} funtion in the KinMS fit, that assumes the ISM mass is distributed according to the input $^{12}$CO(2-1) surface-brightness profile (thus also removing the inconsistency between the ISM mass model and $^{12}$CO(2-1) surface brightness model noted in Section~\ref{ssec:massouter}). This test yields $M_{\rm BH}=2.1_{-0.77}^{+0.54}\times10^6$~\Msun, $\Gamma=0.90_{-0.07}^{+0.07}$ and $i=75\fdg05_{-0.54}^{+0.52}$ (also listed in Table~\ref{tab:bhbest}), almost identical and fully consistent with the results from our default best-fitting model, analogous to the behaviour reported by \citet{Nguyen20}. Our best-fitting \Mbh\ is thus not sensitive to the ISM distribution, as long as the total ISM mass is accurately estimated.

\subsubsection{CO-to-H$_2$ conversion factor}\label{ssec:xco} 

NGC~3593 is classified as a starburst galaxy \citep{Hunter89}, but has a CND SFR of $\lesssim3$~\Msun~year$^{-1}$ \citep{Ho97}, much lower than the typical SFR of starbursts ($10$--$100$~\Msun~year$^{-1}$). This suggests that our adoption of the starbusrt CO-to-H$_2$ conversion factor ($X_{\rm CO}=(1.0\pm0.3)\times10^{20}$~cm$^{-2}$~(K~\kms)$^{-1}$; e.g.\ \citealt{Kuno00, Kuno07, Bolatto13}) in Section~\ref{ssec:massouter} may be inappropriate and may underestimate the total molecular gas mass by a factor of $\approx2$. We therefore test this hypothesis by adopting instead the conversion factor of the Milky Way ($X_{\rm CO}=(2.0\pm0.3)\times10^{20}$~cm$^{-2}$~(K~\kms)$^{-1}$; e.g.\ \citealt{Bolatto13}), that is arguably better suited to targets hosting molecular gas reservoirs with ${\rm SFR}\lesssim3$~\Msun~year$^{-1}$. The best-fitting KinMS model then yields $M_{\rm BH}=1.51_{-0.54}^{+0.83}\times10^6$~\Msun, $\Gamma=0.82_{-0.12}^{+0.12}$ and $i=75\fdg35_{-0.48}^{+0.50}$ (also listed in Table~\ref{tab:bhbest}), suggesting the choice of the conversion factor, and hence the derived molecular gas mass, has a significant effect on the SMBH mass determination (possibly slightly beyond the statistical uncertainties of our default model).

To test whether the ISM could altogether negate the need for a SMBH, we also tested more extreme values of the conversion factor $X_{\rm CO}$, artificially increasing the total gas mass (via $X_{\rm CO}$ and thus $M_{\rm H_2}$) and adjusting $\Gamma$ (otherwise fixing the stellar mass model and $M_{\rm \ion{H}{i}}$) to fit the CND kinematics. For the best-fitting model to allow $M_{\rm BH}=0$, a total gas mass $M_{\rm H_2}+M_{\rm \ion{H}{i}}\gtrsim3\times10^9$~M$_\odot$ is required, more than $10$ times the galaxy's nominal gas mass. Such high conversion factors ($>10$ times the norm) are out of the range typically expected for the metallicity and luminosity of NGC~3593 \citep{Bolatto13}, thus confirming yet again the need for a central SMBH.

\subsubsection{Asymmetric drift correction}\label{ssec:asymdrift} 

\citet{Barth16a} and \citet{Boizelle19} have suggested that pressure support may significantly bias SMBH mass estimates obtained from modelling the kinematics of molecular gas, due to potentially high turbulent velocity dispersions. In NGC~3593, except for a few regions with $^{12}$CO(2-1) LOS velocity dispersions $\sigma\approx25$--$33$~\kms\ (coincident with clumps of high emission; see panel~D of Fig.~\ref{fig:co21moms} and panel~C of Fig.~\ref{fig:co21moms2}), the molecular gas velocity dispersion is generally low ($\sigma\approx10$--$15$~\kms). Compared to the observed rotation velocities $V_{\rm rot}$, $\sigma/V_{\rm rot}<0.1$ in the entire CND, suggesting that the rotation velocities are similar to the circular velocities (i.e.\ $V_{\rm rot}\approx V_{\rm circ}$). Any asymmetric drift correction is thus likely to have a minimal impact on our results.

To more rigorously test this, given such small $\sigma/V_{\rm rot}$, we assume that the random motions in the CND are approximately equal in the radial ($\sigma_{\rm r}$) and vertical ($\sigma_{\rm z}$) directions (i.e.\ $\sigma_{\rm r}=\sigma_{\rm z}$) and that both are equal to the observed velocity dispersion $\sigma$ ($\langle V_{\rm r}V_{\rm z}\rangle=0$). We can then calculate the ratio of the velocity dispersions in the radial and azimuthal directions $\sigma_{\rm r}/\sigma_\phi=\sigma/\sigma_\phi$ using the epicycle approximation (Eq.~4-33 of \citealt{Binney87}), and the asymmetric drift correction can then be estimated as 
\begin{equation*}
  V_{\rm circ}^2-V_{\rm rot}^2=\sigma^2\left[-r\dfrac{d\ln\Sigma_{{\rm H}_2}}{dr}-r\dfrac{d\ln\sigma^2}{dr}-\left(1-\dfrac{\sigma^2}{\sigma_\phi^2}\right)\right]~.
\end{equation*}
This yields corrections $<20\%$, so we conclude that any asymmetric drift correction (and associated uncertainties) will be small compared to other potential sources of error in this low-mass object. In turn, this suggests that the thin disc assumption is good enough to describe the nuclear molecular gas of NGC~3593, consistent with our earlier assumption of thin disc of the {\tt SkySampler} (see Section~\ref{ssec:massouter}).

\section{Central Cores}\label{sec:cores}

\subsection{Nuclear star cluster and massive core of cold molecular gas}\label{ssec:nsccmc}

\begin{figure}
  \centering\includegraphics[scale=0.105]{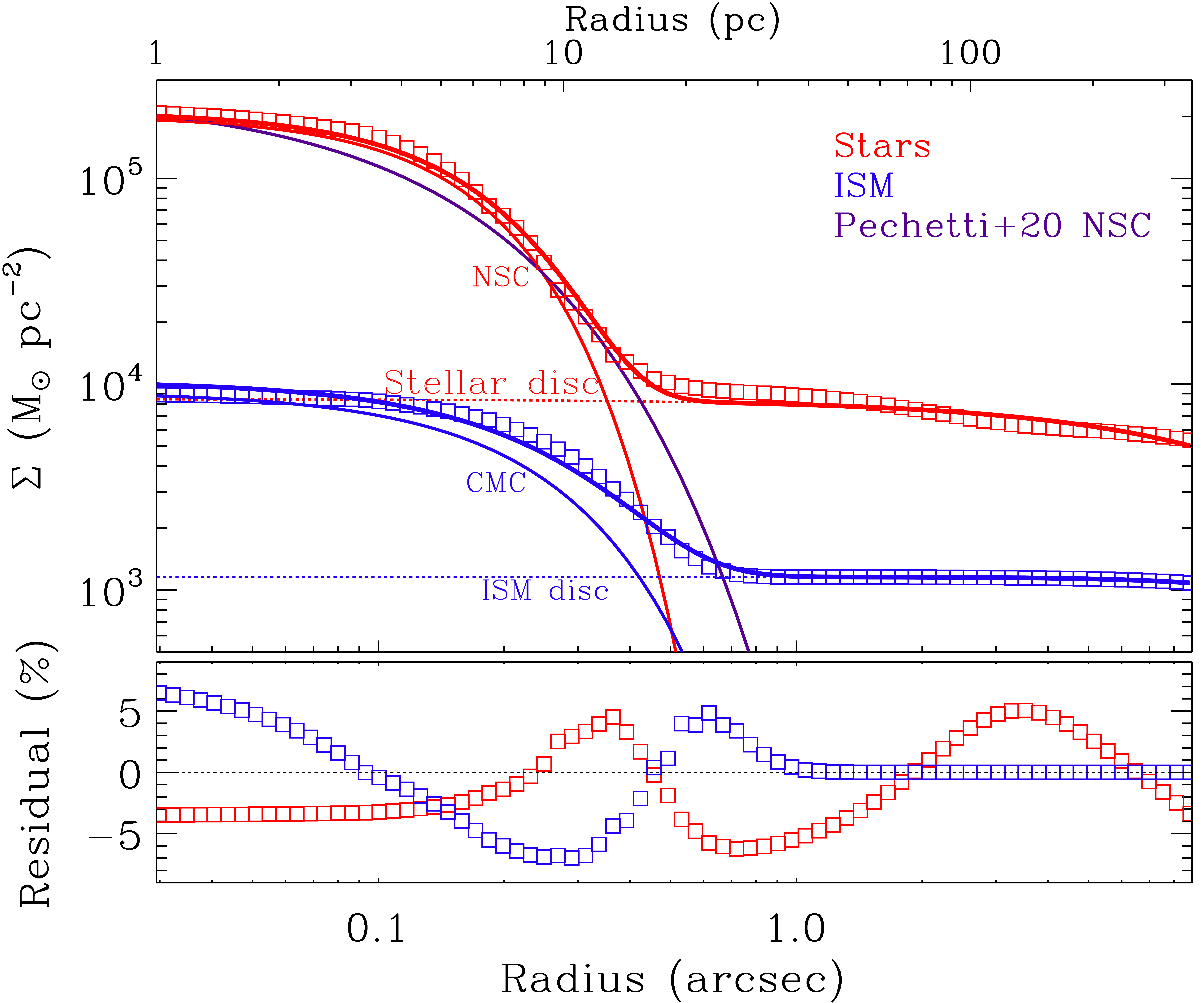}
  \caption{{\bf Top:} Radial mass surface-density profile (open squares) of the stars (red) and ISM (blue) of NGC~3593, overlaid with the best-fitting narrow S\'{e}rsic component (thin solid line; respectively NSC and CMC), broad S\'{e}rsic component (dotted line; respectively stellar disc and ISM disc) and total profile (thick solid line) in matching colour (see Table~\ref{tab:sersics}). The purple thin line is the NSC of \citet{Pechetti20}. {\bf Bottom:} fractional residuals ((data-model)/data).} 
  \label{fig:nsc}   
\end{figure}

We hereby characterise the morphological properties of the CMC and NSC, estimating their masses and sizes and comparing them to estimates from our stellar and ISM (MGE) mass models. We use the \texttt{Image Reduction and Analysis Facility} (\texttt{IRAF}) \texttt{ellipse} task \citep{Jedrzejewski87a} to extract radial mass surface-density profiles of the stars (panel~D of Fig.~\ref{fig:colourmaps}, i.e.\ $F$814W emission corrected for stellar populations and dust; see Section~\ref{ssec:massinner}) and ISM (panel~B of Fig.~\ref{fig:co21moms}, i.e.\ $^{12}$CO(2-1) emission corrected for $X_{\rm CO}$, \ion{H}{i} and dust; Section \ref{ssec:massouter}) in concentric annuli with varying position angles and ellipticities, although keeping both fixed does not change our results. For convenience, we then fit both the stellar and the ISM radial mass surface-density profile with a double-S\'{e}rsic function. The fits are carried out using a non-linear least-squares algorithm (\texttt{IDL} \texttt{MPFIT} function; \citealt{Markwardt09}), and the results are shown in the top panel of Fig.~\ref{fig:nsc}. Before comparing the model and data, at each iteration the double-S\'{e}rsic function of the stars is first convolved by the $F$814W PSF and the double-S\'{e}rsic function of the ISM is first convolved by the synthesised beam of our ALMA observations, thus yielding spatially-deconvolved (i.e.\ intrinsic) model (double-S\'{e}rsic function) parameters. We associate the narrow and the broad component with respectively the NSC and the outer stellar disc for the stellar-mass profile, and with the CMC and the ISM disc for the ISM-mass profile. Our best-fitting double-S\'{e}rsic models are good representations of the data, with fractional residuals smaller than $8\%$ ((data-model)/data; see the bottom panel of Fig.~\ref{fig:nsc}). The best-fitting double-S\'{e}rsic parameters and associated total mass estimates are listed in Table~\ref{tab:sersics}.

Our newly-derived NSC total stellar mass ($M_{\rm NSC}$) is an order of magnitude smaller than that reported by \citet{Pechetti20} ($M\approx1.58\times10^8$~M$_\odot$). This large discrepancy may be caused by several factors. First, \citeauthor{Pechetti20}'s (\citeyear{Pechetti20}) spatially-deconvolved photometric fit using the $F$814W image likely suffers from heavy and uncorrected dust extinction in the nucleus. Indeed, they did not correct for dust but applied a mask and interpolated the northern half of the light distribution based on the southern half. Combined with their assumption of a constant and large \ml\ ($M/L\approx5.5$~${\rm M}_\odot/{\rm L}_{\odot}$), this may lead to a more massive stellar component. Second, \citeauthor{Pechetti20}'s (\citeyear{Pechetti20}) NSC has a larger size ($r_{\rm NSC,e}=5.50\pm0.23$~pc), and thus has more mass at large radii ($r>8$~pc), than our newly-derived NSC ($r_{\rm NSC,e}=5.0\pm1.0$~pc; see Fig.~\ref{fig:nsc}). Third, \citeauthor{Pechetti20}'s (\citeyear{Pechetti20}) adopted distance to NGC~3593 is $\approx11$~Mpc \citep{Karachentsev04}, $>50\%$ greater than our adopted distance, and hence they derive a significantly larger NSC mass (see footnote 2). Here, our stellar-mass map (and associated radial stellar-mass surface-density profile) accounts for all these effects and is thus arguably a more accurate representation of the true stellar-mass distribution. 

For comparison, the total stellar mass and effective radius estimated from the three innermost Gaussian components in Table~\ref{tab:star_mges} are $r_{\rm NSC,e}=0\farcs15\pm0\farcs03$ (or $5.0\pm1.0$~pc) and $M_{\rm NSC}=(1.67\pm0.48)\times10^7$~M$_\odot$, fully consistent with those derived from our radial stellar-mass profile (see Table~\ref{tab:sersics}), as expected. As a consistency check, we also verified that the total mass of the stellar disc inferred here ($(1.3\pm0.4)\times10^{10}$~M$_\odot$; see Table~\ref{tab:sersics}) is in agreement with that of the two counter-rotating stellar discs modelled by \citet{Coccato13} ($\approx1.5\times10^{10}$~M$_\odot$; see Section~\ref{sec:ngc3593}), on which our stellar-mass model is ultimately based (see Sections~\ref{ssec:massouter} and \ref{ssec:massinner}). Assuming this total stellar mass to be that of a discy bulge, the bulge then clearly belongs to the sub-M$^\star$ category and has a stellar mass in agreement with those of other Milky Way-like targets. 

\begin{table}
  \caption{Double S\'ersic Fits.}
  \begin{tabular}{lcccc}
    \hline\hline  
    Comp. & $r_{\rm e}\;({\rm pc})$ & $r_{\rm e} \;({\rm arcsec})$ & $n$ & Mass\;($\times10^7$~\Msun)\\
       (1) & (2) & (3)& (4) & (5) \\
    \hline
    \multicolumn{5}{c}{\bf Stars} \\
    \hline
    NSC  & $5.0\pm1.0$ & $0.15\pm0.03$ & $1.0\pm0.1$ & $1.67\pm0.48$ \\
    Disc   & $567\pm30$ & $16.2\pm0.8$  & $1.4\pm0.2$ & $1,275\pm370$ \\
    \hline
    \multicolumn{5}{c}{\bf ISM} \\
    \hline
    CMC  & $11.2\pm2.8$ & $0.32\pm0.08$ & $1.1\pm0.1$ & $0.54\pm0.12$ \\
    Disc   & $444.5\pm3.5$& $12.7\pm0.1$& $1.0\pm0.1$ & $36.5\pm8.7$ \\
    \hline
  \end{tabular}
  \parbox[t]{0.472\textwidth}{\textit{Notes:} Columns~1 to 5 list each component's name, effective (half-light) radius in parsec and arcsecond, S\'{e}rsic index and total mass, respectively.}
  \label{tab:sersics}
\end{table}

We also compare the CMC size and mass inferred from the first S\'{e}rsic component (Table~\ref{tab:sersics}) to those derived from the central Gaussian of the ISM MGE model (Table~\ref{tab:gas_mges}), that are $M_{\rm CMC}=5.2\times10^6$~\Msun\ and $r_{\rm CMC,e}=0\farcs39$ (or 13.3 pc). The CMC mass and size derived from the double-S\'{e}rsic fit are thus in good agreement with those derived from the MGE approach, if a little smaller ($\approx5\%$ larger and $20\%$ smaller, respectively).

The CMC is co-spatial with the NSC, both being located at the centre of the more extended and fainter CND. The CMC is however more extended than the NSC (see Table~\ref{tab:sersics}), suggesting that radiation from the NSC is at least partially shielded by dust and H$_2$. However, high-spatial resolution observations of other targets have often revealed central CO depressions or holes \citep[e.g.][]{Barth16a, Barth16b, Davis17, Boizelle19, North19, Smith19, Nguyen20}. The high incidence of these holes may be caused by the true absence of molecular gas or by changing excitation conditions \citep{Imanishi18, Izumi18}, but most importantly the size of these holes is $\sim R_{\rm SOI}$, that is not spatially-resolved by our ALMA observations of NGC~3593. We therefore cannot rule out the presence of such a hole at the centre of NGC~3593. Higher-angular resolution observations with $\theta_{\rm FWHM}\lesssim\theta_{\rm R_{\rm SOI}}$ are required to better resolve the CMC, and thus establish whether it is a genuine CMC or instead harbours a central hole currently unnoticed due to beam smearing.

\subsection{\Mbh--\Mnsc\ scaling relation}\label{ssec:bhnsc}

\begin{figure}
  \centering\includegraphics[scale=0.15]{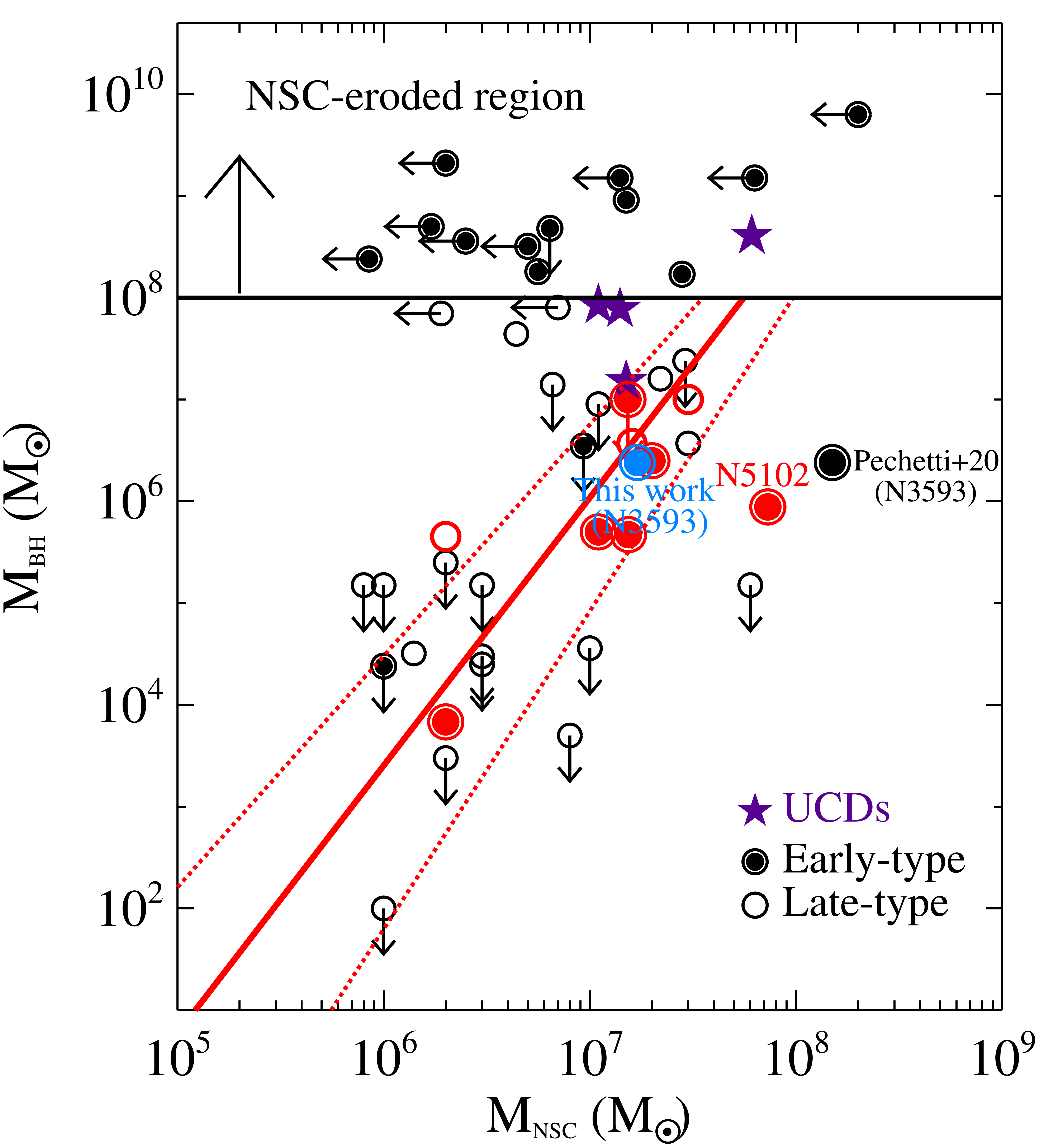}
  \caption{\citeauthor{Graham20}'s (\citeyear{Graham20}) \Mbh--\Mnsc\ scaling relation (red solid line) and its $1\sigma$ uncertainties (red dotted lines). Purple stars and red circles show the inner stellar masses of four well-known UCDs \citep{Ahn17, Ahn18, Afanasiev18, Voggel18} and nine nearby NSCs \citep{Graham09, Schodel09, Lauer12, Lyubenova13, denBrok15, Nguyen17, Nguyen18, Davis20} with dynamical \Mbh\ measurements, respectively. Black circles are NSCs taken from \citet{Neumayer12}. Our own measurement in NGC~3593 is indicated by a cyan circle. Open and filled circles indicate late- and early-type galaxies, respectively. For galaxies with $M_{\rm BH}\gtrsim10^8$~\Msun, the NSCs erode at the expense of the BHs \citep{Bekki10, Neumayer20}.} 
  \label{fig:bhnsc}   
\end{figure}

\citet{Graham20} recently discussed a new correlation between \Mbh\ and \Mnsc\ in low-mass galaxies and UCDs, that combines the \Mnsc--\Mbulge\ and \Mbh--\Mbulge\ (or equivalently \Mnsc--$\sigma_\star$ and \Mbh--$\sigma_\star$) correlations \citep[e.g.][]{Scott13, Graham16, Capuzzo-Dolcetta17, DavisBL19, Sahu19a, Sahu19b}:
\begin{equation*}
  \log\left(\frac{M_{\rm NSC}}{{\rm M}_\odot}\right)=(0.38\pm0.06)\log\left(\frac{M_{\rm
        BH}}{10^{7.89}\,{\rm M}_\odot}\right)+(7.70\pm0.20)~.
\end{equation*}
This \Mbh--\Mnsc\ scaling relation, shown in Fig.~\ref{fig:bhnsc}, works well to predict \Mbh\ or \Mnsc\ in nearby low-mass galaxies and UCDs if either of the two masses is known. All objects are well-studied and nearby, and thus harbour bona fide SMBHs and NSCs with reliably-measured masses. Only the low-mass lenticular galaxy NGC~5102 is an obvious outlier \citep{Caldwell87, Davidge08, Nguyen18, Nguyen19}, owing to its very massive and extended NSC ($r_{\rm NSC,e}\approx26$~pc, $M_{\rm NSC}=7.3\times10^7$~\Msun; \citealt{Nguyen18}). An analogous inconsistency is found for NGC~3593 when we adopt the \citeauthor{Pechetti20}'s (\citeyear{Pechetti20}) NSC mass. However, our own \Mbh\ (Section~\ref{ssec:results}) and \Mnsc\ (Section~\ref{ssec:nsccmc}) estimates are consistent with the \citet{Graham20} \Mbh--\Mnsc\ scaling relation.

The truncation of the \Mbh--\Mnsc\ scaling relation at $M_{\rm BH}\gtrsim10^8$~\Msun\ (and resulting \Mnsc\ upper limits; see Fig.~\ref{fig:bhnsc}) comes from the fact that while NSCs are common in low-mass galaxies ($\approx75\%$ of galaxies with $5\times10^8\lesssim M_\star\lesssim10^{11}$~\Msun\ have NSCs; \citealt{Boker02, Cote06, Seth08a, Seth08b}), more massive SMBHs in more massive galaxies (with $R_{\rm SOI}\gtrsim10$~pc, larger than the NSCs themselves) start to erode their NSCs. Most of the UCDs with dynamically-measured SMBHs are in this ``NSC-eroded region'' and do not follow the \Mbh--\Mnsc\ correlation.

The emerging \Mbh--\Mnsc\ correlation in low-mass galaxies provides a new tool to predict the SMBH mass function at the low-mass end \citep[e.g.][]{Shankar04, Graham07, Vika09, Kelly13}. This in turn will provide critical insight on the origin of UCDs as stellar remnant nuclei of threshed galaxies \citep[e.g.][] {Mieske13, Seth14}, and will improve the accuracy of the predictions of the expected number of tidal disruption events \citep{Stone16, Stone17} and the SMBH--SMBH and SMBH--stellar black hole merging rates, all direct consequences of hierarchical galaxy formation \citep{Voggel19}.

\section{Discussion}\label{sec:discussion}

\subsection{\Mbh--galaxy properties scaling relations in the low-mass regime}\label{ssec:scalingrelation}

\begin{figure*}
  \centering
  \includegraphics[scale=0.55]{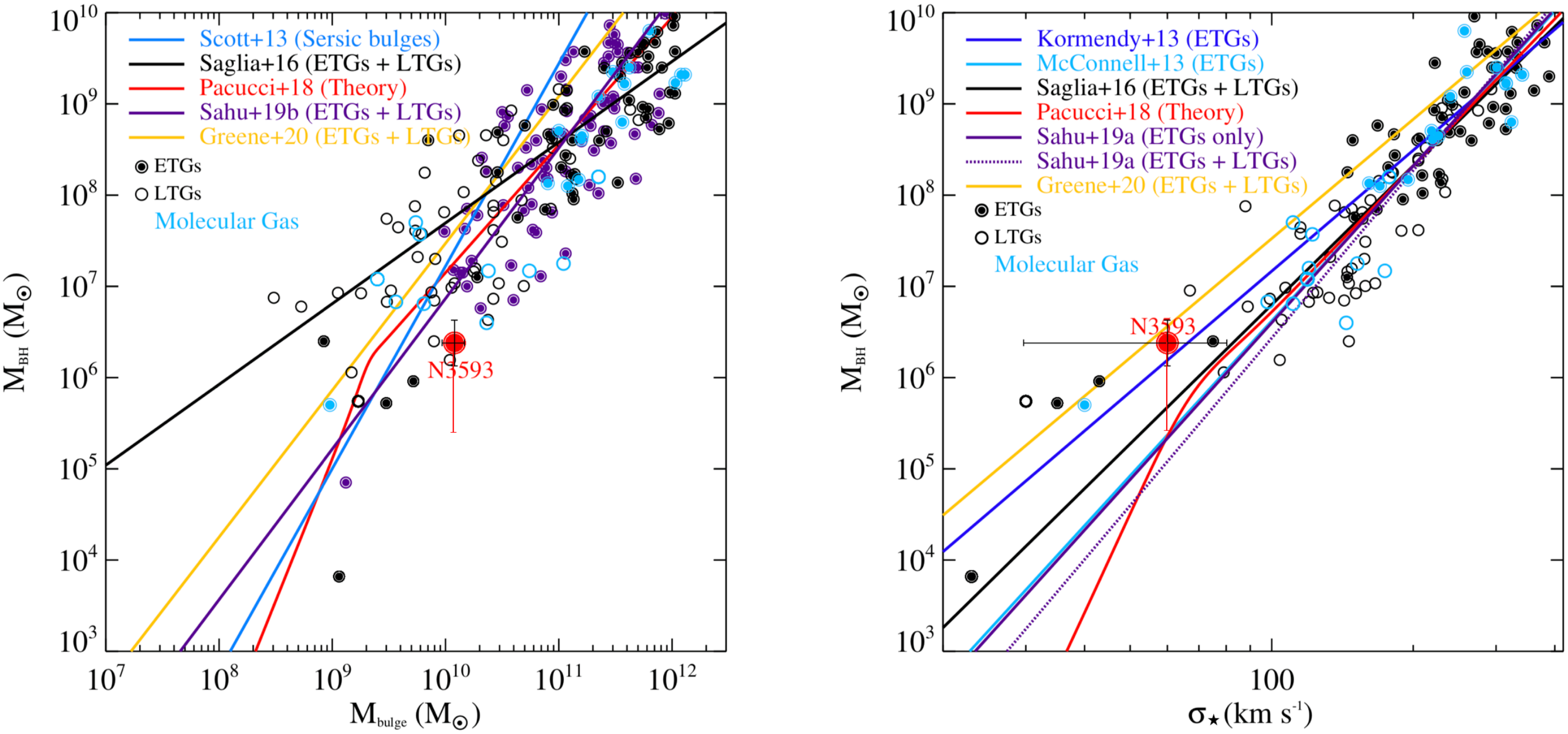}
    \caption{Our NGC~3593 \Mbh\ measurement (red filled circle) in the context of the \Mbh--\Mbulge\ (left panel) and \Mbh--$\sigma_\star$ (right panel) scaling relations. The error bars arising from the $3\sigma$ statistical uncertainties of our default best-fitting model are plotted in black, while those that combine the systematic uncertainties associated with the stellar and/or ISM mass model are plotted in red. $27$ molecular-gas dynamical measurements using both ALMA \citep{Davis17, Davis18, Davis20, Onishi15, Barth16a, Barth16b, Boizelle19, Boizelle20, Combes19, Cohn21, Smith19, Smith21, Nagai19, North19, Nguyen20, Nguyen21a} and CARMA \citep{Davis13, Onishi17} are plotted in cyan, with bulge masses taken from \citet{McConnell13}, \citet{Krajnovic13}, \citet{Salo15}, \citet{Savorgnan16}, \citet{Nguyen17}, \citet{Sani18}, \citet{Nguyen18} and \citet{Nguyen19}. Other measurements taken from \citet{Saglia16} and \citet{Sahu19b} are plotted in black and purple, respectively. Early- and late-type galaxies are indicated with filled and open circles, respectively. The empirical scaling relations of \citet{Scott13}, \citet{Kormendy13}, \citet{McConnell13}, \citet{Saglia16}, \citet{Sahu19a, Sahu19b}, and \citet{Greene20} are overlaid, colour-coded according to the legend. We also overlay the theoretical prediction of a bimodality from \citeauthor{Pacucci18} (\citeyear{Pacucci18}; red line).}
      \label{fig:scaling}   
\end{figure*}

We now consider our NGC~3593 \Mbh\ measurement in the context of various \Mbh--\Mbulge\ \citep[e.g.][]{Scott13, Saglia16, Sahu19b, Greene20} and \Mbh--$\sigma_\star$ \citep[e.g.][]{Kormendy13, McConnell13, Saglia16, Sahu19a, Sahu19b, Greene20} scaling relations, as shown in the left and the right panel of Fig.~\ref{fig:scaling}, respectively, and focusing on the low-mass regime. For the mass of the bulge of NGC~3593, we adopt the disc stellar mass listed in Table~\ref{tab:sersics} (thus excluding the NSC; see also Section~\ref{ssec:nsccmc}), assumed to be that of a discy bulge. For the stellar velocity dispersion of the bulge of NGC~3593, we adopt $\sigma_\star\approx60$~\kms\ from \citet{Bertola96}.

Our best-fitting NGC~3593 \Mbh\ inferred from molecular-gas dynamical modelling is fully consistent with the empirical \Mbh--\Mbulge\ correlation of galaxies with cored central surface-brightness profiles (i.e.\ bulges)  from \citet{Scott13}, the empirical \Mbh--\Mbulge\ correlation of both ETGs and LTGs from \citet{Sahu19b}, and the theoretical \Mbh--\Mbulge\ scaling relation of \citet{Pacucci18}. It is however offset negatively by about half an order of magnitude from the correlations of \citet{Saglia16} and \citet{Greene20}. Our \Mbh\ measurement is also offset negatively by about one order of magnitude from the correlations of \citet{Kormendy13} and \citet{McConnell13}, that are however not shown in Fig.~\ref{fig:scaling} as they are constructed primarily from high-mass galaxies with cuspy profiles (i.e. galaxies without central cores). 

With a mass of $\approx2.4\times10^6$~\Msun, the SMBH of NGC~3593 is at the low-mass end of the \citet{Combes19} sample, and it is the second lowest \Mbh\ measured so far using molecular gas kinematics (the lowest currently being $M_{\rm BH}\approx5\times10^5$~\Msun\ in NGC~404; \citealt{Davis20}). Our measurement of the NGC~3593 SMBH mass also fills in a gap for ETGs between low-mass ($M_{\rm bulge}\leq5\times10^{10}$~\Msun\ and $M_{\rm BH}\leq10^7$~\Msun; e.g.\ \citealt{Nguyen18}) and high-mass ($M_{\rm bulge}>5\times10^{10}$~\Msun\ and $M_{\rm BH}>10^7$~\Msun) targets. It is also becoming clear that the \Mbh--\Mbulge\ correlation has a break at a specific ``transition'' mass \citep{Pacucci17, Nguyen19}. More measurements of $\lesssim10^7$~\Msun\ SMBHs from our MBHBM$^\star$ Project will therefore help to confirm this break and pinpoint the exact transition mass, while simultaneously helping to calibrate \Mbh--galaxy scaling relations across the higher galaxy mass range ($M_{\rm bulge}\gtrsim10^{11}$~\Msun).  

Concerning the \Mbh--$\sigma_\star$ correlation, the mass of the SMBH of NGC~3593 is favouring the \citet{Greene20} correlation over others \citep{Kormendy13, McConnell13, Saglia16, Pacucci18, Sahu19a, Sahu19b}. Indeed, it is offset positively by one half to one order of magnitude from these correlations, similarly to other measurements at low $\sigma_\star$ by \citet{denBrok15}, \citet{Nguyen18} and \citet{Nguyen19}. The reason for these systematic offsets is currently unknown. However, higher-spatial resolution observations of the stars and warm ionised gas with {\it James Webb Space Telescope}, and of the molecular gas \citep[e.g.\ CO;][]{Nguyen20, Nguyen21a} and atomic gas \citep[e.g.\ {\rm [CI](1-0)};][]{Nguyen21a} with ALMA, that can probe the kinematics closer to or within the predicted SOIs, will help shed light on this issue before the era of extremely large ground-based optical telescopes and the next phase of ALMA itself.

\subsection{Extended morphology: inflow or outflow?}\label{ssec:morpho}

As discussed previously, the morphology of the molecular gas in the CND of NGC~3593 suggests the presence of a nuclear disc with a two-arm spiral pattern, i.e.\ bi-symmetric arms extending from the centre to a ring-like structure farther out (see panels~A and B of Fig.~\ref{fig:co21moms}). Such features may be associated with shocks, inflows/outflows and/or filaments in the molecular gas disc, and they have been observed at both millimetre/sub-millimetre wavelengths with ALMA (cold molecular gas; e.g.\ \citealt{Combes13, Combes14, Espada17}) and in the NIR with telescopes such as VLT and Gemini (hot H$_2$; e.g.\ \citealt{Riffel13, Davies14, Diniz15}).

The high-intensity regions visible in the bi-symmetric arms and the extended disc could be caused by the compression of gas along the two-arm spiral pattern. Such features (nuclear spirals and rings) are believed to form in the centres of non-axisymmetric potentials, that cause the gas to lose angular momentum and shock \citep[e.g.][]{Maciejewski00, Emsellem01, Shlosman01, Maciejewski04b, Maciejewski04a, Shlosman05, Fathi07}. The gas then flows into the nucleus along the shocks, thus forming the observed nuclear spirals and rings \citep[e.g.][]{Fathi06, Casasola11, Combes13, Combes14}. However, there is no evidence of a bar in the optical morphology (Fig.~\ref{fig:hstimage}) or the $F$450W--$F$814W colour map (panel~A of Fig.~\ref{fig:colourmaps}) of NGC~3593. Thus, the formation of the observed two-arm spiral pattern in NGC~3593 is unlikely to be driven by a barred potential.

\begin{figure}
  \centering\includegraphics[scale=0.37]{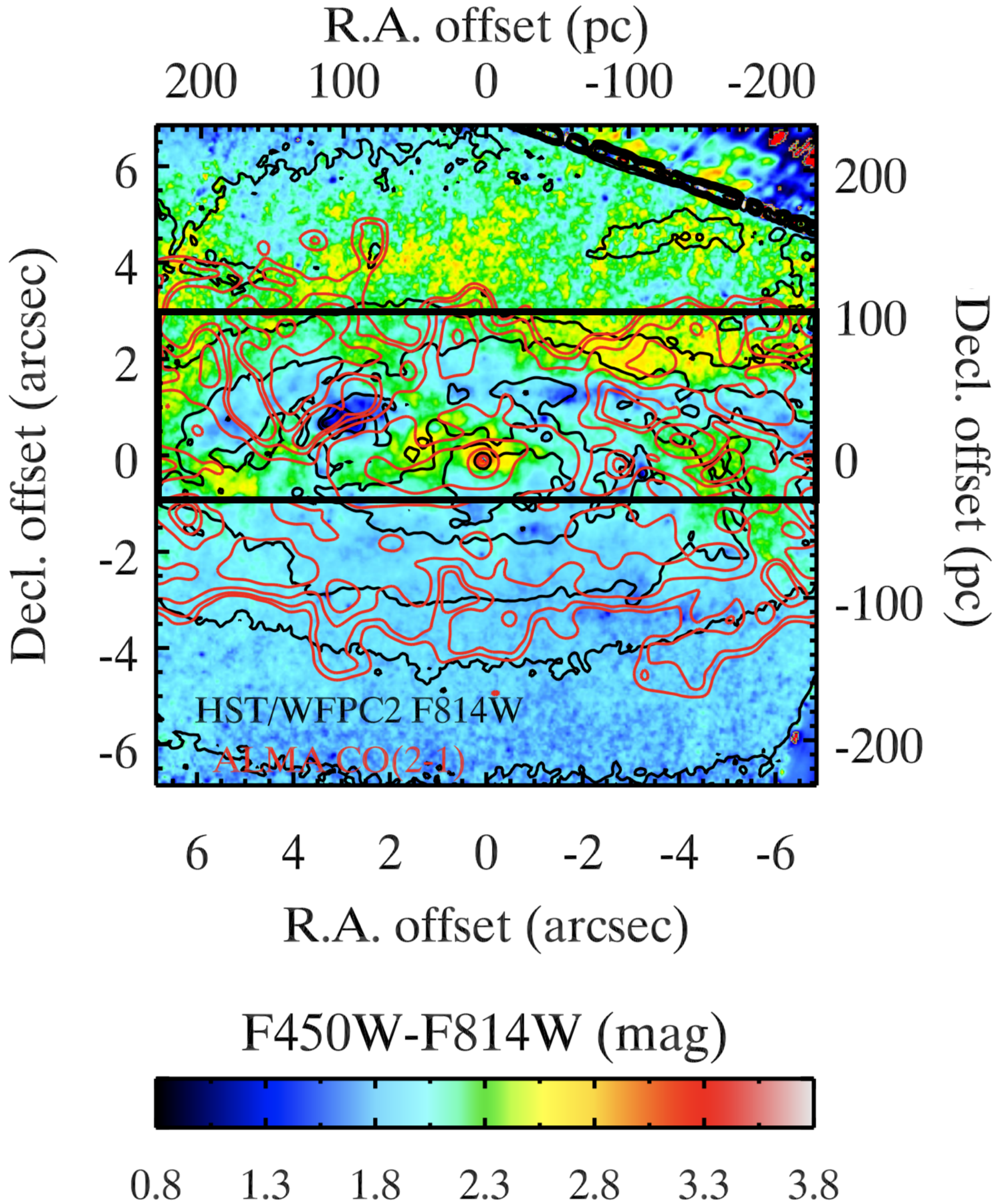}
  \caption{\hst\ WFPC2 $F$450W--$F$814W colour map of the nucleus of NGC~3593 (see also panel~A of Fig.~\ref{fig:colourmaps}), overlaid with $^{12}$CO(2-1) (red) and $F$814W (black) iso-intensity contours. The black rectangle indicates the region with co-spatial molecular gas and dust extinction discussed in the text.} 
  \label{fig:gasdust}   
\end{figure} 

Alternatively, gas from beyond the ring may fall toward the galaxy centre via dynamical friction, potentially also giving rise to a non-axisymmetric potential and thus to the formation of a two-arm spiral. In this scenario, some of the angular momentum from the infalling gas is transferred to the molecular gas disc, creating a shock front and the clumpy high surface-brightness regions \citep[e.g.][]{Malkan98}. The spatial coincidence of these high-intensity, high-velocity dispersion and high-velocity residual regions (see e.g.\ panels~D--L of Fig.~\ref{fig:bestfitmaps2} and panels~D--F of Fig.~\ref{fig:bestfitmaps}) supports this scenario of externally-accreting material, that has not yet fully settled into the host galaxy potential to form a regular and relaxed disc. Supporting evidence in the form of larger-scale filaments and/or warps is however lacking. Future observations with both shorter ($\theta_{\rm FWHM}\gtrsim1\arcsec$) and longer ($\theta_{\rm FWHM}\lesssim0\farcs1$) baselines would be valuable to fully unravel the origin of the nuclear two-arm spiral pattern.

Similarly, both theoretical and observational studies have discussed the role of nuclear spiral arms as bridges to transport molecular gas from kpc scales to nuclei via the removal of angular momentum, possibly also feeding central SMBHs \citep[e.g.][]{Wada02, Maciejewski04a, Maciejewski04b, Fathi06, Fathi11, Fathi13, Casasola08, Hopkins10, vandeVen10, Combes13, Combes14}. On the other hand, nuclear spiral arms can perturb the ambient gas, creating turbulence. This can lead to shock-driven streaming motions and outflows \citep{vandeVen10}, that could explain the features seen in the residual velocity map of the CND of NGC~3593 (panel~F of Fig.~\ref{fig:bestfitmaps2} and \ref{fig:bestfitmaps}).

Although our ALMA data reveal a two-arm spiral structure in the nucleus of NGC~3593, it is weak/flocculent so its dynamical impact on the circumnuclear region remains unclear.  We are therefore unable to firmly conclude that the CND non-circular motions identified through our dynamical modelling can only be explained by either inflows or outflows associated with the observed two-arm spiral structure.

\subsection{Origin of the molecular gas: external accretion?}\label{gasorigin}

The association between the nuclear dust lanes seen in \hst\ optical images and the morphology of the $^{12}$CO(2-1) emission detected in the CND with ALMA confirms the spatial coincidence of dust and molecular gas (for ETGs, see e.g.\ \citealt{Prandoni07, Prandoni10, Young11, Alatalo13, Nyland17}). Fig.~\ref{fig:gasdust} shows the $F$450W--$F$814W colour image overlaid with the $^{12}$CO(2-1) and $F$814W iso-intensity contours. While dust and molecular gas are co-spatial within a $\approx4\arcsec\times14\arcsec$ ($\approx140\times490$~pc$^2$) region indicated by a black rectangle in Fig.~\ref{fig:gasdust}, the faint molecular gas south of the major axis does not trace obvious dust structures. In addition, there is significant dust extinction beyond the detected $^{12}$CO(2-1) emission to the north of the major axis ($>4\arcsec$ or 140 pc), but within the ALMA antennae primary beam (see panel~A of Fig.~\ref{fig:co21moms}), suggesting the presence of more but fainter molecular gas below our detection threshold ($\approx1$~mJy~beam$^{-1}$~per 10~\kms\ binned channel). 

Despite its lenticular type, NGC~3593 harbours a large amount of not only molecular gas ($M_{\rm H_2}=(2.8\pm1.2)\times10^8$~\Msun; Section~\ref{ssec:massouter}) but also atomic gas ($M_{\rm \ion{H}{i}}=1.3\times10^8$~\Msun; \citealt{Pogge93}),  suggesting a recent replenishment of its neutral gas reservoir \citep{Young14, Babyk19, Davis19a}. The origin of neutral gas in ETGs can be either internal (stellar mass loss or gas remaining from the galaxy formation process; e.g.\ \citealt{Davis19b}) or external (most likely accretion during minor mergers of gas-rich dwarfs; see e.g.\ \citealt{Storchi-Bergmann19} for a review).

Given the absence of a stellar bar that could funnel gas to the centre (see Section~\ref{ssec:morpho}), and as the galaxy also hosts lots of young stars that quickly return gas to the ISM \citep[e.g.][]{Davis16, Davis19b}, stellar mass loss appears to be a viable scenario in NGC~3593. As the younger of the two stellar disc is counter-rotating, this gas would itself also be naturally counter-rotating. Having said that, \citet{Davis19a} developed a toy model including a variety of merger \citep[e.g.][]{Couto13, Fischer15, Riffel15, Couto16, Couto17, Raimundo17} and feedback processes to estimate the gas-rich merger rate and thus the mean gas fraction of ETGs in the local universe. They found that only $25\%$ of all low-redshift ETGs harbouring cold gas in clusters are the remnants of transformed LTGs. The remaining $75\%$ could be the results of gas-rich mergers.

Both internal (stellar mass loss) and external (gas-rich minor merger) mechanisms are thus likely to simultaneously take place in NGC~3593. The former is evidenced by the intensive ongoing star formation within the CND, and by the galaxy's early state of transformation from a spiral into an ETG (as a lenticular galaxy). The latter is evidenced by the co-spatial but counter-rotating stellar, ionised-gas and molecular-gas discs, as misalignment of the angular momenta of cold gas and stars is expected in a merger \citep[e.g.][]{Young02, Young08, Crocker11, Davis11, Lagos15}.

\section{Conclusions}\label{sec:conclusions}

We have presented new ALMA observations of $^{12}$CO(2-1) emission in the nucleus of the lenticular galaxy NGC~3593, that in combination with \hst\ optical images and dynamical modelling reveal the presence of a central SMBH. We summarise our results as follows:

\begin{enumerate} 

\item NGC~3593 hosts a highly inclined ($i\approx75\degr$) $^{12}$CO(2-1) CND extending over a region of $\approx30\arcsec\times10\arcsec$ ($\approx1,050\times350$~pc$^2$) elongated along the galaxy major axis, with a two-arm/bi-symmetric spiral pattern surrounded by a ring-like structure of $\approx10\arcsec$ ($\approx350$~pc) radius. The molecular gas distribution and kinematics reveal the CND to be largely dynamically settled, and coincident and co-rotating with the known ionised-gas and secondary stellar disc (all counter-rotating with respect to the primary stellar disc).

\item The $^{12}$CO(2-1) kinematics beyond the nucleus allow us to constrain the outer mass (i.e.\ stellar and ISM mass surface density) profile of the galaxy accurately, useful to appropriately scale the inner stellar-mass profile, that itself has a large impact on our dynamical \Mbh~measurement.

\item Our default best-fitting dynamical model requires a central SMBH mass $M_{\rm BH}=2.40^{+1.87}_{-1.05}\times10^6$~\Msun\, and a stellar-mass scaling factor $\Gamma=0.89^{+0.06}_{-0.03}$, suggesting that our stellar-mass model and choice of a Chabrier IMF are reasonable (statistical uncertainties only, at the $3\sigma$ level). Considering all potential systematic uncertainties associated with the stellar and/or ISM mass model, the SMBH must have a mass in the range $3.0\times10^5$--$4.3\times10^6$~M$_\odot$.

\item The inferred SMBH mass is consistent with the empirical \Mbh--\Mbulge\ correlation of cored galaxies \citep{Scott13} and the recent compilation of \citet{Sahu19b}, but it is almost one order of magnitude below the \citet{Kormendy13} and \citet{Saglia16} scaling relations for more massive galaxies/black holes. Regarding the \Mbh--$\sigma_\star$ correlation, NGC~3593 is consistent with the correlation of \citet{Greene20}, but it is about half an order of magnitude above that of \citet{McConnell13} and one order of magnitude above those of \citet{Saglia16} and \citet{Sahu19b}.

\item Our accurate stellar-mass model yields improved constraints on the NSC, with a total stellar mass \Mnsc\ approximately $10$ times smaller than that derived purely photometrically by \citet{Pechetti20}, thus making our new \Mbh\ and \Mnsc\ consistent with the recent \Mbh--\Mnsc\ scaling relation of \citet{Graham20}.

\item We detect a CMC co-spatial with the NSC and well described by a S\'{e}rsic profile with an effective radius $r_{\rm CMC,e}=11.2\pm2.8$~pc, a S\'ersic index $n_{\rm CMC}=1.1\pm0.1$ and a total ISM mass $M_{\rm CMC}=(5.4\pm1.2)\times10^6$~\Msun.

\item We have identified a few regions of non-circular gas motions in the $^{12}$CO(2-1) CND, likely associated with the two-arm spiral pattern, but it is unclear whether they are leading to any outflow or inflow. The two-arm spiral could have been formed by gas accretion from the outer gas reservoirs ($r\gtrsim15\arcsec$ or 525~pc) via dynamical friction.

\item The significant molecular (and atomic) gas reservoir in a lenticular galaxy like NGC~3593, counter-rotating with respect to the primary stellar disc, suggests a primarily external gas origin via a gas-rich minor merger, possibly associated with internal stellar mass loss in the younger/more recent stellar component.

\end{enumerate}

\section*{ACKNOWLEDGEMENTS}

The authors would like to thank the anonymous referee for their careful reading and useful comments, that helped to improve the paper greatly. D.D.N. would like to thank the International University - Vietnam National University in Ho Chi Minh City, the National Astrononical Observatory of Japan (NAOJ) and the National Institute of Natural Sciences (NINS) for supporting this work. M.B. was supported by the consolidated grants Astrophysics at Oxford ST/H002456/1 and ST/K00106X/1 from the United Kingdom Research Councils. S.T. acknowledges funding from the European Research Council (ERC) under the European Union's Horizon 2020 research and innovation programme under grant agreement No~724857 (Consolidator Grant ArcheoDyn). T.A.D. acknowledges support from Science and Technology Facilities Council (STFC) grant ST/S00033X/1. M.C. expresses his gratitude for a Royal Society University Research Fellowship (RSURF). T.I. and S.B. are supported by Japan Society for the Promotion of Science (JSPS) KAKENHI grant number 17K14247 and 19J00892, respectively. Basic research in radio astronomy at the U.S.\ Naval Research Laboratory is supported by 6.1 Base Funding. The authors also thank Mark D.\ Smith of the University of Oxford for his enlightening discussions on using the {\tt SkySampler} tool.

This paper makes use of the following ALMA data: ADS/JAO.ALMA\#2017.1.00964.S. ALMA is a partnership of ESO (representing its member states), NSF (USA) and NINS (Japan), together with NRC (Canada) and NSC and ASIAA (Taiwan) and KASI (Republic of Korea), in cooperation with the Republic of Chile. The Joint ALMA Observatory is operated by ESO, AUI/NRAO, and NAOJ. The National Radio Astronomy Observatory is a facility of the National Science Foundation operated under cooperative agreement by Associated Universities, Inc. We thank the ALMA operators and staff and the ALMA help desk for diligent feedback and invaluable assistance during the processing of these data. 

{\it Facilities:} ALMA and \hst\ WPFC2.

{\it Software:} \texttt{IDL}, \texttt{CASA}, \texttt{Python}, \texttt{astropy}, 
\texttt{emcee}, \texttt{KinMS}, \texttt{MgeFit}, \texttt{IRAF}, and
\texttt{Kinemetry}.

\section*{Data Avaibility}

The data underlying this article will be shared on reasonable request to the corresponding author. Alternatively, the pipeline calibrated ALMA data is available from the archive with the project code 2017.1.00964.S


\appendix

\section{Supplementary Tables}

\begin{table}
  \caption{Model parameters best fitting the outer part of the $^{12}$CO(2-1) disc.} 
    \begin{tabular}{lcccccc} 
    \hline\hline       
    Parameter & Search range & Best fit & $3\sigma$ uncertainty \\
    (1) & (2) & (3) & (4) \\  
    \hline 
     {\it Stellar mass model:} & & & \\
    $\log(\Sigma_{\star,4\arcsec}/{\rm M}_\odot\,{\rm pc}^{-2})$ & $(1\to8)$ & $4.05$ & $-0.40$, $+0.51$\\[1mm]
    {\it Gas CND:} & & & \\  
    $f$ (Jy~\kms) 								      & $(10^2\to5\times10^3$) & $1049.72$ & $-3.70$, $+2.23$\\
    $i$ ($\degr$) 								      & $(70\to90)$ & $75.00$ & $-2.15$, $+3.73$\\
    $\sigma_0$ (\kms) 							      & $(1\to50)$ & $16.08$ & $-1.17$, $+6.41$\\[1mm]   
    {\it Nuisance:} & & & \\
    $x_{\rm c}$ ($\arcsec$) 						      & $(-1.0\to1.0)$ & $+0.04$ & $-0.42$, $+0.36$\\
    $y_{\rm c}$ ($\arcsec$) 						      & $(-1.0\to1.0)$ & $-0.12$ & $-0.33$, $+0.35$\\
    $v_{\rm off}$ (\kms) 							      & $(-50\to50)$ & $32.75$ & $-2.29$, $+4.58$\\   
    \hline
  \end{tabular}
 \parbox[t]{0.475\textwidth}{\textit{Notes:} Table columns list respectively each parameter name, search range, best fit and uncertainty at the $3\sigma$ confidence level ($0.14$--$99.86\%$ of the PDF). Uncertainties at the $1\sigma$ confidence level ($16$--$84\%$) are shown at the top of the histograms in Fig.~\ref{fig:posterio_pvd}. The parameters $x_{\rm c}$, $y_{\rm c}$ and $v_{\rm off}$ are nuisance parameters defined relative to the adopted galaxy centre $(11^{\rm h}14^{\rm m}37\fs1$, $+12\degr49\arcmin05\farcs6$, $629$~\kms).}
  \label{tab:fit}
\end{table}

\begin{table}
  \caption{Best-fitting model parameters and associated statistical uncertainties.} 
  \begin{tabular}{lcccr} 
    \hline\hline       
    Parameter & Best fit & $1\sigma$ error&  $3\sigma$ error\\
              &  & ($16$--$84\%$) & ($0.14$--$99.86\%$)  \\  
    (1) & (2) & (3) & (4)  & \\  
    \hline
    \multicolumn{4}{l}{{\bf $F$814W mass model without masking:}} \\  
    $\log(M_{\rm BH}/{\rm M}_\odot)$ &  $6.28$  & $-0.09$, $+0.05$ & $-0.30$, $+0.22$ \\
    $\Gamma$                       	        &  $0.86$  & $-0.03$, $+0.05$ & $-0.10$, $+0.16$ \\
    $i$ ($\degr$)                     		&$74.48$ & $-0.18$, $+0.20$ & $-0.58$, $+0.63$ \\[1mm]
  
    \multicolumn{4}{l}{{\bf $F$450W mass model with masking:}}  \\  
    $\log(M_{\rm BH}/{\rm M}_\odot)$ & $6.10$ & $-0.07$, $+0.06$ & $-0.25$, $+0.23$ \\
    $\Gamma$                       & $0.83$ & $-0.05$, $+0.05$ & $-0.20$, $+0.19$ \\
     $i$ ($\degr$)                     & $74.67$ & $-0.15$, $+0.15$ & $-0.48$, $+0.50$ \\[1mm]
    
     \multicolumn{4}{l}{{\bf NSC mass model ($\Sigma_{\star,j=1,2,3}$ from Table~\ref{tab:star_mges}) $\times~1.10$:}} \\ 
    $\log(M_{\rm BH}/{\rm M}_\odot)$ & $5.83$ & $-0.08$, $+0.08$ & $-0.25$, $+0.25$ \\
    $\Gamma$                       & $0.91$ & $-0.04$, $+0.05$ & $-0.12$, $+0.14$ \\
     $i$ ($\degr$)                    & $75.15$ & $-0.34$, $+0.32$ & $-1.18$, $+1.07$ \\[1mm]
   
    \multicolumn{4}{l}{{\bf NSC mass model ($\Sigma_{\star,j=1,2,3}$ from Table~\ref{tab:star_mges}) $\times~1.15$:}} \\ 
    $\log(M_{\rm BH}/{\rm M}_\odot)$ & $5.70$ & $-0.06$, $+0.05$ & $-0.19$, $+0.17$ \\
    $\Gamma$                       & $0.60$ & $-0.03$, $+0.04$ & $-0.09$, $+0.11$ \\
    $i$ ($\degr$)                     & $75.42$ & $-0.33$, $+0.35$ & $-1.25$, $+1.32$ \\[1mm]
    
     \multicolumn{4}{l}{{\bf NSC mass model ($\Sigma_{\star,j=1,2,3}$ from Table~\ref{tab:star_mges}) $\times~1.30$:}} \\ 
    $\log(M_{\rm BH}/{\rm M}_\odot)$ & $\ge5.5$ & $-0.11$, $+0.07$ & $-0.33$, $+0.20$ \\
    $\Gamma$                       & $0.16$ & $-0.03$, $+0.04$ & $-0.09$, $+0.11$ \\
    $i$ ($\degr$)                     & $75.71$ & $-0.33$, $+0.35$ & $-1.25$, $+1.32$ \\[1mm]
    
    \multicolumn{4}{l}{{\bf \citeauthor{Pechetti20}'s (\citeyear{Pechetti20}) NSC mass model:}} \\  
    $\log(M_{\rm BH}/{\rm M}_\odot)$ & $6.49$ & $-0.05$, $+0.02$ & $-0.15$, $+0.07$ \\
    $\Gamma$                       & $0.89$ & $-0.02$, $+0.03$ & $-0.07$, $+0.10$ \\
     $i$ ($\degr$)                     & $74.34$ & $-0.17$, $+0.15$ & $-0.61$, $+0.55$ \\[1mm]
   
    \multicolumn{4}{l}{{\bf Constant \ml\ model:}}  \\  
    $\log(M_{\rm BH}/{\rm M}_\odot)$ & $6.19$ & $-0.05$, $+0.04$ & $-0.20$, $+0.12$ \\
    $\Gamma$                       & $0.80$ & $-0.05$, $+0.05$ & $-0.15$, $+0.16$ \\
     $i$ ($\degr$)                     & $73.86$ & $-0.17$, $+0.25$ & $-0.57$, $+0.78$ \\[1mm]
  
    \multicolumn{4}{l}{{\bf ISM disc model:}} \\  
    $\log(M_{\rm BH}/{\rm M}_\odot)$ & $6.32$ & $-0.06$, $+0.06$ & $-0.20$, $+0.21$ \\
    $\Gamma$                       & $0.90$ & $-0.02$, $+0.02$ & $-0.07$, $+0.07$ \\
     $i$ ($\degr$)                     & $75.05$ & $-0.16$, $+0.15$ & $-0.54$, $+0.52$ \\[1mm]
   
    \multicolumn{4}{l}{{\bf Milky Way $X_{\rm CO}$ model:}} \\  
    $\log(M_{\rm BH}/{\rm M}_\odot)$ & $6.18$ & $-0.06$, $+0.06$ & $-0.18$, $+0.19$ \\
    $\Gamma$                       & $0.82$ & $-0.04$, $+0.04$ & $-0.12$, $+0.12$ \\
     $i$ ($\degr$)                     & $75.35$ & $-0.15$, $+0.16$ & $-0.48$, $+0.50$ \\
    \hline
  \end{tabular}
   \parbox[t]{0.472\textwidth}{\textit{Notes:} Same as Table~\ref{tab:fit2}, but with the CND and other nuisance parameters fixed to their default best-fitting values (as listed in Table~\ref{tab:fit2}).}
  \label{tab:bhbest}
\end{table}

\section{Supplementary Figures}

\begin{figure*}
  \centering\includegraphics[scale=0.57]{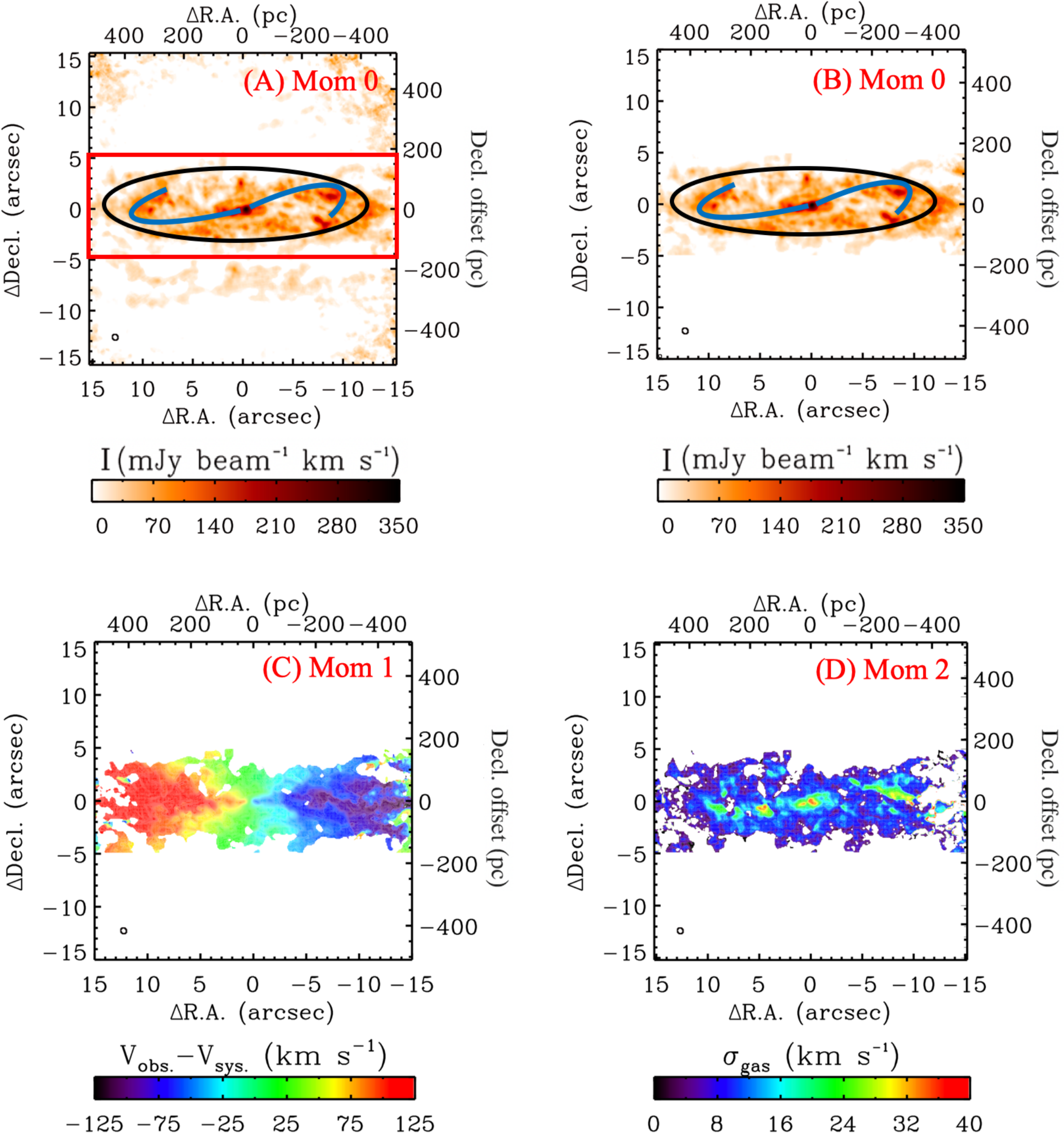}
  \caption{Moment maps of the $^{12}$CO(2-1) molecular gas emission in the central region of NGC~3593. All emission (integrated intensity of the $^{12}$CO(2-1) emission) detected within the primary beam ($r\approx13\arcsec$ or $\approx450$~pc) is shown in panel~A, while panels~B--D only show the central $30\arcsec\times10\arcsec$ ($\approx1050\times350$~pc$^2$; red rectangle in panel~A), roughly corresponding to the CND and including $>85\%$ of the total emission. This is the region used for dynamical modelling and measuring the central \Mbh\ (see Section~\ref{sec:bh}). Emission outside this region is masked out in all analyses. In panels~A and B, the black ellipse and two blue arcs illustrate the outer ring-like structure and two-arm/bi-symmetric spiral pattern of the CND. This pattern extends along the high-integrated intensity ridges co-spatial with the high-velocity dispersion clumps (panel~D). The synthesised beam of $0\farcs33\times0\farcs29$ ($11.6\times10.2$~pc$^2$) is shown as a tiny black ellipse in the bottom-left corner of each panel. The maps are centred on (R.A., decl.)$\,=(11^{\rm h}14^{\rm m}37\fs1$, $+12\degr49\arcmin05\farcs6$).} 
  \label{fig:co21moms} 
\end{figure*}

\begin{figure}
 \centering\includegraphics[scale=0.27]{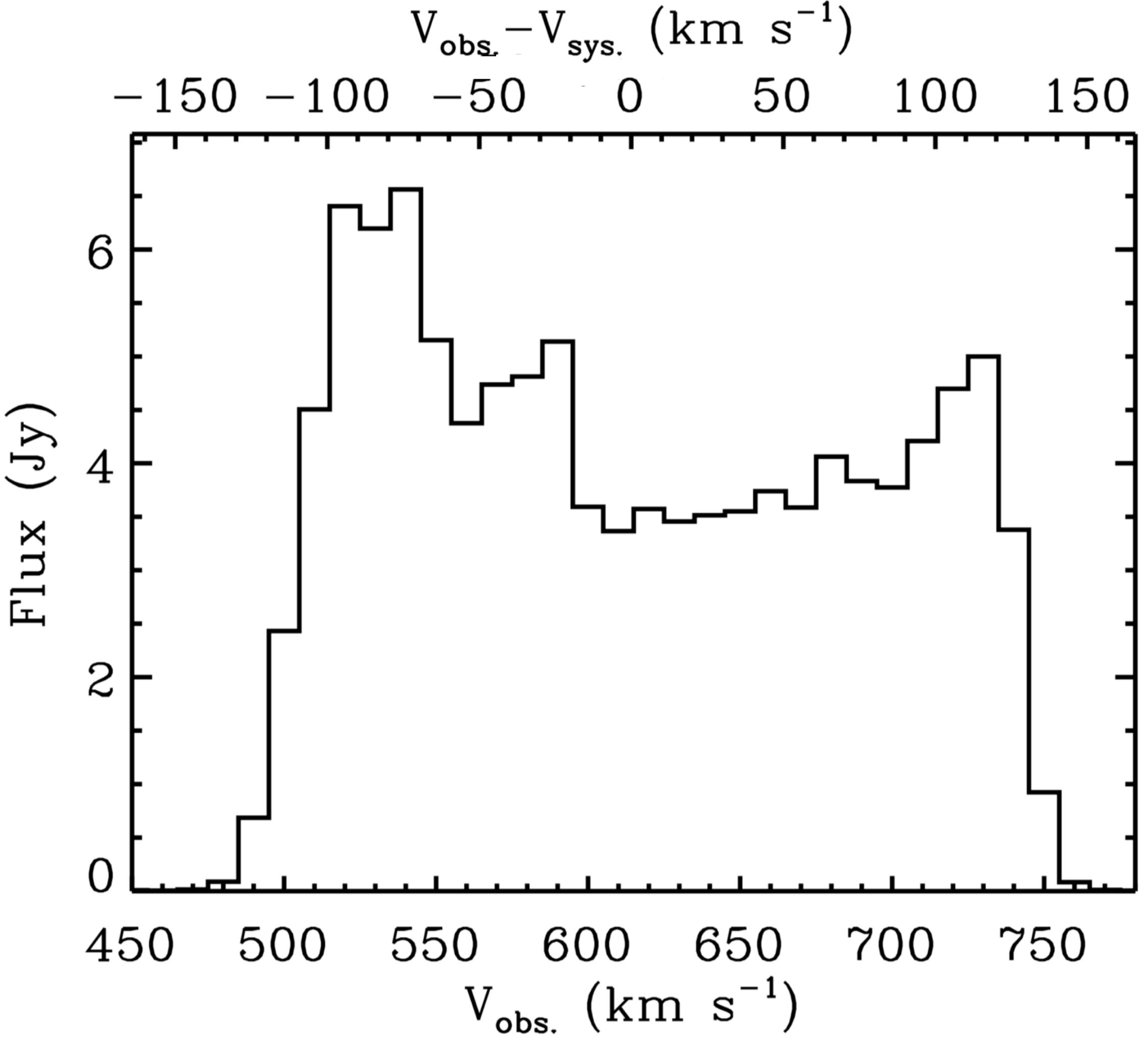}
  \caption{Integrated $^{12}$CO(2-1) spectrum of NGC~3593 extracted  within a region of $30\arcsec\times10\arcsec$ ($\approx1,050\times350$~pc$^2$; see the red rectangle in panel~A of Fig.~\ref{fig:co21moms}), that includes $>85\%$ of the detected emission. The classic symmetric double-horn signature of a rotating disc is clearly revealed.}
      \label{fig:co21spec}  
\end{figure}

\begin{figure}
  \centering\includegraphics[scale=0.3]{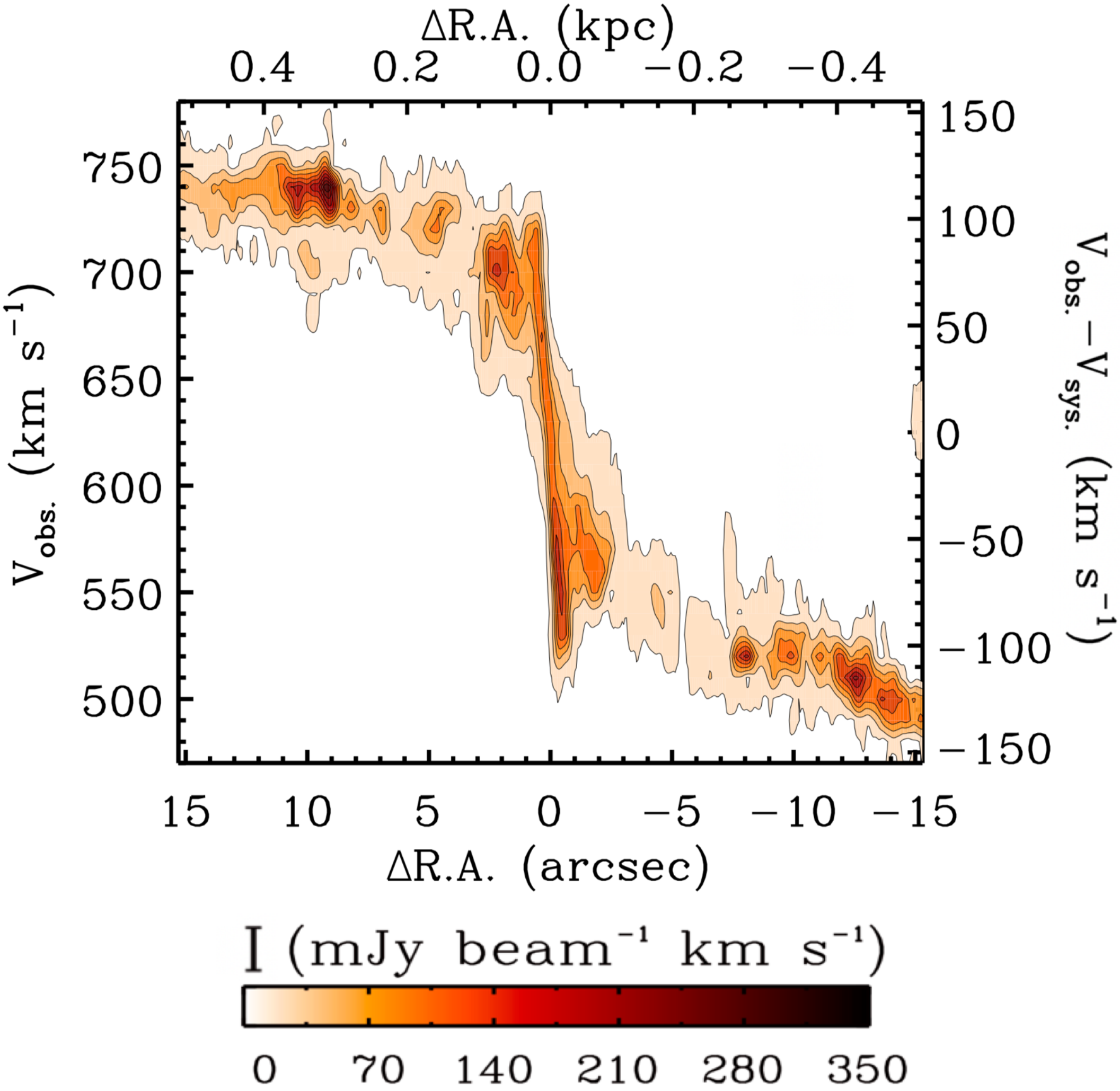}
  \caption{PVD of the $^{12}$CO(2-1) emission of NGC~3593 extracted along the kinematic major axis with a slit of width $0\farcs3$ ($10.5$~pc).} 
  \label{fig:co21pvd0}  
\end{figure}

\begin{figure*}
  \centering\includegraphics[scale=0.84]{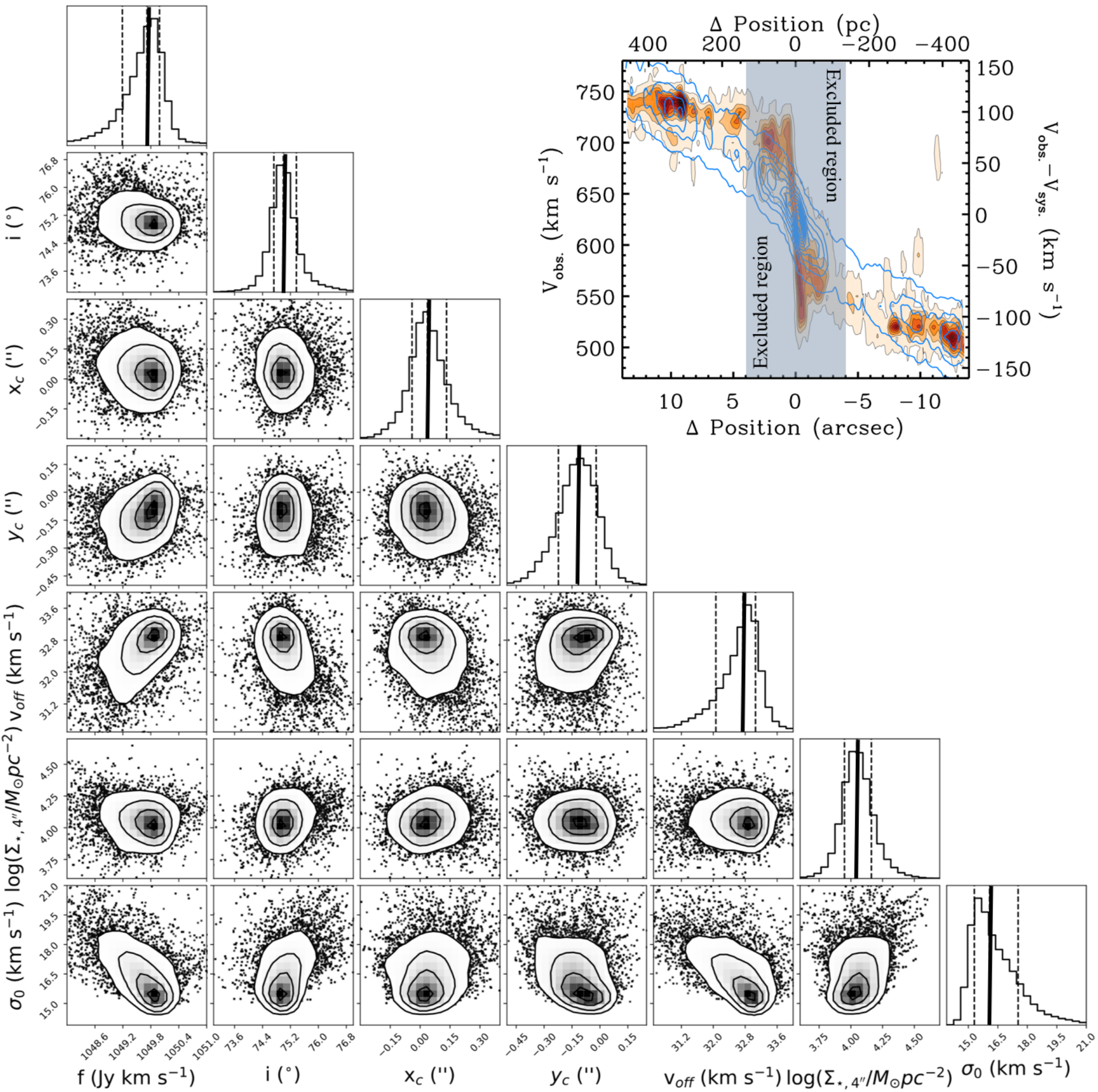} 
  \caption{{\bf Corner plot:} Posterior distributions of the KinMS model parameters fitted to the $^{12}$CO(2-1) kinematics in the outer part ($4\arcsec\leq r\lesssim30\arcsec$ or $140\leq r\lesssim1,050$~pc) of the disc, explored in a Bayesian framework using the MCMC technique. The top panel of each column shows the marginalised PDF of the associated parameter, with the median (also the best fit as the difference is $<3\%$; vertical solid line) and $16$--$84\%$ (i.e.\ $1\sigma$; vertical dashed lines) confidence levels overlaid. The lower panels show the 2D marginalisations with the other model parameters. Out from the centre, the thick solid contours indicate the $0.5\sigma$ ($31$--$69\%$), $1\sigma$ ($16$--$84\%$), $2\sigma$ ($2.3$--$97.7\%$), and $3\sigma$ ($0.14$--$99.86\%$) confidence levels. See Table~\ref{tab:fit} for a quantitative description of the likelihoods of all fitting parameters. $\Sigma_{\star,4\arcsec}$ (only) is presented on a logarithmic scale. {\bf Top-insert plot:} PVD of the $^{12}$CO(2-1) emission of NGC~3593 extracted along the kinematic major axis using a slit of width $0\farcs3$ ($10.5$~pc; orange scale and grey contours), overlaid with that of the model best fitting the outer part ($4\arcsec\leq r\lesssim30\arcsec$; blue contours) of the molecular gas distribution and kinematics (see Section~\ref{ssec:massouter}). The inner $4\arcsec$ region excluded in the fit is shaded in gray.}  
      \label{fig:posterio_pvd}   
\end{figure*}

\begin{figure*}
  \includegraphics[scale=0.73]{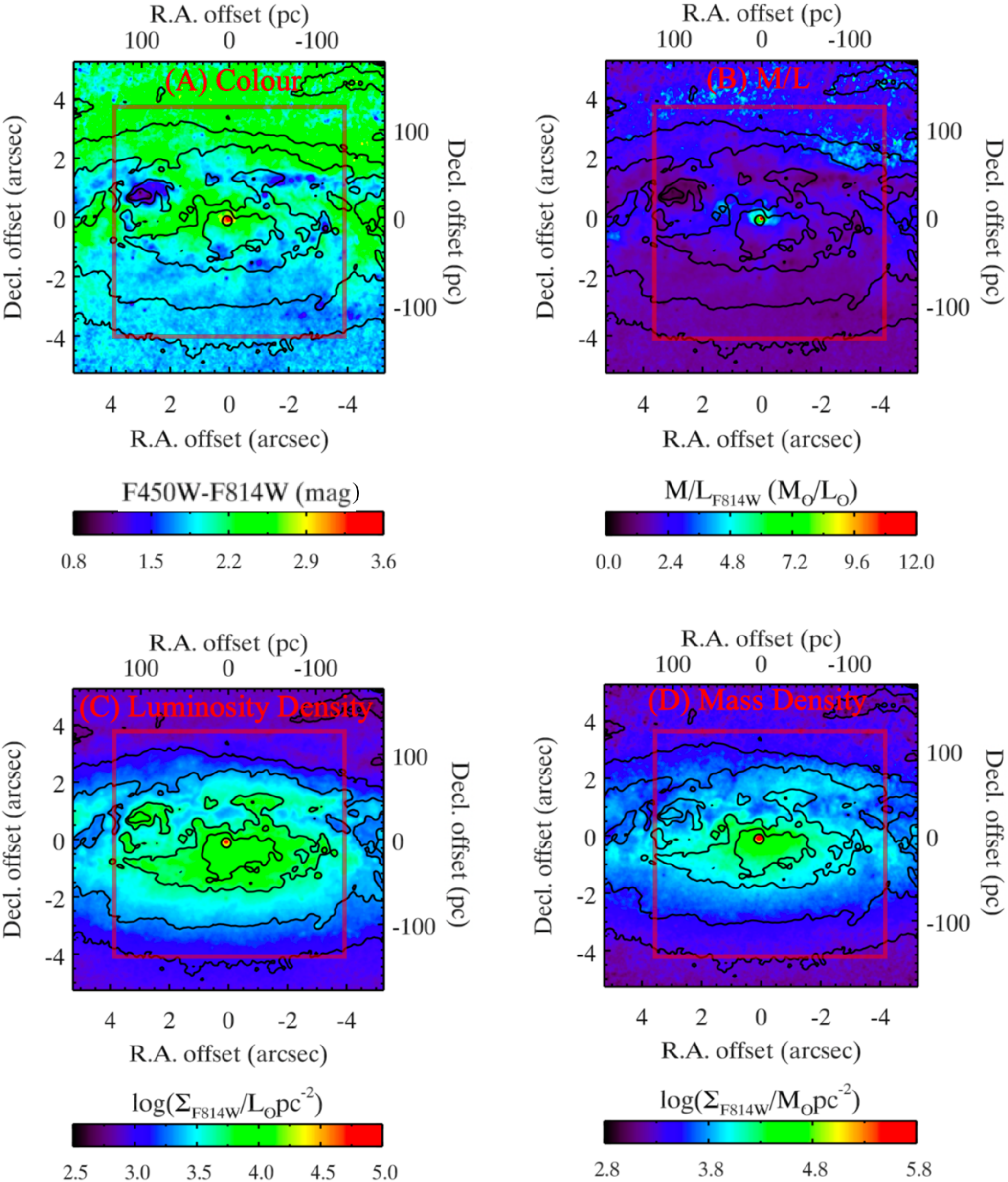} 
  \caption{Inner stellar-mass model of NGC~3593. {\bf Panel~A:} \hst/WFPC2 $F$450W--$F$814W colour map. {\bf Panel~B:} $F$814W-band mass-to-light ratio (\ml$_{F{\rm814W}}$) map, constructed using the \citet{Roediger15} colour--\ml\ relation. {\bf Panel~C:} $F$814W-band luminosity surface-density map. {\bf Panel~D:} stellar-mass surface-density map, generated by multiplying the $F$814W-band luminosity surface-density map (panel~C) with the \ml$_{F{\rm814W}}$ map (panel~B). This map is fit with an MGE model (Fig.~\ref{fig:massmodelcontour}), and the resulting inner stellar-mass surface-density model is truncated and scaled to the outer stellar-mass surface-density model ($4\arcsec\le r\lesssim30\arcsec$; see Section~\ref{ssec:massouter}) to create the combined stellar-mass surface-density model (see Section~\ref{ssec:mass} and Fig.~\ref{fig:massmodel1d}). In all panels, contours show the $F$814W-band isophotes at $\mu_{F\rm{814W}}=16.0$, $16.5$, $17.0$, $18.0$ and $19.0$~mag~arcsec$^{-2}$. The red square in each panel delineates the inner mass region ($r<4\arcsec$; see Section~\ref{ssec:massinner}).}  
  \label{fig:colourmaps}   
\end{figure*}

\begin{figure*}
  \centering\includegraphics[scale=0.4]{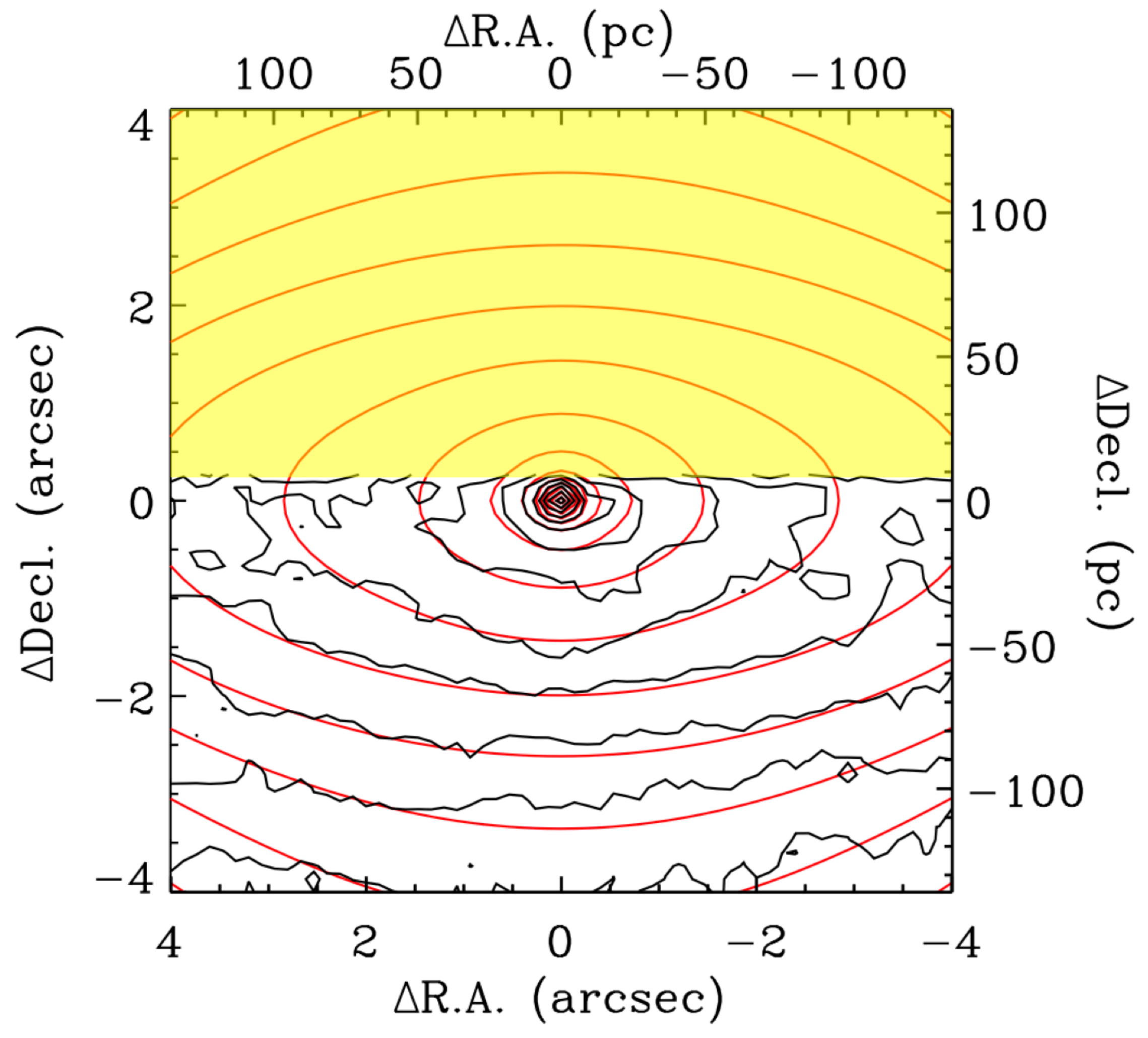} 
  \caption{Comparison between the stellar-mass surface-density map of NGC~3593 (black contours; panel~D of Fig.~\ref{fig:colourmaps}) and its best-fitting MGE parametrisation (red contours). The yellow region shows the pixels strongly affected by dust extinction, and thus excluded from the fit, on the northern side of the galaxy. Only the inner region ($4\arcsec\times4\arcsec$ or $140\times140$~pc$^2$) is shown, corresponding to the red squares in the panels of Fig.~\ref{fig:colourmaps}.}
  \label{fig:massmodelcontour}   
\end{figure*}

\begin{figure*}
  \centering\includegraphics[scale=0.40]{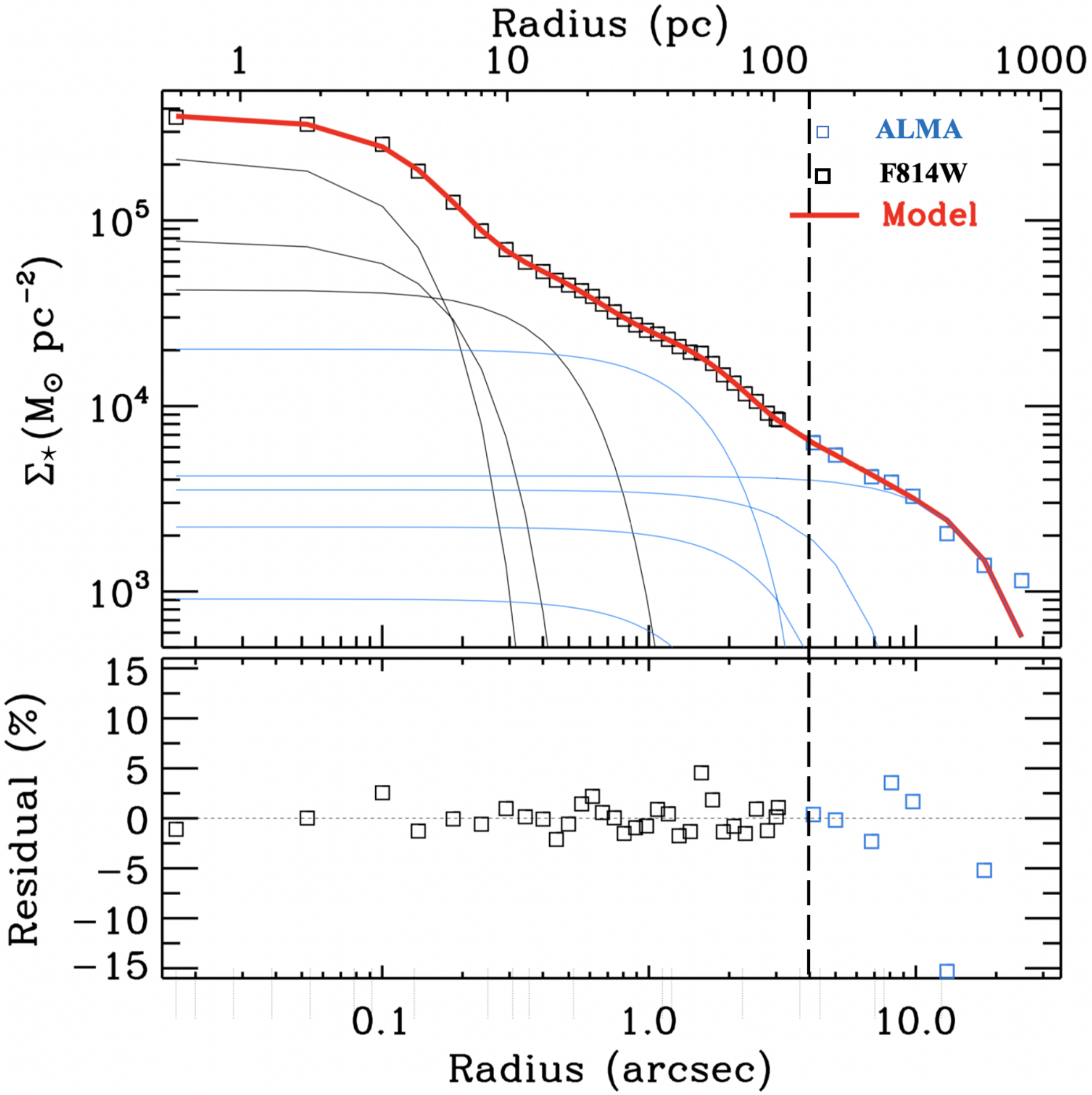} 
  \caption{{\bf Top:} Major-axis cut of the combined stellar-mass (stellar-mass surface density) model, constrained using \hst\ (black open squares) and ALMA (blue open squares) data (separated by a black vertical dashed line at a radius $r=4\arcsec$), overlaid with the best-fitting MGE model (thick red solid line), itself the summation of multiple Gaussian components listed in Table~\ref{tab:star_mges} (thin black and blue lines for the inner and outer components, respectively). {\bf Bottom:} fractional residuals ((data-model)/data).} 
      \label{fig:massmodel1d}   
\end{figure*}

\begin{figure*}
  \centering\includegraphics[scale=0.525]{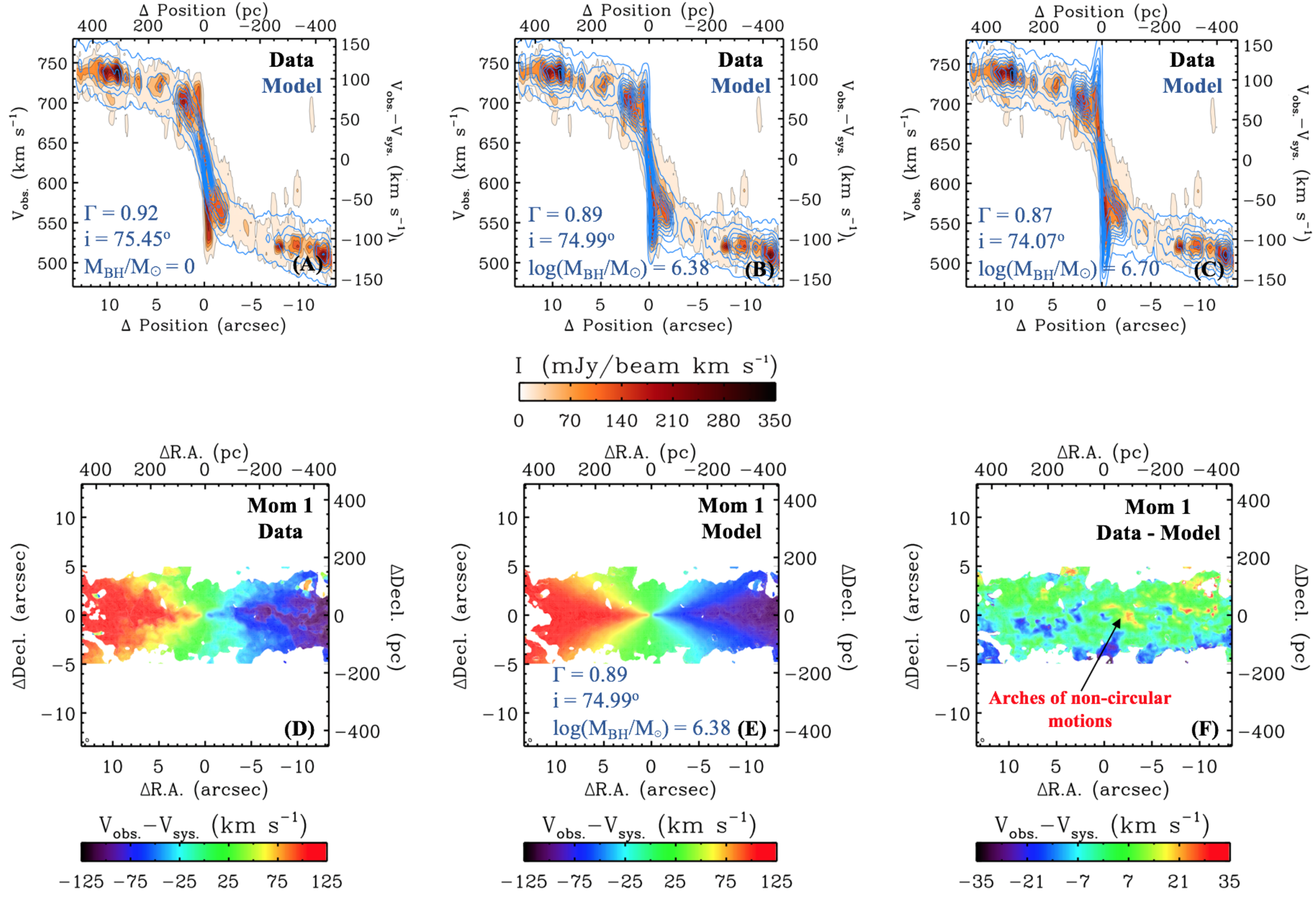}\\
  \vspace{1mm}
  \centering\includegraphics[scale=0.525]{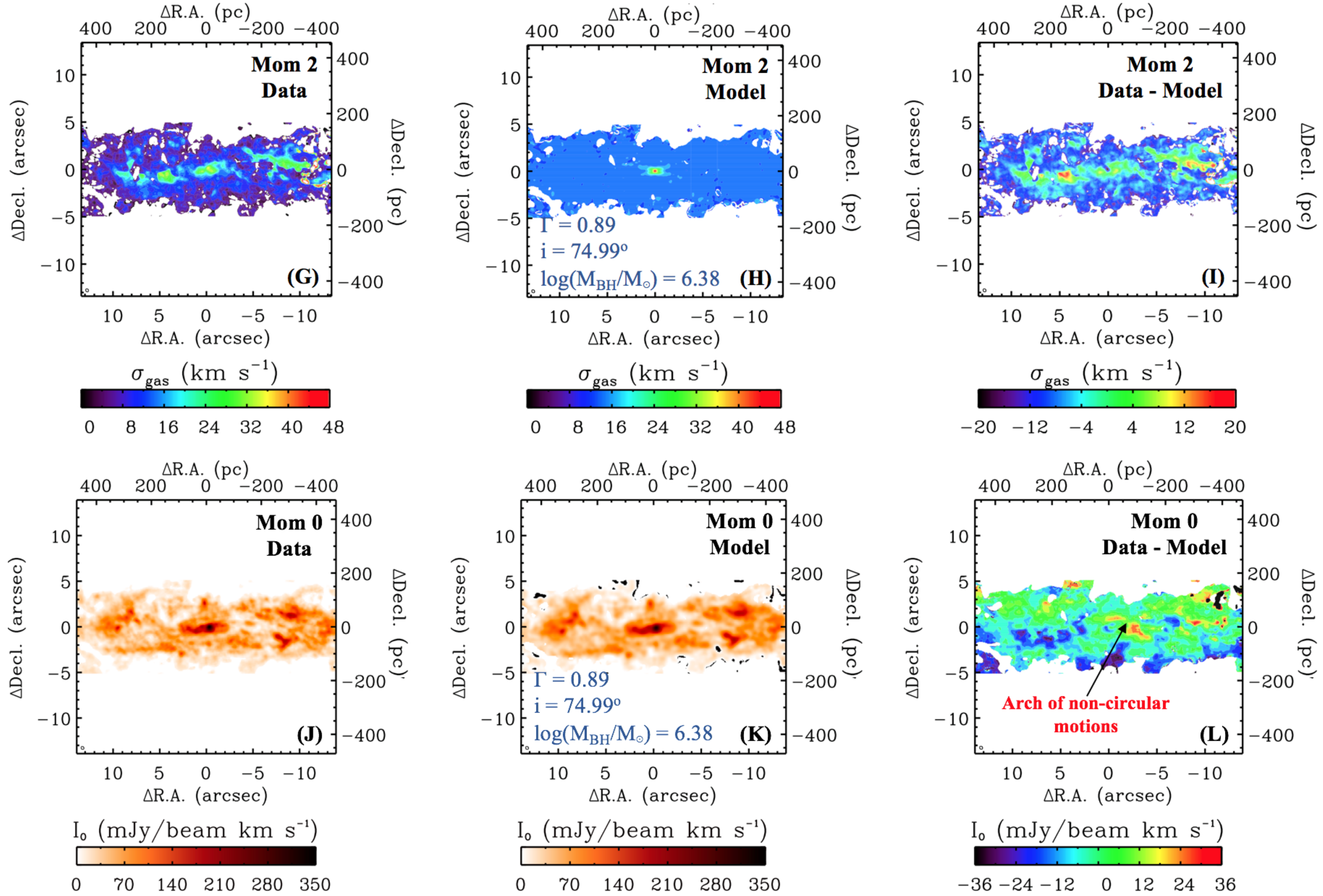} 
  \caption{{\bf First row:} PVD of the $^{12}$CO(2-1) emission of NGC~3593 extracted along the kinematic major-axis (orange scale and grey contours), overlaid with the modelled PVD (blue contours) of the best-fitting model with no SMBH (panel~A; \Mbh$\,=0$, $\Gamma=0.92$ and $i=74\fdg45$), the best-fitting model with a SMBH (panel~B; \Mbh$\,=2.40\times10^6$~\Msun, $\Gamma=0.89$ and $i=74\fdg99$), and a model with an overly massive SMBH (panel~C; \Mbh$\,=5.02\times10^6$~\Msun, $\Gamma=0.87$ and $i=74\fdg07$). {\bf Second row:} Mean LOS velocity map of the observed $^{12}$CO(2-1) emission of NGC~3593 (panel~D) and the best-fitting model with a SMBH (panel~E), and residuals between the two (panel~F; data-model). {\bf Third/fourth row:} As for the second row, but for the LOS velocity dispersion (panels~G, H, and I) and the integrated intensity (panels~J, K, and L) of the $^{12}$CO(2-1) emission of NGC~3593. The synthesised beam of $0\farcs33\times0\farcs29$ ($11.6\times10.2$~pc$^2$) is shown as a tiny black ellipse in the bottom-left corner of each map.}
  \label{fig:bestfitmaps2}   
\end{figure*}

\begin{figure*}
  \centering\includegraphics[scale=0.84]{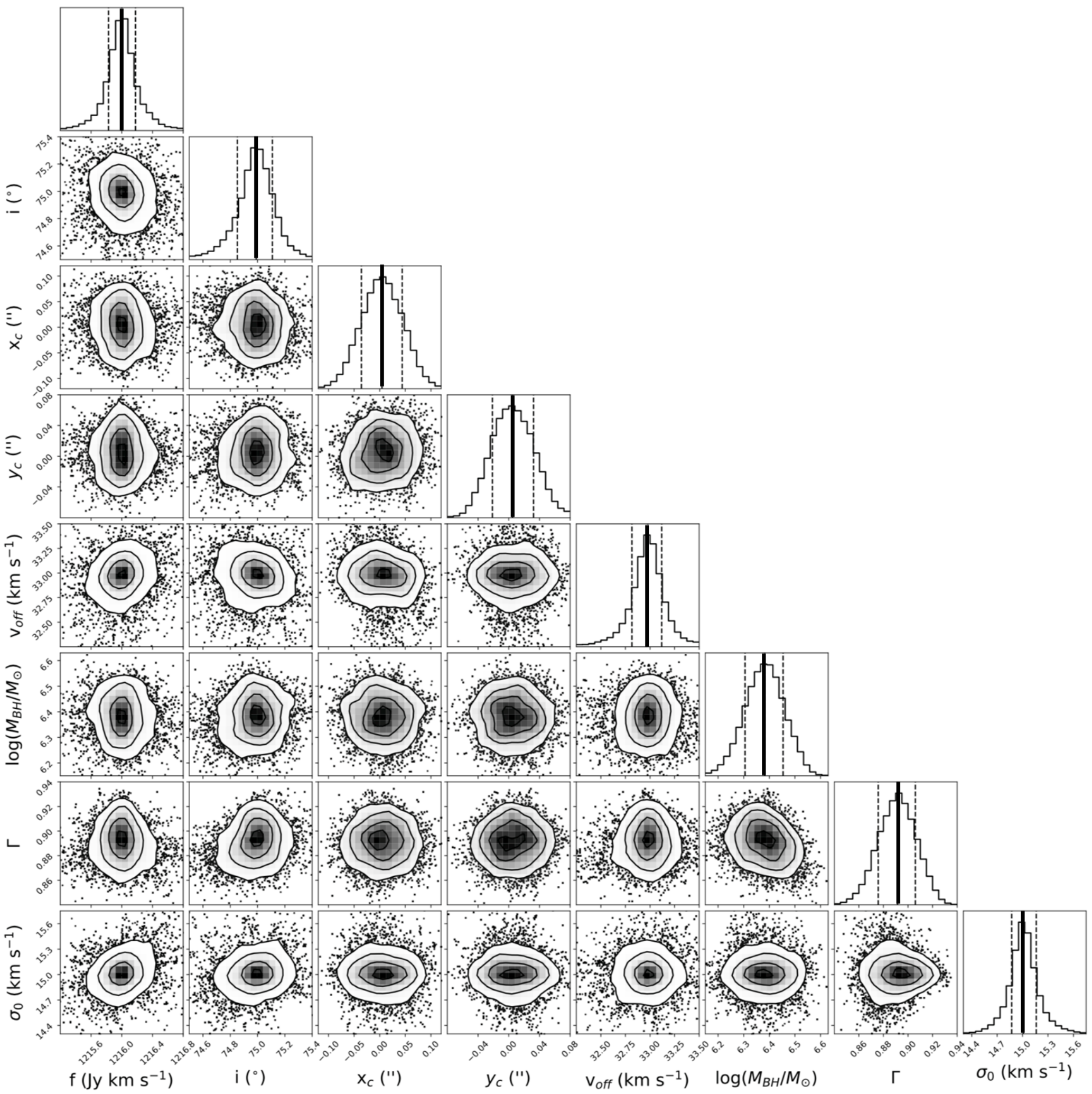} 
  \caption{As Fig.~\ref{fig:posterio_pvd}, but for the KinMS model parameters fitted to the $^{12}$CO(2-1) kinematics in the inner part ($4\arcsec\times4\arcsec$ or $140\times140$~pc$^2$) of the disc. See Table~\ref{tab:fit2} for a quantitative description of the likelihoods of all fitting parameters. \Mbh\ (only) is presented on a logarithmic scale. For illustrations of how well this model describes the data, see Figs~\ref{fig:bestfitmaps2} and \ref{fig:bestfitmaps}.}  
      \label{fig:posterial_full}   
\end{figure*}

\bsp	
\label{lastpage}
\end{document}